\documentclass[12pt]{iopart}
\usepackage{fullpage,graphicx,epsf,ulem}
\usepackage{iopams}
\usepackage{amsfonts,amssymb}
\usepackage{amscd}
\usepackage{cite}
\usepackage{color}
\usepackage{graphicx}
\usepackage{epsfig}
\usepackage[english]{babel}
\usepackage{amsfonts}
\usepackage{soul}
\begin{document}

\def\th{\theta}
\def\l{\label}
\def\p{\partial}
\def\be{\begin{equation}}
\def\ee{\end{equation}}
\def\bea{\begin{eqnarray}}
\def\eea{\end{eqnarray}}
\def\bef{\begin{figure}}
\def\eef{\end{figure}}
\def\bml{\begin{mathletters}}
\def\eml{\end{mathletters}}
\def\l{\label}
\def\b{\beta}
\def\no{\nonumber}
\def\fr{\frac}
\def\o{\omega}
\def\O{\omega}
\def\p{\partial}
\def\n{\nabla}
\def\a{\alpha}
\def\b{\beta}
\def\eps{\epsilon}
\def\g{\gamma}
\def\d{\delta}
\newcommand{\dd}{\mbox{d}}
\title{Kuramoto model of synchronization: Equilibrium and nonequilibrium
aspects}
\author{Shamik Gupta$^{1}$, Alessandro Campa$^2$, and Stefano Ruffo$^3$}
\address{$^1$ Laboratoire de Physique Th\'{e}orique et Mod\`{e}les
Statistiques (CNRS UMR 8626), Universit\'{e} Paris-Sud, Orsay, France}
\address{$^2$ Complex Systems and Theoretical Physics Unit, Health and Technology
Department, Istituto Superiore di Sanit\`{a}, and INFN Roma1, Gruppo
Collegato Sanit\`{a}, Viale Regina Elena 299,
00161 Roma, Italy}
\address{$^3$ Dipartimento di Fisica e Astronomia and CSDC,
Universit\`{a} di Firenze, INFN and CNISM, via G. Sansone, 1 50019 Sesto
Fiorentino, Italy}
\ead{shamikg1@gmail.com,alessandro.campa@iss.infn.it,stefano.ruffo@gmail.com}

\begin{abstract}
The phenomenon of spontaneous
synchronization, particularly within the framework of the Kuramoto
model, has been a subject of intense research over the years.
The model comprises oscillators with distributed natural
frequencies interacting through a mean-field coupling, and serves as a paradigm to study synchronization. In
this review, we put forward a general framework in which we discuss in a
unified way known results with more recent developments obtained for a
generalized Kuramoto model that includes inertial effects and noise. We describe the model from a
different perspective, highlighting the long-range nature of the
interaction between the oscillators, and emphasizing the equilibrium and out-of-equilibrium
aspects of its dynamics from a statistical physics point of view. In this review, we first introduce the
model and discuss both for the noiseless and noisy dynamics and for unimodal frequency distributions the synchronization
transition that occurs in the stationary state. We then introduce the generalized model, and analyze its dynamics using tools
from statistical mechanics. In particular, we discuss its
synchronization phase diagram
for unimodal frequency distributions. Next, we describe
deviations from the mean-field setting of the Kuramoto model. To this
end, we consider the generalized Kuramoto dynamics on a one-dimensional
periodic lattice on the sites of
which the oscillators reside and interact with one another with a
coupling that decays as an inverse power-law of their separation along the lattice. For two specific cases, 
namely, in the absence of noise and inertia, and in the case when the natural
frequencies are the same for all the oscillators, we discuss how the
long-time transition to synchrony is governed by the
dynamics of the mean-field mode (zero Fourier mode) of the spatial distribution of
the oscillator phases.
\end{abstract}
\date{\today}
\maketitle
\tableofcontents
\section{Introduction}
\l{intro}
A remarkable phenomenon quite ubiquitous in nature is that of collective
synchronization, in which a large population of coupled oscillators spontaneously
synchronizes to oscillate at a common frequency,
despite each constituent having a different natural frequency
of oscillation \cite{Pikovsky:2001}. One witnesses such a spectacular cooperative effect in
many physical and biological systems over length and time scales
that span several orders of magnitude. Some common examples are  
metabolic synchrony in yeast cell suspensions \cite{Bier:2000},
synchronized firings of cardiac pacemaker cells \cite{Winfree:1980},
flashing in unison by groups of fireflies \cite{Buck:1988}, voltage
oscillations at a common frequency in an array of current-biased
Josephson junctions \cite{Wiesenfeld:1998}, phase synchronization in
electrical power distribution networks \cite{Filatrella:2008,Rohden:2012,Dorfler:2013},
rhythmic applause \cite{Neda:2000}, animal flocking behavior
\cite{Ha:2010}; see \cite{Strogatz:2003} for a survey.

The Kuramoto model provides a simple theoretical framework to study how
synchronization may emerge spontaneously in the dynamics of a many-body
interacting system \cite{Kuramoto:1975,Kuramoto:1984}. The model
comprises globally-coupled oscillators of distributed natural frequencies that
are interacting via a mean-field coupling through the sine of their phase differences, with
the phases following a first-order dynamics in time. Over
the years, many aspects of the model, including applications cutting
across disciplines, from physical and biological to even social
modelling, have been considered in the literature
\cite{Strogatz:2000,Acebron:2005}.

An early motivation behind studying the Kuramoto model was to
explain the spectacular phenomenon of spontaneous synchronization among fireflies: In parts of south-east Asia, thousands of male fireflies gather in trees at
night and flash on and off in unison. In this respect,
focussing on fireflies of a particular species (the {\it Pteroptyx
mallacae}), a study due to Ermentrout revealed that the approach to
synchronization from an initially unsynchronized state is faster in the
Kuramoto setting than in reality \cite{Ermentrout:1991}.
Ermentrout proposed a
route to reconciliation by elevating the first-order dynamics of the Kuramoto model
to the level of second-order dynamics. Including also a Gaussian noise
term that accounts for the stochastic fluctuations of the natural
frequencies in time \cite{Sakaguchi:1988}, one arrives at a generalized Kuramoto model
including inertia and noise, in which oscillator phases have a
second-order dynamics in time \cite{Acebron:1998,Hong:1999,Acebron:2000}. One can
prove that the resulting dynamics leads to a nonequilibrium stationary
state (NESS) at long times \cite{Gupta:2014}.

Study of NESSs is an active area of research of modern day statistical
mechanics \cite{Privman:1997}. Such states are characterized by a violation of detailed
balance leading to a net non-zero probability current around a closed
loop in the configuration space. One of the primary challenges in this field is to
formulate a tractable framework to analyze nonequilibrium systems on a common footing,
similar to the one due to Gibbs and Boltzmann that has been established for equilibrium
systems \cite{Derrida:2005}.

In a different context than that of coupled oscillators, the dynamics of
the generalized Kuramoto model also describes a long-range
interacting system of particles moving on a unit circle under the
influence of a set of external drive in the form of a quenched external
torque acting on the individual
particles, in the presence of noise. With the noise, but without the external torques, the resulting model is the so-called Brownian mean-field
(BMF) 
model \cite{Chavanis:2011,Chavanis:2013}, introduced as a generalization of the celebrated Hamiltonian
mean-field (HMF) model that serves as a prototype to study statics and dynamics
of long-range interacting systems \cite{Inagaki:1993,Ruffo:1995}. In recent years, there has been a surge in interest in studies of systems with
long-range interactions. In these systems, the
inter-particle potential in $d$ dimensions decays at large separation
$r$ as $r^{-\alpha}$, with $0 \le \alpha \le d$
\cite{Campa:2009,Gupta:2010}. Examples are gravitational systems
\cite{Chavanis:2006}, plasmas \cite{Escande:2010}, two-dimensional
hydrodynamics \cite{Bouchet:2012}, charged and dipolar systems
\cite{Bramwell:2010}, etc.  
Unlike systems with short-range interactions, long-range interacting
systems are generically non-additive, implying that dividing the
system into macroscopic subsystems and summing over their
thermodynamic variables such as energy do
not yield the corresponding variables of the whole system.
Non-additivity leads to many significant thermodynamic and
dynamical consequences, such as negative microcanonical specific heat,
inequivalence of statistical ensembles, and others, which are unusual with short-range
interactions \cite{Campa:2009}. 

In this review, starting with the first-order mean-field dynamics of
the original Kuramoto model, we progressively modify the dynamics by including
first the effects of a Gaussian noise, and then the consequences of an
inertial term that makes the dynamics second order in time. In each
case, we discuss the possible transitions to synchrony that the
resulting stationary state exhibits. Here, we will explicitly consider a
unimodal distribution of the natural frequencies. While the derivation of
the phase diagram in the original model is based on an insightful
self-consistent approach due to Kuramoto, inclusion of Gaussian noise allows to
employ usual tools of statistical mechanics and explicitly study the
evolution of the phase space distribution by using a Fokker-Planck
approach. In both these cases, the transition
between the unsynchronized and the synchronized phase turns out to be
continuous or second order. We conveniently study the dynamics of the generalized Kuramoto
model that includes the effects of both inertia and noise by introducing a reduced parameter space involving
dimensionless moment of inertia, temperature, and width of the frequency
distribution. We point out the relation of the model to the BMF model,
thereby making references to the literature on
long-range interacting systems. We give a rigorous proof that the system at long times
settles into a NESS unless the width of the frequency distribution is zero when it has an equilibrium
stationary state. We highlight that the generalized dynamics exhibits a
nonequilibrium first-order transition from a synchronized phase at low
parameter values to an unsynchronized phase at high values. As a result, the
system as a function of the transition parameters switches over in a
discontinuous way from one phase to another, thereby mimicking an abrupt
off-on switch. This may be contrasted to the case of no inertia when the transition is continuous. In
proper limits, we discuss how one may recover the known continuous phase
transitions in the Kuramoto model and in its noisy extension, and an
equilibrium continuous transition in the BMF model. The present approach
offers a complete and consistent picture of the phase diagram, unifying
previous results with new ones in a common framework. In the last part
of the review, we consider deviations from the mean-field setting of the
Kuramoto model. To this end, we analyze the generalized Kuramoto
dynamics on a one-dimensional periodic lattice on the sites of which the oscillators reside and
interact with a coupling that decays as an inverse power-law of their
separation along the lattice. We consider two specific cases of the
dynamics, namely, in the absence of noise and inertia, and in the case when the natural
frequencies are the same for all the oscillators (giving rise to the so-called $\alpha$-HMF
model). For the latter case, we consider both overdamped and underdamped
dynamics. In particular, we discuss how the
long-time transition to synchrony is governed by the
dynamics of the mean-field mode (zero Fourier mode) of the spatial distribution of
the oscillator phases. In this review, besides extensive numerical simulations, aspects of
phase diagram are derived analytically by performing a linear stability
analysis of the mean-field incoherent stationary state. Moreover, for the
case of the overdamped dynamics of the generalized Kuramoto
model on the lattice with the same natural
frequency for all the oscillators, we present analytical
results also on the linear stability analysis of the mean-field synchronized stationary
state. We end the review with conclusions and
perspectives.

\section{Kuramoto model of globally coupled oscillators}
\l{Kuramoto-bare}
We start with a derivation of the dynamics of the Kuramoto model by
following Ref. \cite{Kuramoto:1975}. Consider
first a single Landau-Stuart oscillator. Its dynamics is given in
terms of the complex variable $Q$ as
\be
\fr{\dd Q}{\dd t}=i\omega Q +(\alpha-\beta |Q|^2)Q,
\l{LS}
\ee
with $\alpha,\beta,\omega \in \mathbb{R}$, and additionally, $\alpha,\beta>0$.
In Ref. \cite{Kuramoto:1975}, it is explained that the oscillator represented by
Eq. (\ref{LS}) is a simple model for self-organized systems like, e.g., reacting chemical
species.
Writing $Q$ in terms of its argument and modulus as $Q=\rho
e^{i\th}$ with $\rho,\th \in \mathbb{R}$ and $\th \in [-\pi,\pi]$, we see from equation (\ref{LS}) that $\theta$ rotates
uniformly in time with angular frequency equal to the parameter $\omega$, while $\rho$ has a stable
value $\rho_{\rm stable}=\sqrt{\alpha/\beta}$, such that
$\dd\rho/\dd t|_{\rho=\rho_{\rm stable}}=0,\dd\rho/\dd t|_{\rho<\rho_{\rm stable}}
>0$, and $\dd\rho/\dd t|_{\rho>\rho_{\rm stable}}<0$. Then, setting $\rho=\rho_{\rm
stable}$ leads to self-sustained limit-cycle oscillations at frequency
$\omega$ with amplitude $\rho_{\rm stable}$; the phase $\theta$ varies
as a function of time as
\be
\fr{\dd \th}{\dd t}=\omega.
\ee

Next, consider a population of $N$ interacting Landau-Stuart oscillators
with varying frequencies $\omega_1,\omega_2,\ldots,\omega_N$ distributed
according to a given probability distribution $g(\omega)$. The dynamics of the $i$th oscillator with angular
frequency $\omega_i$ may be modelled as
\be
\fr{\dd Q_i}{\dd t}=i\omega_i Q_i +(\alpha-\beta |Q_i|^2)Q_i + \sum_{j=1,j\ne
i}^N
K_{ij} Q_j,
\l{LS-II}
\ee
where the real parameter $K_{ij}>0$ describes the coupling between the $i$th and the
$j$th oscillator. In deriving his model while starting from the
dynamics (\ref{LS-II}), Kuramoto considers three
simplifying premises, namely, 
\begin{enumerate}
\item{the limit of an infinite number of oscillators: $N \to \infty$,}
\item{the coupling $K_{ij}~\forall~ i,j$ scaling as $K_{ij}=K/N$ with $K$ finite, implying thereby that every oscillator is
coupled {\it weakly} and with equal strength to every other oscillator, and}
\item{the limit $\alpha,\beta \to \infty$, while keeping $\alpha/\beta$ fixed and
finite, and, moreover, $\omega_i ~\forall~ i$ being finite.}
\end{enumerate}
Writing $Q_i=\rho_i
e^{i\th_i}$, we see that because of the above assumptions,
$\rho_i ~\forall~ i$ while starting from
an initial value will relax over a time of $O(1/\beta)$ to its
limit-cycle value equal to $\sqrt{\alpha/\beta}$. As a result, the long-time
dynamics corresponds to self-sustained
limit-cycle oscillations for each oscillator, which is described by the evolution equation
\be
\fr{\dd \th_i}{\dd t}=\omega_i+\frac{K}{N}\sum_{j=1}^N \sin(\th_j-\th_i).
\l{Kuramoto-eom}
\ee
Equation (\ref{Kuramoto-eom}) is the governing dynamical equation of the
Kuramoto model. 
\subsection{Synchronization transition} 
\l{Kuramoto}
Most investigations of the Kuramoto model have been for a unimodal
$g(\omega)$, i.e., one which is symmetric about the mean
$\langle \omega \rangle$, and which decreases monotonically and continuously
to zero with increasing $|\omega-\langle \omega \rangle|$. We will
denote by $\sigma$ the width of the distribution
$g(\omega)$ (e.g., for a Gaussian
distribution, $\sigma$ is the standard deviation). As mentioned in the
introduction, we will in this review consider specifically such frequency distributions. 
By going to a comoving frame rotating with
frequency $\langle \omega \rangle$ with respect to the laboratory frame, one may from now
on consider in the dynamics (\ref{Kuramoto-eom}) the $\omega_i$'s to
have zero mean without loss of generality; we will implement this in the
rest of the review. 

In his early works, Kuramoto adduced a remarkable self-consistent
analysis to predict the long-time stationary state of the dynamics
(\ref{Kuramoto-eom}). This analysis and its further generalizations have
established that the stationary state is characterized by one of two possible phases, depending on
whether $K$ is below or above a critical value $K_c$, given by \cite{Strogatz:2000,Acebron:2005}
\be 
K_c=\fr{2}{\pi g(0)}.
\l{Kura-bareKc}
\ee
The system for $K < K_c$ is in an unsynchronized or incoherent phase in which the
oscillators exhibit independent oscillations, while for $K > K_c$ in a synchronized phase in which a macroscopic fraction of
oscillators are in synchrony. On tuning $K$, a continuous phase
transition occurs between the two phases. More precisely, the Kuramoto
model being a dynamical system, one does not quite have a
phase transition in the sense of thermodynamics, but rather a
bifurcation for the order parameter, see below. For noisy dynamics, cases of which will be studied later in the review, a phase transition
of course has its usual meaning as in thermodynamics. Here, by synchrony, we mean
that in the limit $N \to \infty$, the
oscillators have time-independent phases in the comoving frame, and have therefore phases that rotate uniformly in time with the same
frequency $\langle \omega \rangle$ in the laboratory frame. The magnitude $r(t)$ and the phase $\psi(t)$ of the complex order parameter, defined as 
\be
{\bf r}(t)=r(t)e^{i\psi(t)}\equiv\fr{1}{N}\sum_{j=1}^N e^{i\th_j(t)},
\l{r-defn}
\ee
measure the amount of synchronization and the average
phase, respectively. 
For $K < K_c$, $r(t)$ while starting
from any initial value relaxes at long times to zero, corresponding to an
incoherent stationary state. On the other hand, for $K > K_c$, $r(t)$ has a
non-zero stationary state value $r_{\rm st}(K) \le 1$ that
increases continuously with $K$, and is such that $r_{\rm st}(K \to
K^+_c)=0$. In the limit $K \to \infty$, all the oscillators are
synchronized and at the same phase, so that $r_{\rm
st}(K \to \infty)=1$. 
In terms of $r(t)$ and $\psi(t)$, the dynamics
(\ref{Kuramoto-eom}) reads 
\be
\fr{\dd \th_i}{\dd t}=\omega_i+K r \sin(\psi-\th_i).
\l{Kuramoto-eom1}
\ee

We now briefly recall a self-consistent analysis due to Kuramoto
\cite{Strogatz:2000} that leads to equation (\ref{Kura-bareKc}). The starting point is to
note that in the stationary state, the single-oscillator distribution
$\rho(\th,\omega,t)$, giving the fraction of oscillators with natural
frequency $\omega$ that has phase $\th$ at time $t$, converges to the
time-independent form $\rho_{\rm st}(\th,\omega)$. Note that $\rho$
satisfies $\rho(\th,\omega,t)=\rho(\th+2\pi,\omega,t)$, and the
normalization $\int_{-\pi}^\pi \dd \th ~\rho(\th,\omega,t)=1 ~\forall~
\omega,t$. When synchronized, the average phase in
the stationary state, $\psi_{\rm st}$, will be a constant that may be set to
zero by choosing properly the origin of the phase axes. The stationary
value $r_{\rm st}$ of $r(t)$, on the other hand, satisfies
\be
r_{\rm st}=\int \dd \th \int \dd\omega ~g(\omega)e^{i\th} \rho_{\rm st}(\th,\omega).
\l{r-st}
\ee
Since $r_{\rm st}$ is real, the above equation implies that the
imaginary part of the right hand side should vanish. As $g(\omega)$
is symmetric: $g(\omega)=g(-\omega)$, the vanishing of the imaginary
part is ensured if $\rho_{\rm st}(-\th,-\omega)=\rho_{\rm st}(\th,\omega)$.
This is indeed the case, as we will see below. On the basis of the above
discussion, one may rewrite equation
(\ref{Kuramoto-eom1}) in the stationary state as 
\be
\fr{\dd\th_i}{\dd t}=\omega_i- K r_{\rm st} \sin \th_i.
\l{Kuramoto-eom2}
\ee

The result (\ref{Kura-bareKc}) follows from a self-consistent equation derived by adopting the following
strategy: At a fixed $K$, and for a given value of $r_{\rm st}$, (i)
obtain the stationary state distribution $\rho_{\rm st}(\th,\omega)$ implied by the dynamics
(\ref{Kuramoto-eom2}), and (ii) require that the distribution when plugged
into the right hand side of equation (\ref{r-st}) reproduces the given value
of $r_{\rm st}$ on the left hand side, thereby yielding the self-consistent equation; we now demonstrate
this procedure. First, it follows from equation (\ref{Kuramoto-eom2}) that the dynamics
of oscillators with $|\omega_i| \le Kr_{\rm st}$ approaches in time a stable fixed point
defined implicitly by
\be
\omega_i=Kr_{\rm st}\sin \th_i,
\ee
so that the $i$th oscillator in this group has the time-independent
phase $\th_i=\sin^{-1}[\omega_i/(Kr_{\rm st})]$; $|\th_i| \le
\pi/2$. This group of oscillators are thus ``locked'' or
synchronized, and has the distribution
\be
\rho_{\rm st}(\th,\omega)=K r_{\rm st} \cos \th ~\delta\Big(\omega- K r_{\rm st} \sin \th\Big)
\Theta (\cos \th ) ;
~~|\omega| \le Kr_{\rm st},
\l{Kura-distr-locked}
\ee
where $\Theta(\cdot)$ is the Heaviside step function. On the other hand, oscillators with $|\omega_i| \ge Kr_{\rm st}$ have
ever drifting time-dependent phases. However, to be consistent with the
fact that we have a time-independent average phase, it is required that $\rho_{\rm
st}(\th,\omega)$ for this group of ``drifting'' oscillators has the form
\be
\rho_{\rm st}(\th,\omega)=\fr{C}{|\omega- K r_{\rm st} \sin \th|};
~~|\omega| > Kr_{\rm st};
\l{Kura-distr-drifting}
\ee
this ensures that oscillators are more crowded at $\th$-values with
lower local velocity $\dd\th_i/{\rm d}t$ than at values with higher local
velocity. The constant $C$ in equation 
(\ref{Kura-distr-drifting}) is fixed by the normalization condition $\int_{-\pi}^\pi \dd\th~ \rho_{\rm st}(\th,\omega)=1
~\forall~ \omega$, yielding
\be
C=\fr{1}{2\pi}\sqrt{\omega^2-(Kr_{\rm st})^2}.
\ee

We now require that the given value of $r_{\rm st}$ coincides with the
one implied by the distributions in equations
(\ref{Kura-distr-locked}) and (\ref{Kura-distr-drifting}). Plugging the
latter forms into equation (\ref{r-st}), we get
\bea
r_{\rm st}&=&\int_{-\pi}^\pi \dd \th \int_{|\omega| > Kr_{\rm st}} \dd
\omega~g(\omega) e^{i\th}\fr{C}{|\omega- K r_{\rm st} \sin
\th|}\nonumber \\
&+&\int_{-\frac{\pi}{2}}^{\frac{\pi}{2}} \dd
\th \int_{|\omega| \le Kr_{\rm st}} \dd
\omega~g(\omega) e^{i\th} K r_{\rm st} \cos \th \,\delta\Big(\omega- K r_{\rm st} \sin \th\Big).
\eea
The first integral on the right hand side vanishes due to the symmetry $g(\omega)=g(-\omega)$
combined with the property that $\rho_{\rm st}(\th+\pi,-\omega)=\rho_{\rm st}(\th,\omega)$ for
the ``drifting'' oscillators, see equation (\ref{Kura-distr-drifting}). The imaginary part of the
second integral vanishes on using $g(\omega)=g(-\omega)$
and $\rho_{\rm st}(-\th,-\omega)=\rho_{\rm st}(\th,\omega)$ for the ``locked'' oscillators,
see equation (\ref{Kura-distr-locked}); the real part, after integration over $\omega$, finally yields
\be
r_{\rm st}=Kr_{\rm st}\int_{-\pi/2}^{\pi/2}\dd \th~\cos^2 \th ~g(Kr_{\rm st}\sin
\th),
\ee
which is the desired self-consistent equation. This equation has the 
trivial solution $r_{\rm st} = 0$, valid for any value of $K$,
corresponding to the incoherent phase with $\rho_{\rm st}(\th,\omega) =
1/(2\pi) ~\forall~ \th,\omega$. There can however be another solution
corresponding to $r_{\rm st} \ne 0$ that satisfies
\be
1=K\int_{-\pi/2}^{\pi/2}\dd \th~ \cos^2 \th ~g(Kr_{\rm st}\sin \th).
\l{Kura-bifurcation}
\ee
This solution bifurcates continuously from the incoherent solution at
the value $K = K_c$ given by equation (\ref{Kura-bareKc}) that follows from
the above equation on taking the limit $r_{\rm st} \to 0^+$. Since for a
unimodal $g(\omega)$, one has a negative second derivative at $\omega = 0$, $g''(0)<0$, one finds by expanding
the integrand in equation (\ref{Kura-bifurcation}) as a powers series in $r_{\rm st}$ that the
bifurcation in this case is supercritical. As a matter of fact, it is not difficult to see that
for a unimodal $g(\omega)$, a solution of equation (\ref{Kura-bifurcation}) exists only for $K \ge K_c$. Indeed,
the right hand side of equation (\ref{Kura-bifurcation}) is equal to $K \pi g(0)/2$ for $r_{\rm
st}=0$, while its partial derivative with respect to $r_{\rm st}$, given by
\be
K^2\int_{-\pi/2}^{\pi/2}\dd \th~ \cos^2 \th \sin \th ~g'(Kr_{\rm st}\sin \th),
\l{kura_r_deriv}
\ee
is negative definite (here and henceforth, prime will denote derivative). On the other hand, for $r_{\rm st}=1$, the right
hand side of equation
(\ref{Kura-bifurcation}) after the change of variable $K\sin \th = u$ can be
written as
\be
\int_{-K}^K \dd u \left( 1- \frac{u^2}{K^2}\right)^{\frac{1}{2}}g(u) ,
\l{kura_r_eq1}
\ee
which is clearly smaller than $1$, tending to $1$ as $K \to \infty$.
Finally, its derivative
with respect to $K$ is
\bea
\int_{-\pi/2}^{\pi/2}\dd \th~ \cos^2 \th ~g(Kr_{\rm st}\sin \th) +
Kr_{\rm st} \int_{-\pi/2}^{\pi/2}\dd \th~ \cos^2 \th \sin \th ~g'(Kr_{\rm st}\sin \th) \nonumber \\
= \int_{-\pi/2}^{\pi/2}\dd \th~ \sin^2 \th  ~g(Kr_{\rm st}\sin \th) ,
\l{kura_k_deriv}
\eea
which is positive.
These properties imply that a solution
$r_{\rm st}$ of equation (\ref{Kura-bifurcation})
exists for $K\ge K_c$, which equals $0$ for $K=K_c$, and which
increases with $K$ and approaches unity as $K\to \infty$.

The linear stability of the incoherent solution, $\rho_{\rm st}(\th,\omega) = 1/(2\pi) ~\forall~ \th,\omega$ \cite{Strogatz:2000},
will be considered in Sec. \ref{seckurlongrange}, where it will appear as a special case of the Kuramoto
model with non-mean-field long-range interactions. The stability
analysis will establish that the incoherent state is neutrally stable
below $K_c$ and unstable above. 
\subsection{The noisy Kuramoto model}
\l{noisyKuramoto}
In order to account for stochastic fluctuations of the $\omega_i$'s in time, the dynamics
(\ref{Kuramoto-eom}) with an additional Gaussian noise term $\eta_i(t)$
on the right hand side was studied by Sakaguchi \cite{Sakaguchi:1988}.
The dynamical equations are 
\be
\fr{\dd \th_i}{\dd t}=\omega_i+K r \sin(\psi-\th_i)+\eta_i(t),
\l{Kura-noise-eom}
\ee
where
\be
\langle \eta_i(t) \rangle=0, ~~\langle \eta_i(t)\eta_j(t')
\rangle=2D\delta_{ij}\delta(t-t'),
\ee
with the parameter $D$ standing for the noise strength, while here and
from now on, angular
brackets will denote averaging with respect to noise realizations. In
presence of $\eta_i(t)$, the continuous transition of the bare model is
sustained, with $K_c$ shifted to \cite{Sakaguchi:1988}
\be
K_c(D) = 2 \Big[\int_{-\infty}^{\infty} \dd \omega~
\fr{g(D\omega)}{\omega^2+1}\Big]^{-1}.
\l{Kura-noise-Kc}
\ee
On taking the limit $D \to 0$ in the above equation, one recovers
the transition point (\ref{Kura-bareKc}) for the bare model. In the
following, we briefly sketch the derivation of equation
(\ref{Kura-noise-Kc}), following Ref. \cite{Sakaguchi:1988}.

The starting point is to write down a Fokker-Planck equation for the
time evolution of the distribution $\rho(\th,\omega,t)$, which follows
straightforwardly from the dynamics (\ref{Kura-noise-eom}) as
\be
\fr{\partial \rho}{\partial t}=-\fr{\partial }{\partial
\th}\Big[\Big(\omega+K r \sin(\psi-\th)\Big)\rho\Big]+D\fr{\partial^2\rho}{\partial \th^2}.
\l{fokplaeqnoisy}
\ee
As before, in the stationary state, we set $\psi_{\rm st}=0$, and obtain from the above equation the result
\bea
\rho_{\rm st}(\th,\omega)&=&\exp\Big(\fr{-Kr_{\rm st}+\omega
\theta+Kr_{\rm st}\cos \th}{D}\Big)\rho_{\rm st}(0,\omega)\nonumber \\
&\times&\Big[1+\fr{(e^{-2\pi \omega/D}-1)\int_0^\th \dd \th' ~e^{(-\omega
\th'-Kr_{\rm st}\cos \th')/D}}{\int_{-\pi}^{\pi} \dd \th'~e^{(-\omega
\th'-Kr_{\rm st}\cos \th')/D}}\Big],
\l{Kura-noise-distr}
\eea
where $\rho_{\rm st}(0,\omega)$ is fixed by the normalization
$\int_{-\pi}^\pi \dd \th ~\rho_{\rm st}(\th,\omega)=1 ~\forall~ \omega$. For $r_{\rm st} = 0$ the above
expression reduces to the incoherent state $\rho_{\rm st}(\th,\omega) = 1/(2\pi) ~\forall~ \th,\omega$.
Substituting equation (\ref{Kura-noise-distr}) into equation (\ref{r-st}), one obtains a self-consistent
equation for $r_{\rm st}$. As for the Kuramoto model, it has the trivial solution $r_{\rm st}=0$, corresponding to
the incoherent state. In finding the other solution, one observes that the imaginary part
of the right hand side of equation (\ref{r-st}) is zero due to the
symmetry $g(\omega)=g(-\omega)$ together with the property $\rho_{\rm st}(-\th,-\omega)=\rho_{\rm
st}(\th,\omega)$, see equation (\ref{Kura-noise-distr}); thus, only the real part contributes. Expanding the resulting equation
in powers of $Kr_{\rm st}/D$, 
and taking the limit $r_{\rm st} \to 0^+$ \cite{Sakaguchi:1988} yield the critical coupling strength $K_c(D)$
given by equation (\ref{Kura-noise-Kc}).
\subsubsection{Linear stability analysis of the incoherent stationary state}
\l{linear-stability-kura-noise}
The stability analysis of the incoherent state $\rho_{\rm st}(\th,\omega) = 1/(2\pi) ~\forall~ \th,\omega$ is performed by
studying the linearized Fokker-Planck equation obtained from equation
(\ref{fokplaeqnoisy}) after expanding $\rho(\theta,\omega,t)$ as 
\be
\rho(\theta,\omega,t) = \frac{1}{2\pi} + \delta \rho(\theta,\omega,t);
~~|\delta \rho| \ll 1.
\l{linearnoisy}
\ee
Writing explicitly the expression for $r$, the resulting linear equation is
\bea
\l{fokplalinearnoisy}
\frac{\partial}{\partial t} \delta \rho(\theta,\omega,t) = -\omega \frac{\partial}{\partial \theta} \delta \rho(\theta,\omega,t)
+ D \frac{\partial^2}{\partial \theta^2} \delta \rho(\theta,\omega,t)
\nonumber \\
+\frac{K}{2\pi} \int_{-\pi}^{\pi} \dd \theta' \, \int \dd \omega' \, g(\omega') \cos (\theta'-\theta) \delta \rho(\theta',\omega',t).
\eea
With the Fourier expansion
\be
\l{fouriernoisy}
\delta \rho(\theta,\omega,t) = \sum_{k=-\infty}^{+\infty}
\widehat{\delta \rho}_k(\omega,t)e^{ik\theta},
\ee
equation (\ref{fokplalinearnoisy}) gives
\be
\l{fokplalinearnoisyfourier}
\fl \frac{\partial}{\partial t} \widehat{\delta \rho}_k(\omega,t) = -ik\omega \widehat{\delta \rho}_k(\omega,t)
- D k^2 \widehat{\delta \rho}_k(\omega,t)
+\frac{K}{2}\left(\delta_{k,1} + \delta_{k,-1}\right) \int \dd \omega' \, g(\omega')\widehat{\delta \rho}_k(\omega',t).
\ee
For $k\ne \pm 1,$ the integral term vanishes, and we have
\be
\l{fokplalinearnoisyfourierkne1}
\frac{\partial}{\partial t} \widehat{\delta \rho}_k(\omega,t) = -ik\omega \widehat{\delta \rho}_k(\omega,t)
- D k^2 \widehat{\delta \rho}_k(\omega,t),
\ee
so that with $\omega$ varying in the support of $g(\omega)$, one has a
continuous spectrum of stable modes that decay exponentially in time with rate $Dk^2$. For $k=\pm 1$, after posing
\be
\l{expokeq1}
\widehat{\delta \rho}_{\pm 1}(\omega,t) = \widetilde{\delta \rho}_{\pm 1}(\omega,\lambda)e^{\lambda t},
\ee
we have
\be
\l{fokplalinearnoisyfourierk1}
\left[ \lambda \pm i\omega +D\right] \widetilde{\delta \rho}_{\pm 1}(\omega,\lambda) =
\frac{K}{2} \int \dd \omega' \, g(\omega')\widetilde{\delta \rho}_{\pm 1}(\omega',\lambda).
\ee
This equation also admits a continuous spectrum of stable modes, given by $\lambda = \mp i \omega_0 -D$ for each
$\omega_0$ in the support of $g(\omega)$. The modes, normalized so that
the right hand side of equation (\ref{fokplalinearnoisyfourierk1}) is equal to 1, are given by
\be
\widetilde{\delta \rho}_{\pm 1}(\omega,\mp i \omega_0 -D) = \mp i {\mathcal P} \frac{1}{\omega - \omega_0}
+ c_{\pm 1}(\omega_0) \delta (\omega - \omega_0) ,
\l{eigen_cont_pm1_noisy}
\ee
with
\be
c_{\pm 1}(\omega_0) g(\omega_0) = \frac{2}{K} \pm i {\mathcal P}
\int \dd \omega \, \frac{g(\omega)}{\omega -\omega_0} ,
\l{eigen_factor_noisy}
\ee
where ${\mathcal P}$ denotes the principal value. However, unlike for $k
\ne \pm 1$, there is also a discrete spectrum for $\lambda \pm i \omega
+ D \ne 0$. From equation (\ref{fokplalinearnoisyfourierk1}), we have
\be
\widetilde{\delta \rho}_{\pm 1}(\omega,\lambda) =
 \frac{K}{2(\lambda \pm i \omega + D)}
 \int \dd \omega' \,
 \widetilde{\delta \rho}_{\pm 1}(\omega',\lambda) g(\omega').
\l{fokplalinearnoisyfourierk1_discr}
\ee
In order to have a non-trivial solution of the above equation, the integral on the right hand
side must not vanish. We can impose that this integral is equal to $1$, since equation (\ref{fokplalinearnoisyfourierk1}) is linear.
We then obtain the dispersion relation
\be
\frac{K}{2}\int_{-\infty}^{+\infty} \dd \omega \,
\frac{g(\omega)}{\lambda \pm i \omega +D} = 1.
\l{disper_relat_noisy}
\ee
Decomposing $\lambda$ into real and imaginary parts, $\lambda = \lambda_r +i \lambda_i$, we obtain from
equation (\ref{disper_relat_noisy}) that
\bea
\frac{K}{2}\int_{-\infty}^{+\infty} \dd \omega \,
g(\omega)\frac{\lambda_r + D}{(\lambda_r + D)^2 + (\lambda_i \pm
\omega)^2} = 1,
\l{disper_relatr_noisy} \\
\frac{K}{2}\int_{-\infty}^{+\infty} \dd \omega \,
g(\omega)\frac{\lambda_i \pm \omega}{(\lambda_r + D)^2 + (\lambda_i \pm \omega)^2} = 0.
\l{disper_relati_noisy}
\eea
With the change of variable $\lambda_i \pm \omega =x$, the integral in
the second equation can be transformed to
\be
\int_0^{+\infty} \dd x \, \left[g(\pm x \mp \lambda_i) - g(\mp x \mp\lambda_i)\right]
\frac{x}{(\lambda_r + D)^2 + x^2}.
\l{disper_relati_noisy_b}
\ee
One may check that for unimodal $g(\omega)$, the above expression can be equal to $0$ only for
$\lambda_i = 0$. So equation (\ref{disper_relatr_noisy}) becomes
\be
\frac{K}{2}\int_{-\infty}^{+\infty} \dd \omega \,
g(\omega)\frac{\lambda + D}{(\lambda + D)^2 + \omega^2} = 1 ,
\l{disper_relatr_noisy_b}
\ee
with $\lambda$ real. This equation shows that only solutions $\lambda > -D$ are possible; when such
a solution is not present, there is no discrete spectrum, and the incoherent state is stable.
However, stability holds also when there is a solution $\lambda <0$, since we have seen that
all the eigenvalues of the continuous spectrum have a negative real part.
The change of variable $\omega = (\lambda + D)y$ transforms equation
(\ref{disper_relatr_noisy_b}) to
\be
\frac{K}{2}\int_{-\infty}^{+\infty} \dd y \,
g\left[(\lambda + D)y\right]\frac{1}{1 + y^2} = 1.
\l{disper_relatr_noisy_c}
\ee
The left hand side tends to $0$ as $\lambda \to \infty$, while its derivative with respect to
$\lambda$ is
\be
\frac{K}{2}\int_{-\infty}^{+\infty} \dd y \,
g'\left[(\lambda + D)y\right]\frac{y}{1 + y^2},
\l{disper_relatr_noisy_der}
\ee
which is negative. Therefore, a solution for $\lambda$ exists only when the value of the left hand side of
equation (\ref{disper_relatr_noisy_c}) for $\lambda = -D$ is larger than $1$. In particular, we have stability
when this solution is negative; the threshold for stability is thus given by
\be
\frac{K}{2}\int_{-\infty}^{+\infty} \dd y \,
g(Dy)\frac{1}{1 + y^2} = 1 ,
\l{disper_relatr_noisy_thres}
\ee
that gives the critical value (\ref{Kura-noise-Kc}).

\section{Generalized Kuramoto model with inertia and noise}
\l{chap2}
In the
generalized dynamics, an additional dynamical variable, namely, angular
velocity, is assigned to each oscillator, thereby elevating the
first-order dynamics of the Kuramoto model to the level of second-order
dynamics; the equations of motion are 
\cite{Acebron:1998,Hong:1999,Acebron:2000}:
\bea
\fr{\dd\th_i}{\dd t}=v_i,\nonumber \\
\l{eom} \\
m\fr{\dd v_i}{\dd t}=-\gamma v_i+\widetilde{K}
r\sin(\psi-\th_i)+\gamma\omega_i+\widetilde{\eta}_i(t). \nonumber
\eea
Here, $v_i$ is the angular velocity of the $i$th oscillator, $m$ is the moment of inertia of the oscillators, $\gamma$ is the
friction constant, $\widetilde{K}$ is the strength of the coupling
between the oscillators,
while $\widetilde{\eta}_i(t)$ is a Gaussian noise with
\be
\langle \widetilde{\eta}_i(t) \rangle=0, ~~\langle
\widetilde{\eta}_i(t)\widetilde{\eta}_j(t')
\rangle=2\widetilde{D}\delta_{ij}\delta(t-t').
\l{etatide-prop}
\ee
In the limit of overdamped motion ($m \to 0$ at a fixed $\gamma \ne 0$),
the dynamics (\ref{eom}) reduces to
\be
\gamma \fr{\dd\th_i}{\dd t}=\widetilde{K}
r\sin(\psi-\th_i)+\gamma\omega_i+\widetilde{\eta}_i(t).
\l{eom-overdamped1}
\ee
Then, defining $K \equiv \widetilde{K}/\gamma$ and
$\eta_i(t) \equiv \widetilde{\eta}_i(t)/\gamma$ so that
$D=\widetilde{D}/\gamma^2$, the dynamics (\ref{eom-overdamped1}) for $D=0$
becomes that of the Kuramoto model, equation (\ref{Kuramoto-eom}), and for $D
\ne 0$ that of its noisy version, the dynamics
(\ref{Kura-noise-eom}).

In Appendix A, we illustrate how the
dynamics (\ref{eom}) without the noise term, studied in
\cite{Tanaka:1997}, arises in a completely different context, namely, in electrical
power distribution networks comprising synchronous
generators (representing power plants) and motors (representing
customers) \cite{Filatrella:2008,Rohden:2012}; the dynamics arises in the approximation in which every node of the network is
connected to every other.
\subsection{The model as a long-range interacting
system}
\l{longrangemodel}
We now discuss that in a different context than that of coupled oscillators,
the dynamics (\ref{eom}) describes a long-range
interacting system of particles moving on a unit circle, with each
particle acted upon by a quenched external torque $\widetilde{\omega}_i \equiv \gamma
\omega_i$.

Much recent exploration of the static and dynamic properties
of long-range interacting systems has been pursued within the framework
of an analytically tractable  prototypical model called the Hamiltonian
mean-field (HMF) model \cite{Inagaki:1993,Ruffo:1995}. The model comprises $N$ particles of mass
$m$ moving on a unit circle and interacting through a long-range interparticle
potential that is of the mean-field type: every particle is coupled to every other with equal strength.
This system can also be seen as a set of $XY$-rotators that reside on a lattice and interact through
ferromagnetic coupling. The structure and dimensionality of the lattice need not be specified, since the coupling
between each pair of rotators is the same (mean-field system). Since the
configuration of an $XY$-rotator is defined
by a single angle variable, one might also view the rotators as spin vectors. 
However, one should be aware that this
identification is not completely correct. This is because the dynamics
of spins is defined differently, through Poisson brackets, or,
equivalently, through the derivative of the Hamiltonian with respect to
the spins, that yields the effective magnetic field acting on the individual spins.
On the other hand, the $XY$-rotators are more
correctly identified with particles confined to a circle, with angle and
angular momentum as canonically conjugate variables, and with the dynamics generated by
the Hamilton equations for these variables \cite{Gupta:2011}. Apart from academic interest, the model provides a
tractable reference to study physical systems like gravitational sheet
models \cite{Tsuchiya:1994} and the free-electron laser
\cite{Barre:2004}. 

The Hamiltonian of the HMF model is \cite{Ruffo:1995}
\be
H=\sum_{i=1}^{N}\fr{p_i^2}{2m}+\fr{\widetilde{K}}{2N}\sum_{i,j=1}^{N}\left[1-\cos(\th_i-\th_j)\right],
\l{HMF-H}
\ee
where $\th_i \in [-\pi,\pi]$ gives the position of the $i$th particle
on the circle, while $p_i=mv_i$ is its conjugated angular
momentum, with $v_i$ being the angular velocity. 
The time evolution of the system within a microcanonical ensemble follows the deterministic Hamilton equations of motion given by
\bea
&&\fr{\dd \th_i}{\dd t}=v_i, \nonumber \\
\l{hameq} \\
&&m\fr{\dd v_i}{\dd t}=\widetilde{K}
r\sin(\psi-\th_i). \nonumber
\eea
Here, the quantities $r$ and $\psi$ are as defined in equation (\ref{r-defn}). The dynamics conserves the total energy and momentum.
Under the evolution (\ref{hameq}), the system at long times settles into
an equilibrium stationary state in which, depending on the energy
density $\eps=H/N$, the system could be in one of two possible phases.
Namely, for $\eps$ smaller than a critical value $\eps_c=3\widetilde{K}/4$, the
system is in a clustered phase in which the particles are close together
on the circle, while for energies larger than $\eps_c$, the particles
are uniformly distributed on the circle, characterizing a homogeneous
phase \cite{Campa:2009}. A continuous phase transition between the two phases is effectively characterized by the
quantity $r(t)$ defined in equation (\ref{r-defn}), that in the present
context may be interpreted as the specific magnetization of the system.
The phase transition may be interpreted as one from a high-energy paramagnetic
phase (similar to the incoherent phase in the setting of coupled
oscillators) to a low-energy ferromagnetic phase (similar to the
synchronized phase).

One may generalize the microcanonical dynamics (\ref{hameq}) to include
the effect of an interaction with an external heat bath at temperature
$T$. The resulting
Brownian mean-field (BMF) model has thus a canonical ensemble dynamics
given by \cite{Chavanis:2011,Chavanis:2013}.
\bea
&&\fr{\dd \th_i}{\dd t}=v_i, \nonumber \\
\l{bmfeq} \\
&&m\fr{\dd v_i}{\dd t}=-\gamma v_i+\widetilde{K}
r\sin(\psi-\th_i)+\widetilde{\eta}_i(t), \nonumber
\eea
where $\widetilde{\eta}_i(t)$ is defined in equation (\ref{etatide-prop}). A
fluctuation-dissipation relation expresses the strength $\widetilde{D}$
of the noise in terms of the temperature $T$ and the friction constant $\gamma$ as
\be
\widetilde{D}=\gamma k_BT.
\l{D-relation}
\ee
We will set the Boltzmann constant $k_B$ to unity in the rest of the paper. 
The canonical dynamics (\ref{bmfeq}) also leads to a long-time
equilibrium stationary state in which a generic configuration $C$ with energy
$E(C)$ occurs with the usual Gibbs-Boltzmann weight: $P_{\rm eq}(C)
\propto e^{-E(C)/T}$. The phase transition in the HMF model within the
microcanonical ensemble occurs within the canonical ensemble on tuning
the temperature across the critical value $T_c=\widetilde{K}/2$. This latter
critical value may be derived very
simply by following standard
procedure \cite{Campa:2009}. Due to full rotational $O(2)$ symmetry, we
may take the direction along which particles
cluster to be along $x$ without
loss of generality. Then, the $x$-component of the magnetization $r$,
namely, $r_x=r \cos \psi$, satisfies in equilibrium the equation
\be
r_x= \fr{\int_{-\pi}^{\pi} \dd\th \cos \th ~e^{\widetilde{K}(r_x/T) \cos
\th}}{\int_{-\pi}^{\pi} \dd\th ~e^{\widetilde{K}(r_x/T)\cos \th}}.
\l{rx-HMF}
\ee
Close to the critical point ($T \to T_c$), expanding the above equation to leading
order in $r_x$, we get
\bea
r_x\Big(2\pi-\fr{\widetilde{K}}{T}\int_{-\pi}^{\pi} \dd\th ~\cos^2 \th\Big)=0.
\eea
With $r_x \ne 0$, we get the critical temperature as $T_c=\widetilde{K}/2$.

Let us now envisage the following situation: A set of quenched external
torques $\{\widetilde{\omega}_i\equiv \gamma \omega_i\}$ acts on each of the particles,
thereby pumping energy into the system. In this case, the second
equation in the canonical
dynamics (\ref{bmfeq}) has an additional term $\widetilde{\omega}_i$ on
the right hand side. The resulting dynamics becomes exactly identical to the dynamics
(\ref{eom}) of the generalized Kuramoto model. 
\subsection{Previous studies}
\l{previous}
Introducing inertia alone into the Kuramoto dynamics (equation (\ref{eom}) without the noise term)
has significant consequences. Tanaka {\it
et al.} showed, mainly on the basis of numerical simulations, that finite large inertia leads to the synchronization transition
becoming of first order, occurring in
an abrupt way on tuning $K$ \cite{Tanaka:1997}. Analysis of the dynamics
(\ref{eom}) in the continuum limit, based on a suitable
Fokker-Planck-like equation, was pursued in Refs. \cite{Acebron:1998,Acebron:2000}. It was shown that for a Lorentzian $g(\omega)$, either larger
inertia or larger frequency spread (measured in terms of the width of
the Lorentzian $g(\omega)$) makes the system harder to synchronize,
leading to an incoherent stationary state.
\subsection{Dynamics in a reduced parameter space}
\l{reducedspace}
We start our analysis of the dynamics (\ref{eom}) by noting that the effect of $\sigma$ may be made explicit by replacing $\omega_i$
in the second equation by $\sigma \omega_i$.
We thus consider from now on the dynamics (\ref{eom}) with
the substitution $\omega_i \rightarrow \sigma \omega_i$. In the resulting
model, $g(\omega)$ therefore has zero mean and unit width. Also, we will
consider in the dynamics (\ref{eom}) the parameter $\widetilde{D}$ to
have the scaling (\ref{D-relation}).

For $m \ne 0$, using dimensionless quantities 
\bea
\l{dmsless1}
&&\overline{t}\equiv t\sqrt{\widetilde{K}/m}, \\
\l{dmsless2}
&&\overline{v}_i\equiv
v_i\sqrt{m/\widetilde{K}}, \\
\l{dmsless3}
&&1/\sqrt{\overline{m}}\equiv
\gamma/\sqrt{\widetilde{K}m}, \\
\l{dmsless4}
&&\overline{\sigma} \equiv
\gamma \sigma/\widetilde{K},\\
\l{dmsless5}
&&\overline{T} \equiv
T/\widetilde{K},\\ 
\l{dmsless6}
&&\overline{\eta}_i(\overline{t})\equiv
\widetilde{\eta}_i(t)/\widetilde{K},
\eea
the
equations of motion become
\bea
\fr{\dd\th_i}{\dd\overline{t}}=\overline{v}_i, \nonumber \\
\l{eom-scaled} \\ 
\fr{\dd\overline{v}_i}{\dd\overline{t}}=-\fr{1}{\sqrt{\overline{m}}}\overline{v}_i+r\sin(\psi-\th_i)+\overline{\sigma}\omega_i+\overline{\eta}_i(\overline{t}),
\nonumber 
\eea
where 
\be
\langle \overline{\eta}_i(\overline{t})\overline{\eta}_j(\overline{t}')
\rangle=2(\overline{T}/\sqrt{\overline{m}})\delta_{ij}\delta(\overline{t}-\overline{t}').
\ee
For $m=0$, using dimensionless time 
\be
\overline{t}\equiv
t(\widetilde{K}/\gamma),
\ee
and with $\overline{\sigma}$ and $\overline{T}$ defined as above, the dynamics becomes the overdamped motion
\bea
&&\fr{\dd\th_i}{\dd\overline{t}}=r\sin(\psi-\th_i)+\overline{\sigma}\omega_i+\overline{\eta}_i(\overline{t}),
\l{eom-overdamped}
\eea
where 
\be
\langle \overline{\eta}_i(\overline{t})\overline{\eta}_j(\overline{t}')
\rangle=2\overline{T}\delta_{ij}\delta(\overline{t}-\overline{t}').
\ee
We have thus reduced the dynamics
(\ref{eom}) involving five parameters,
$m,\gamma,\widetilde{K},\sigma,T$, to the dynamics
(\ref{eom-scaled}) (or (\ref{eom-overdamped}) in
the overdamped limit) that involves only three
dimensionless parameters, $\overline{m},\overline{T},\overline{\sigma}$. 
From now on, we consider the dynamics in this reduced parameter space, dropping 
overbars for simplicity of notation.

With $\sigma=0$ (i.e. $g(\omega)=\delta(\omega)$ 
\cite{Acebron:1998},\cite{Acebron:2000}),
the dynamics (\ref{eom-scaled}) is that of the BMF model with an equilibrium stationary
state. For other $g(\omega)$, the
dynamics (\ref{eom-scaled}) violates detailed balance, leading to a
nonequilibrium stationary state (NESS) \cite{Gupta:2014}. In section
\ref{detailedbalance},
we give a rigorous proof of this statement.
\subsection{Non-equilibrium first-order synchronization phase
transition}
\l{phasetransition}
In this section, we report on a very interesting non-equilibrium phase
transition that occurs in the stationary state of the dynamics
(\ref{eom-scaled}). 
As discussed above, the three relevant parameters of the dynamics are
$m,T,\sigma$. In this three-dimensional space of parameters, let us
first locate the phase transitions in the Kuramoto model and in its noisy extension, discussed
in section \ref{Kuramoto-bare}. 
\begin{itemize}
\item{The phase transition of the Kuramoto dynamics ($m=T=0$, $\sigma
\ne 0$) corresponds now to a continuous transition from a
low-$\sigma$ synchronized to a high-$\sigma$ incoherent
 phase across the critical point $\sigma_c(m=0,T=0)=\pi g(0)/2$.}
 \item{Extending the Kuramoto dynamics to $T \ne 0$, the above mentioned
 critical point becomes a second-order critical line on the $(T,\sigma)$-plane, given by solving
$2=\int_{-\infty}^{\infty}d\omega ~g(\omega)
[T/(T^2+\omega^2\sigma^2_c(m=0,T))]$.}
\item{The transition in the BMF dynamics ($m,T \ne 0, \sigma=0$)
corresponds now to a continuous transition occurring at the critical
temperature $T_c=1/2$.}
\end{itemize}
\begin{figure}[here!]
\centering
\includegraphics[width=0.6\textwidth]{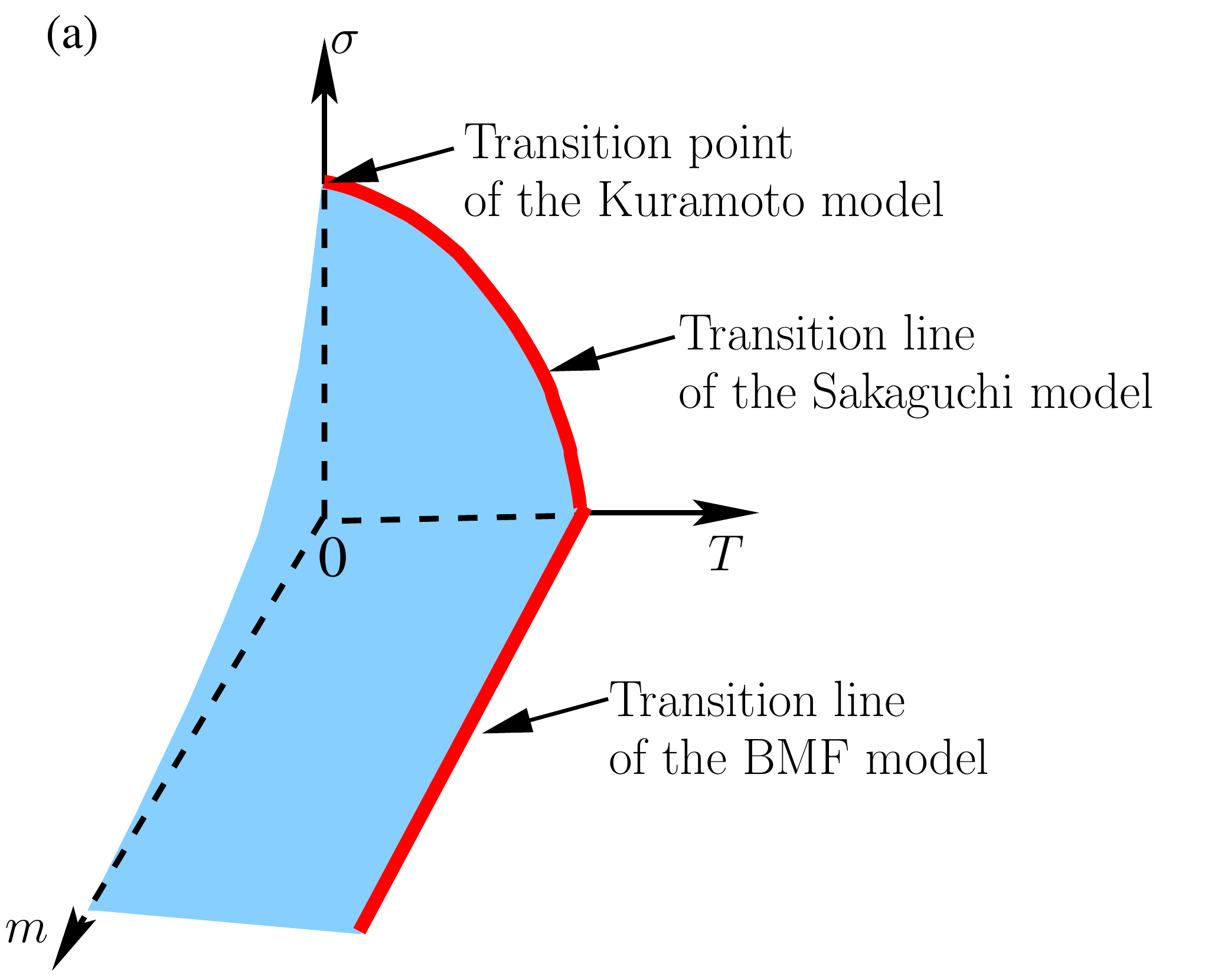}
\includegraphics[width=100mm]{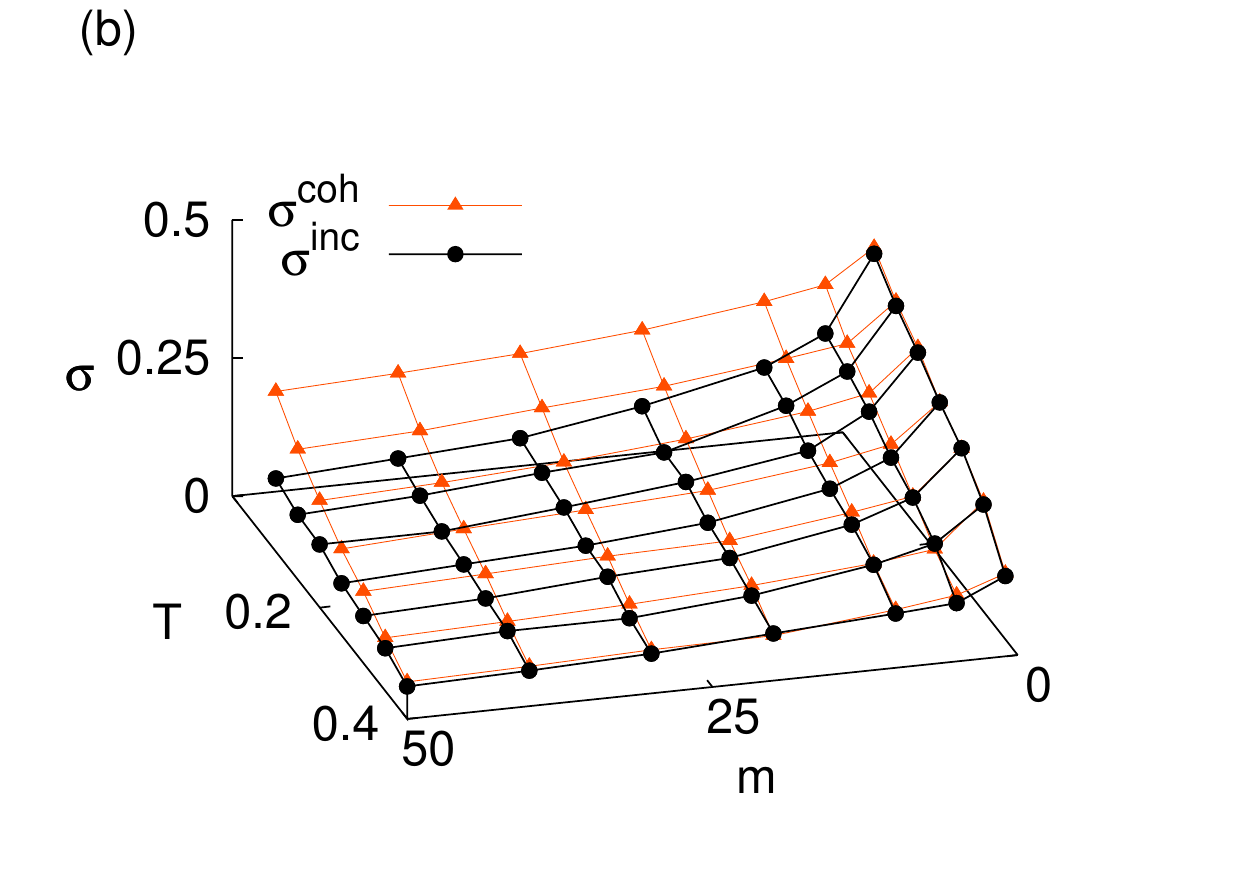}
\caption{(a) The figure shows the schematic phase diagram of model
(\ref{eom-scaled}) in terms of dimensionless moment of inertia $m$, temperature $T$, and width of the
frequency distribution $\sigma$. Here, the shaded blue
surface is a first-order transition surface, while the thick red lines are
second-order critical lines. The system is synchronized inside the region
bounded by the surface, and is incoherent outside. The transitions of
known models are also marked in the figure. The blue surface in (a) is
bounded from above and below by the dynamical stability thresholds
$\sigma^{\rm coh}(m,T)$ and $\sigma^{\rm inc}(m,T)$ of respectively the
synchronized and the incoherent phase, which are estimated in $N$-body
simulations from hysteresis plots (see Fig. \ref{fig:hys-mvary} for an example); the
surfaces $\sigma^{\rm coh}(m,T)$ and $\sigma^{\rm inc}(m,T)$ for
$N=500$ in the case of a Gaussian $g(\omega)$ with zero mean and unit
width are shown in panel (b).}
\label{fig:phdiag}
\end{figure}

\begin{figure}[here!]
\centering
\includegraphics[width=100mm]{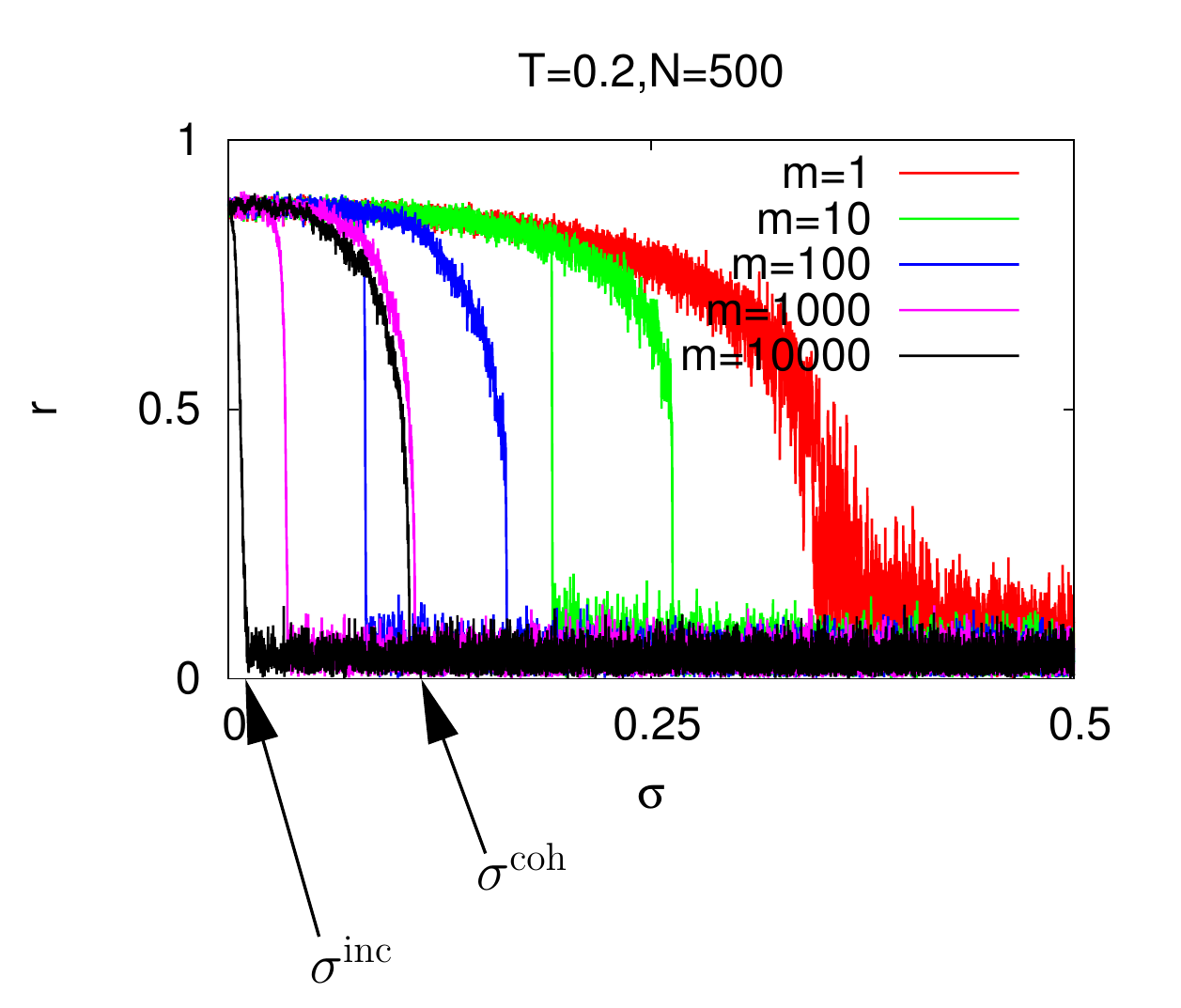}
\caption{For the model (\ref{eom-scaled}), the figure shows $r$ vs. adiabatically tuned
$\sigma$ for different $m$ values at $T=0.2<T_c=1/2$, showing also the stability thresholds, $\sigma^{\rm inc}(m,T)$ and
$\sigma^{\rm coh}(m,T)$, for $m=10000$. The data are obtained from
simulations with $N=500$. For a given $m$, the branch of the plot to
the right (left) corresponds to $\sigma$ increasing (decreasing); for
$m=1$, the two branches almost overlap. The data are for the
Gaussian $g(\omega)$ given by equation (\ref{gomega-gaussian}).}
\l{fig:hys-mvary}
\end{figure}

\begin{figure}[here]
\centering
\includegraphics[width=100mm]{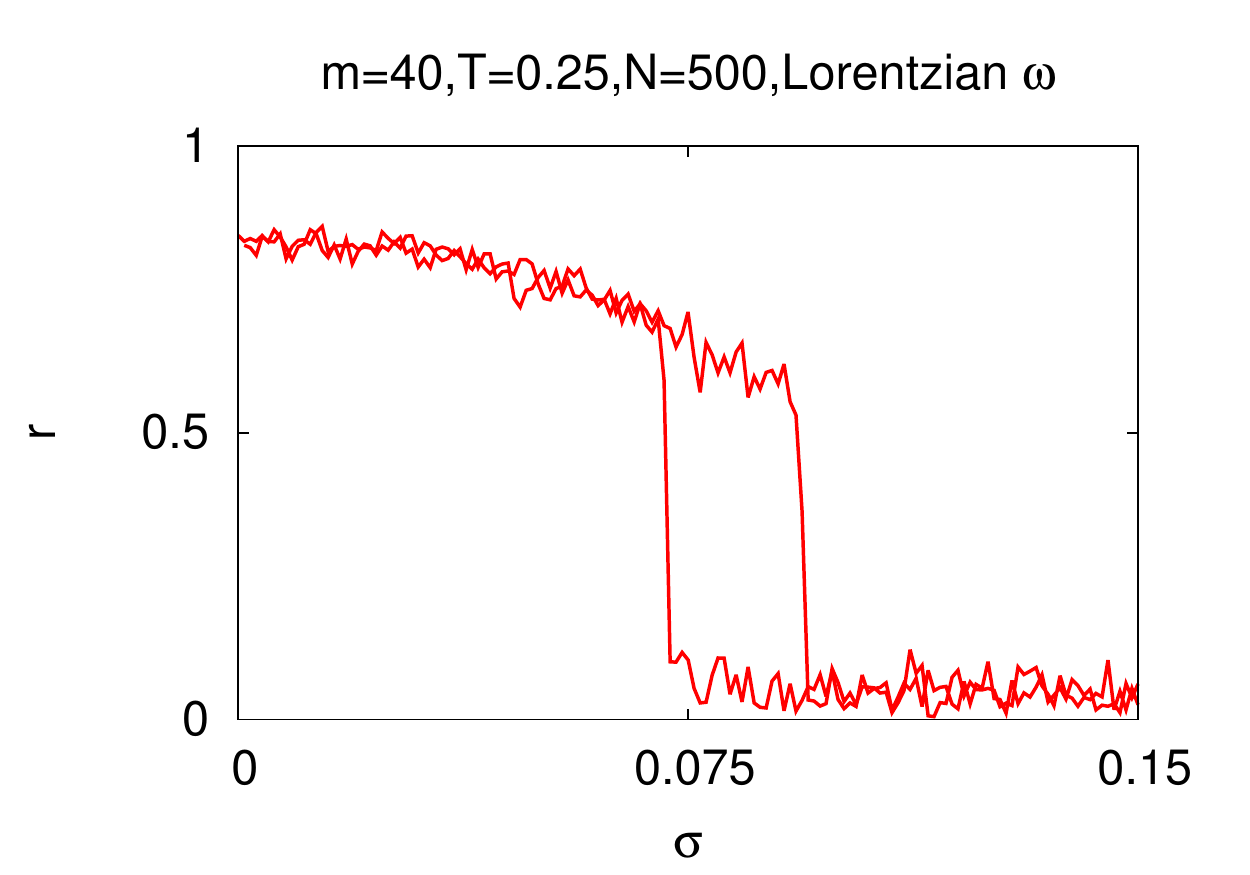}
\caption{For the model (\ref{eom-scaled}), the figure shows $r$ vs. adiabatically tuned
$\sigma$ at $m=40,T=0.25$. The data are obtained from
simulations with $N=500$. 
The branch of the plot to
the right (left) corresponds to $\sigma$ increasing (decreasing). The data are for a
Lorentzian $g(\omega)$ with zero mean and unit width. 
}
\l{fig:hys-lor}
\end{figure}

\begin{figure}[here]
\centering
\includegraphics[width=100mm]{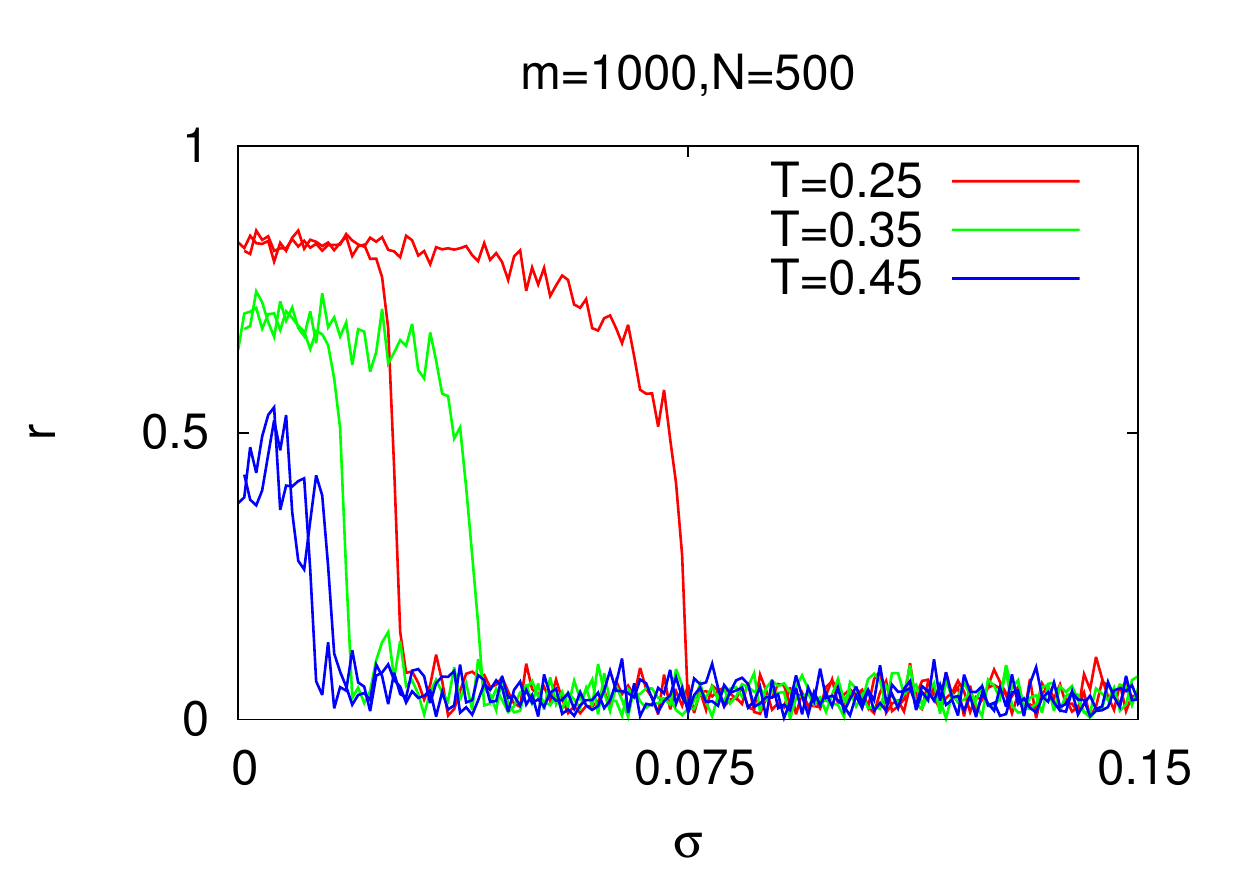}
\caption{For the model (\ref{eom-scaled}), the figure shows $r$ vs. adiabatically tuned
$\sigma$ for different temperatures $T \le T_c=1/2$ at a fixed moment of
inertia $m=1000$. The data are obtained from
simulations with $N=500$. 
For a given $T$, the branch of the plot to
the right (left) corresponds to $\sigma$ increasing (decreasing); for $T
\ge 0.45$, the two branches almost overlap. The data are for the
Gaussian $g(\omega)$ given by equation (\ref{gomega-gaussian}).}
\l{fig:hys-Tvary}
\end{figure}

\begin{figure}[here!]
\centering
\includegraphics[width=100mm]{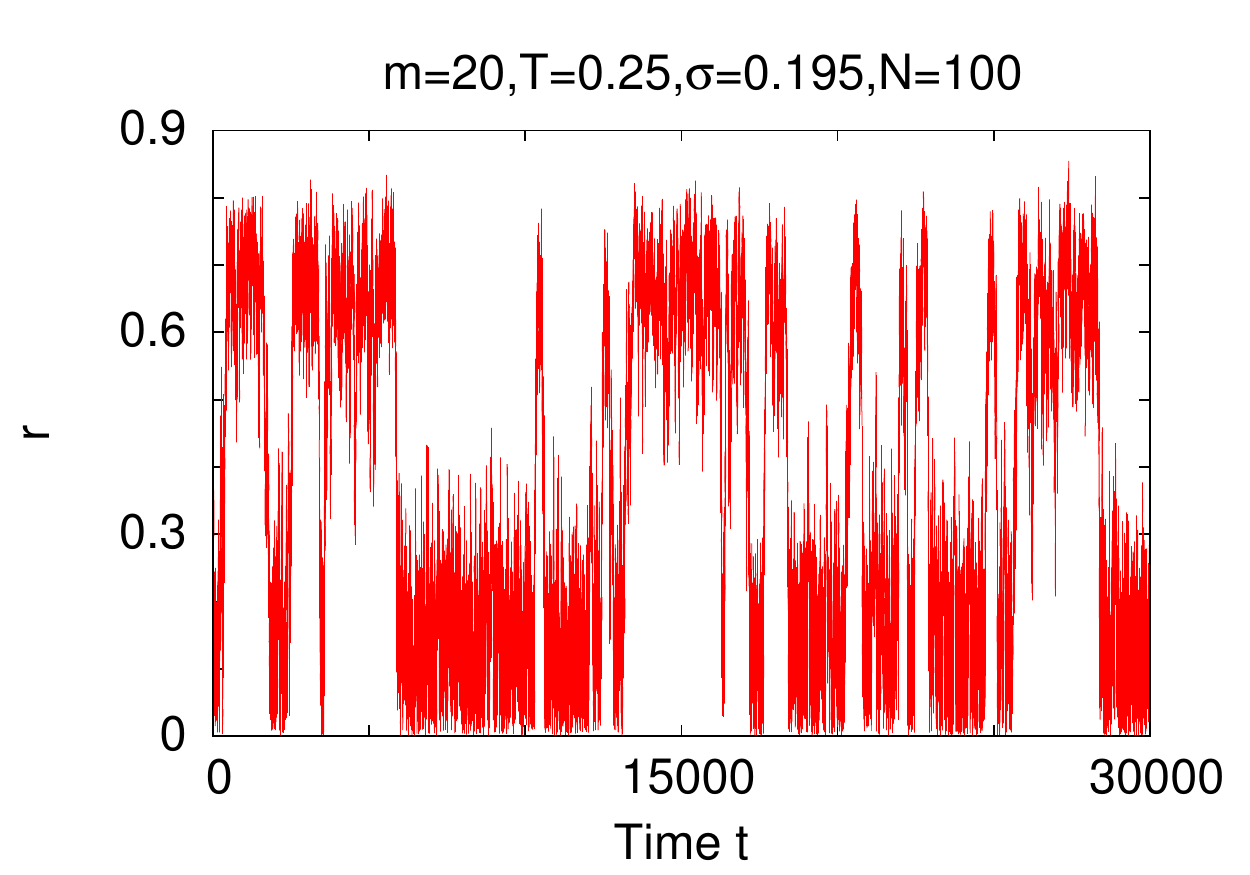} \\
\includegraphics[width=100mm]{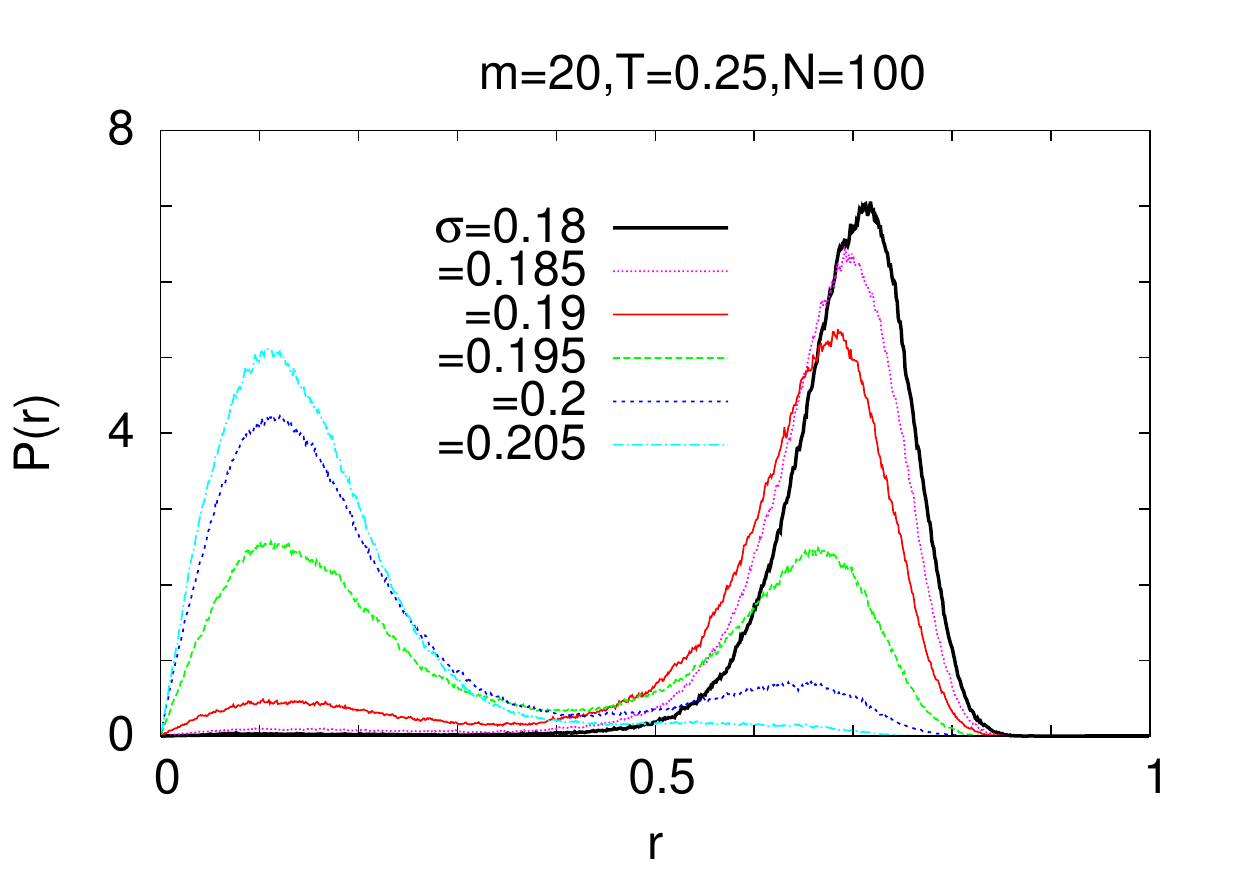}
\caption{For the dynamics (\ref{eom-scaled}) at $m=20,T=0.25,N=100$, and the Gaussian
$g(\omega)$ given by equation (\ref{gomega-gaussian}), (a) shows at 
$\sigma=0.195$, the numerically estimated first-order phase transition
point, $r$ vs. time in the stationary state, while (b)
shows the distribution $P(r)$ at several $\sigma$'s 
around $0.195$. The data are obtained from
simulations with $N=100$.}
\l{fig:r-vs-t-pr}
\end{figure}

The complete phase diagram is shown schematically in Fig.
\ref{fig:phdiag}(a), where the thick red second-order critical lines denote the continuous
transitions mentioned above. On the other hand, for $m,\sigma,T$ all
non-zero, we demonstrate below that the synchronization transition becomes first order,
occurring across the shaded blue transition surface. This surface is bounded by the
second-order critical lines on the $(T,\sigma)$ and $(m,T)$ planes, and by a first-order transition line on
the $(m,\sigma)$-plane. Let us remark that all phase
transitions for $\sigma \ne 0$ are in NESSs, and are interpreted to be of dynamical origin, accounted for by stability
considerations of stationary solutions of equations (for example,
the Kramers equation discussed below) for temporal evolution of phase space
distributions. More rigorously, to qualify as thermodynamics phases, one
needs to show that the different phases extremize a free energy-like quantity (e.g., a large
deviation functional \cite{Touchette:2009}). Such a demonstration in this
nonequilibrium scenario is a daunting task, while for $\sigma=0$, the phases have actually been shown to minimize the equilibrium free energy
\cite{Campa:2009}. 

In order to demonstrate the first-order nature of the transition, we performed
$N$-body simulations for a representative $g(\omega)$, i.e., the
Gaussian given by equation (\ref{gomega-gaussian}).
For given $m$ and $T$, we prepared an initial state with all oscillators at $\th=0$
and frequencies $v_i$'s sampled from a Gaussian distribution with zero mean and
standard deviation $\propto T$. We then let the system equilibrate at
$\sigma=0$, and subsequently increase $\sigma$ adiabatically to high values and back in a cycle. The
simulations involved integrations of the $2N$ coupled equations of motion
(\ref{eom-scaled}), see Appendix B
for details. Figure
\ref{fig:hys-mvary} shows the behavior of $r$ for several $m$'s at a fixed $T$
less than the BMF transition point $T_c=1/2$, where one may observe sharp jumps and
hysteresis behavior expected of a first-order transition. With decrease of $m$, the jump in $r$ becomes less
sharp, and the hysteresis loop area decreases, both features being consistent with the
transition becoming second-order-like as $m \to 0$, see Fig. 
\ref{fig:phdiag}(a). For $m=10000$, we mark in Fig. \ref{fig:hys-mvary} the approximate stability thresholds for the 
incoherent and the synchronized phase, denoted respectively by $\sigma^{\rm inc}(m,T)$ and
 $\sigma^{\rm coh}(m,T)$. The actual phase
transition point $\sigma_c(m,T)$ lies in between the two thresholds. Let
us note from the figure that both the thresholds decrease and approach zero with
the increase of $m$. A qualitatively similar behavior is observed for a
Lorentzian $g(\omega)$, see Fig. \ref{fig:hys-lor}. Figure
\ref{fig:hys-Tvary} shows hysteresis plots for a Gaussian $g(\omega)$ at a
fixed $m$ and for several values of $T \le T_c$, where one observes
that with $T$ approaching $T_c$, the hysteresis loop area decreases, jumps in $r$ become less sharp and
occur between smaller and smaller values that approach zero. Moreover, the
$r$ value at $\sigma=0$ decreases as $T$ increases towards $T_c$,
reaching zero at $T_c$. Disappearance of the hysteresis loop with
increase of $T$ similar to that in Fig. \ref{fig:hys-Tvary} was reported in Ref.
\cite{Hong:1999}.
Our findings suggest that the thresholds $\sigma^{\rm inc}(m,T)$ and
 $\sigma^{\rm coh}(m,T)$ coincide on the second-order critical lines, as
 expected, and moreover, they asymptotically come close together and approach zero as $m \to \infty$
at a fixed $T$. 
For given $m$ and $T$, and for $\sigma$ in between $\sigma^{\rm
inc}(m,T)$ and
 $\sigma^{\rm coh}(m,T)$, Fig. \ref{fig:r-vs-t-pr}(a) for $r$ as a function of time in the
stationary state shows bistability, whereby
the system switches back and forth between incoherent ($r \approx 0$)
and synchronized ($r > 0$) states.  The distribution $P(r)$ depicted in Figure
\ref{fig:r-vs-t-pr}(b) is bimodal with a peak around
either $r \approx 0$ or $r>0$ as $\sigma$ varies between $\sigma^{\rm
inc}(m,T)$ and $\sigma^{\rm coh}(m,T)$. Figure \ref{fig:r-vs-t-pr} lends further credence to the phase transition being first order.
\subsection{Analysis in the continuum limit: The Kramers equation}
\l{analysis}
We now turn to an analytical characterization of the dynamics
(\ref{eom-scaled}) in the continuum limit $N \to \infty$. To this end,
we define the
single-oscillator distribution $f(\th,v,\omega,t)$ that 
gives at time $t$ and for each $\omega$ the fraction of oscillators with phase $\th$
and angular velocity $v$. The distribution is $2\pi$-periodic in
$\theta$, and obeys the normalization 
\be
\int_{-\pi}^{\pi} \dd\th \int_{-\infty}^{\infty} \dd v
~f(\th,v,\omega,t)=1,
\ee while evolving 
following the Kramers equation \cite{Acebron:2000,Gupta:2014}
\be
\frac{\partial f}{\partial t}=-v\frac{\partial f}{\partial
\th}+\frac{\partial}{\partial
v}\Big(\frac{v}{\sqrt{m}}-\sigma \omega-r\sin(\psi-\th)\Big)f+\frac{
T}{\sqrt{m}}\frac{\partial^2 f}{\partial v^2}, 
\l{Kramers}
\ee
where 
\be
re^{i\psi}=\int \dd\th \dd v \dd\omega
~g(\omega)e^{i\th}f(\th,v,\omega,t).
\ee

Let us briefly sketch the derivation of equation (\ref{Kramers}), while the
details may be found in Ref. \cite{Gupta:2014}. We will along the way
also indicate how one may prove rigorously that the dynamics
(\ref{eom-scaled}) does not satisfy detailed balance unless $\sigma=0$.
For simplicity of presentation, we first
consider the case of a discrete bimodal $g(\omega)$, and then in the end
extend our discussion to a general $g(\omega)$. Then, consider a given
realization of $g(\omega)$ in which there are $N_{1}$ oscillators with
frequencies $\omega_{1}$ and $N_{2}$ oscillators with frequencies $\omega_{2}$, where $N_{1}+N_{2}=N$. 
Let us then define the $N$-oscillator distribution function
$f_{N}(\theta_{1},v_{1},\dots,\theta_{N_{1}},v_{N_{1}},\theta_{N_{1}+1},v_{N_{1}+1},\dots,\theta_{N},v_{N},t)$
as the probability density at time $t$ to observe the system around the
values $\{\theta_{i},v_{i}\}_{1\le i\le N}$. In the following, we use
the shorthand notations $z_{i}\equiv(\theta_{i},v_{i})$ and $\mathbf{z}=(z_{1},z_{2},\dots,z_{N})$. Note that
$f_{N}$ satisfies the normalization 
\be
\int\Big(\prod_{i=1}^{N}\dd z_{i}\Big)f_{N}(\mathbf{z},t)=1.
\ee
The distribution $f_N$ evolves in time according to the following Fokker-Planck equation that may be derived 
straightforwardly from the equations of motion
(\ref{eom-scaled}):
\begin{eqnarray}
\frac{\partial f_{N}}{\partial t}
&=&-\sum_{i=1}^{N}\Big[v_{i}\frac{\partial
f_{N}}{\partial\theta_{i}}-\frac{1}{\sqrt{m}}\frac{\partial(v_{i}f_{N})}{\partial
v_{i}}\Big]-\sigma\sum_{j=1}^{N}\Big(\Omega^{T}\Big)_{j}\frac{\partial
f_{N}}{\partial
v_{j}}+\frac{T}{\sqrt{m}}\sum_{i=1}^{N}\frac{\partial^{2}f_{N}}{\partial
v_{i}^{2}}\nonumber \\
&-&\frac{1}{2N}\sum_{i,j=1}^{N}\sin(\theta_{j}-\theta_{i})\Big[\frac{\partial
f_{N}}{\partial v_{i}}-\frac{\partial f_{N}}{\partial v_{j}}\Big], \label{eq:fp-eqn}
\end{eqnarray}
where the $N\times 1$ column vector $\Omega$ has its first $N_{1}$
entries equal to $\omega_{1}$ and the following $N_{2}$
entries equal to $\omega_{2}$, and where the superscript $T$ denotes matrix transpose
operation: 
\be
\Omega^T\equiv\left[\omega_{1} ~\omega_{1}
\dots~\omega_{1}~\omega_{2}\dots ~\omega_{2}\right].
\ee
\subsubsection{Proof that the dynamics does not
satisfy detailed balance unless $\sigma=0$}
\l{detailedbalance}
Let us rewrite the Fokker-Planck equation (\ref{eq:fp-eqn}) as
\begin{eqnarray}
&&\frac{\partial f_{N}(\mathbf{x})}{\partial t}
=-\sum_{i=1}^{2N}\frac{\partial[A_{i}(\mathbf{x})f_{N}(\mathbf{x})]}{\partial
x_{i}}+\frac{1}{2}\sum_{i,j=1}^{2N}\frac{\partial^{2}[B_{i,j}(\mathbf{x})f_{N}(\mathbf{x})]}{\partial x_{i}\partial x_{j}},\label{eq:fp-compact}
\end{eqnarray}
where 
\begin{equation}
x_i=\left\{ 
\begin{array}{ll}
               \theta_{i};i=1,2,\dots,N,  \\
               v_{i-N};i=N+1,\dots,2N,  
               \end{array}
        \right. \\ 
               \label{eq:x-defn} 
\end{equation}
and  
\begin{eqnarray}
\mathbf{x} & =\{x_{i}\}_{1\le i\le2N}.\label{eq:boldx-defn}
\end{eqnarray}
Here, the drift vector $A_{i}(\mathbf{x})$
is given by
\begin{equation}
A_{i}(\mathbf{x}) =\left\{
\begin{array}{ll}
                 v_{i};i=1,2,\dots,N, \\
                 -\frac{1}{\sqrt{m}}v_{i-N}
                 +\frac{1}{N}\sum_{j=1}^{N}\sin(\theta_{j}-\theta_{i-N})+\sigma\Big(\Omega^{T}\Big)_{i-N};\\
                 i=N+1,\dots,2N,
                 \end{array}
           \right. \\
                 \label{eq:Ai-defn}
\end{equation}
while the diffusion matrix $B_{i,j}(\mathbf{x})$ is 
\begin{equation}
B_{i,j}(\mathbf{x})  =\left\{
\begin{array}{ll}
              \frac{2T}{\sqrt{m}}\delta_{ij};i,j>N, \\
              0, ~{\rm Otherwise.}
              \end{array}
           \right. \\
           \label{eq:Bij-defn}
\end{equation}

The dynamics described by the Fokker-Planck equation of the form
(\ref{eq:fp-compact})
satisfies detailed balance if and only if the following conditions
are satisfied \cite{Gardiner:1983}:
\begin{eqnarray}
&&\epsilon_{i}\epsilon_{j}B_{i,j}(\epsilon\mathbf{x}) =B_{i,j}(\mathbf{x}),\label{eq:detailed-balance-cond1}\\
&&\epsilon_{i}A_{i}(\epsilon\mathbf{x})f_{N}^{s}(\mathbf{x})
=-A_{i}(\mathbf{x})f_{N}^{s}(\mathbf{x})+\sum_{j=1}^{2N}\frac{\partial
[B_{i,j}(\mathbf{x})f_{N}^{s}(\mathbf{x})]}{\partial x_{j}},\label{eq:detailed-balance-cond2}
\end{eqnarray}
where $f_{N}^{s}(\mathbf{x})$ is the stationary solution of equation
(\ref{eq:fp-compact}).
Here, $\epsilon_{i}=\pm1$ denotes the parity with
respect to time reversal of the variables $x_{i}$'s: Under time reversal,
we have $x_{i} \rightarrow\epsilon_{i}x_{i}$, where $\epsilon_{i}=-1$
(respectively, $+1$) depending on whether $x_{i}$ is odd (respectively,
even) under time reversal. For example, $\theta_{i}$'s
are even, while $v_{i}$'s are odd.

Using equation (\ref{eq:Bij-defn}), the condition
(\ref{eq:detailed-balance-cond1}) is trivially satisfied, while to check
 the condition given by (\ref{eq:detailed-balance-cond2}), we formally solve
 this equation for $f_{N}^{s}(\mathbf{x})$ and check if the solution solves equation
(\ref{eq:fp-compact}) in the stationary state. From equation (\ref{eq:detailed-balance-cond2}),
we see that for $i=1,2,\dots,N$, the condition reduces to 
\begin{eqnarray}
\epsilon_{i}A_{i}(\epsilon\mathbf{x})f_{N}^{s}(\mathbf{x}) &
=-A_{i}(\mathbf{x})f_{N}^{s}(\mathbf{x}).
\end{eqnarray}
The above equation, using equation (\ref{eq:Ai-defn}), is obviously satisfied. For $i=N+1,\dots,2N$,
we have 
\begin{eqnarray}
v_{k}f_{N}^{s}(\mathbf{x}) & =-\frac{T\partial
f_{N}^{s}(\mathbf{x})}{\partial v_{k}};k=i-N,\label{eq:cond2-eq1}
\end{eqnarray}
solving which we get 
\begin{eqnarray}
f_{N}^{s}(\mathbf{x}) & \propto
d(\theta_{1},\theta_{2},\dots,\theta_{N})\exp\Big[-\frac{1}{2T}\sum_{k=1}^{N}v_{k}^{2}\Big],
\label{eq:stationary-soln1}
\end{eqnarray}
where $d(\theta_{1},\theta_{2},\dots,\theta_{N})$ is a function to be determined.
Substituting the distribution (\ref{eq:stationary-soln1}) into equation (\ref{eq:fp-compact})
and requiring that it is a stationary solution implies that $\sigma$
has to be equal to zero, while 
\be
d(\theta_{1},\theta_{2},\dots,\theta_{N})=\exp\Big(-\frac{1}{2NT}\sum_{i,j=1}^{N}\Big[1-\cos(\theta_{i}-\theta_{j})\Big]\Big).
\ee
Thus, for $\sigma=0$, when the dynamics reduces to
that of the BMF model, we get the stationary
solution as
\begin{equation}
f_{N,\sigma=0}^{s}(\mathbf{z}) \propto\exp\Big[-\frac{H}{T}\Big].\label{eq:bmf-soln}
\end{equation}
where $H$ is the Hamiltonian (\ref{HMF-H}) (expressed in
terms of dimensionless variables introduced above). 
The lack of detailed balance for $\sigma \ne 0$ obviously extends to any distribution
$g(\omega)$.
\subsubsection{Derivation of the Kramers equation}
\l{kramers}
The starting point is to define the reduced
distribution function $f_{s_{1},s_{2}}$, with $s_1=0,1,2,\dots,N_{1}$
and $s_2=0,1,2,\dots,N_{2}$ as \cite{Huang:1987}
\begin{eqnarray}
&&f_{s_{1},s_{2}}(z_{1},z_{2},\dots,z_{s_{1}},z_{N_{1}+1},\dots,z_{N_{1}+s_{2}},t)\nonumber
\\
&&=\frac{N_{1}!}{(N_{1}-s_{1})!N_{1}^{s_{1}}}\frac{N_{2}!}{(N_{2}-s_{2})!N_{2}^{s_{2}}}\int
\dd z_{s_{1}+1}\dots \dd z_{N_{1}}\dd z_{N_{1}+s_{2}+1}\dots
\dd z_{N}f_{N}(z,t).
\label{eq:fs-defn}
\end{eqnarray}
Note that the following normalizations
hold for the single-oscillator distribution functions: 
\be
\int 
\dd z_{1}~f_{1,0}(z_{1},t)=1$, and $\int \dd z_{N_{1}+1}~f_{0,1}(z_{N_{1}+1},t)=1.
\ee

Assuming that 
\begin{enumerate}
\item{$f_{N}$ is symmetric with respect to permutations of dynamical variables within the same group of
oscillators, and}
\item{$f_N$, together with the derivatives $\partial
f_{N}/\partial v_{i} ~\forall~ i$, vanish on the boundaries of the phase
space,}
\end{enumerate}
and then using equation (\ref{eq:fp-eqn}) in equation (\ref{eq:fs-defn}), one
obtains the Bogoliubov-Born-Green-Kirkwood-Yvon
(BBGKY) hierarchy equations for the dynamics (\ref{eom-scaled}) (for
details, see \cite{Gupta:2014}). In
particular, the first equations of the hierarchy are
\begin{eqnarray}
&&\frac{\partial f_{1,0}(\theta,v,t)}{\partial t} +\frac{v\partial
f_{1,0}(\theta,v,t)}{\partial\theta}-\frac{1}{\sqrt{m}}\frac{\partial}{\partial
v}(vf_{1,0}(\theta,v,t))\nonumber \\
&&+\sigma\omega_{1}\frac{\partial f_{1,0}(\theta,v,t)}{\partial v}-\frac{T}{\sqrt{m}}\frac{\partial^{2}f_{1,0}(\theta,v,t)}{\partial v^{2}}\nonumber \\
&&=-\frac{N_{1}}{N}\int \dd
 \theta'\dd v'~\sin(\theta'-\theta)\frac{\partial
 f_{2,0}(\theta,v,\theta',v',t)}{\partial v}\nonumber \\
 &&-\frac{N_{2}}{N}\int \dd
 \theta'\dd v'~\sin(\theta'-\theta)\frac{\partial
 f_{1,1}(\theta,v,\theta',v',t)}{\partial v},\label{eq:fp-1}
\end{eqnarray}
and a similar equation for $f_{0,1}(\theta,v,t)$. In the limit $N \to
\infty$, writing 
\begin{equation}
g(\omega)=
\Big[\frac{N_{1}}{N}\delta(\omega-\omega_{1})+\frac{N_{2}}{N}\delta(\omega-\omega_{2})\Big],
\end{equation}
one can express equation (\ref{eq:fp-1}) in terms of $g(\omega)$.

In order to generalize the above treatment to the case of a continuous
$g(\omega)$, note for this case that the single-oscillator
distribution function is $f(\theta,v,\omega,t)$. The first
equation of the hierarchy is then
\begin{eqnarray}
&&\frac{\partial f(\theta,v,\omega,t)}{\partial t} +\frac{v\partial
f(\theta,v,\omega,t)}{\partial\theta}-\frac{1}{\sqrt{m}}\frac{\partial}{\partial
v}(vf(\theta,v,\omega,t))\nonumber \\
&&+\sigma\omega\frac{\partial f(\theta,v,\omega,t)}{\partial v}-\frac{T}{\sqrt{m}}\frac{\partial^{2}f(\theta,v,\omega,t)}{\partial v^{2}}\nonumber \\
&&=  -\int \dd \omega'\int \dd
\theta'\dd v'~g(\omega')\sin(\theta'-\theta)\frac{\partial
f(\theta,v,\theta',v',\omega,\omega',t)}{\partial v}.\label{eq:final-eqn}
\end{eqnarray}
In the continuum limit $N\to\infty$, one may neglect
oscillator-oscillator
correlations, and approximate $f(\theta,v,\theta',v',\omega,\omega',t)$
as
\bea
f(\theta,v,\theta',v',\omega,\omega',t)
&=&f(\theta,v,\omega,t)f(\theta',v',\omega',t)\nonumber \\
&+&{\rm corrections
~subdominant ~in}~N,
\eea
so that equation (\ref{eq:final-eqn}) reduces to the Kramers equation
(\ref{Kramers}).
\subsection{Stationary solutions of the Kramers equation}
\l{solutionKramers}
The stationary solutions of equation (\ref{Kramers}) are obtained
by setting the left hand side to zero. For $\sigma=0$, the stationary
solution is 
\be
f_{\rm st}(\th,v)\propto \exp[-(v^2/2-r_{\rm
st}\cos \th)/T],
\ee
that corresponds to canonical equilibrium, with $r_{\rm st}$ determined
self-consistently  \cite{Chavanis:2013}, see equation (\ref{rx-HMF}).
For $\sigma \ne 0$, the incoherent stationary state is
\cite{Acebron:2000} 
\be
f^{\rm inc}_{\rm st}(\th,v,\omega)=1/((2\pi)^{3/2}\sqrt{T})\exp[-(v-\sigma \omega
\sqrt{m})^2/(2T)].
\l{inc-state}
\ee

\begin{figure}[here!]
\centering
\includegraphics[width=100mm]{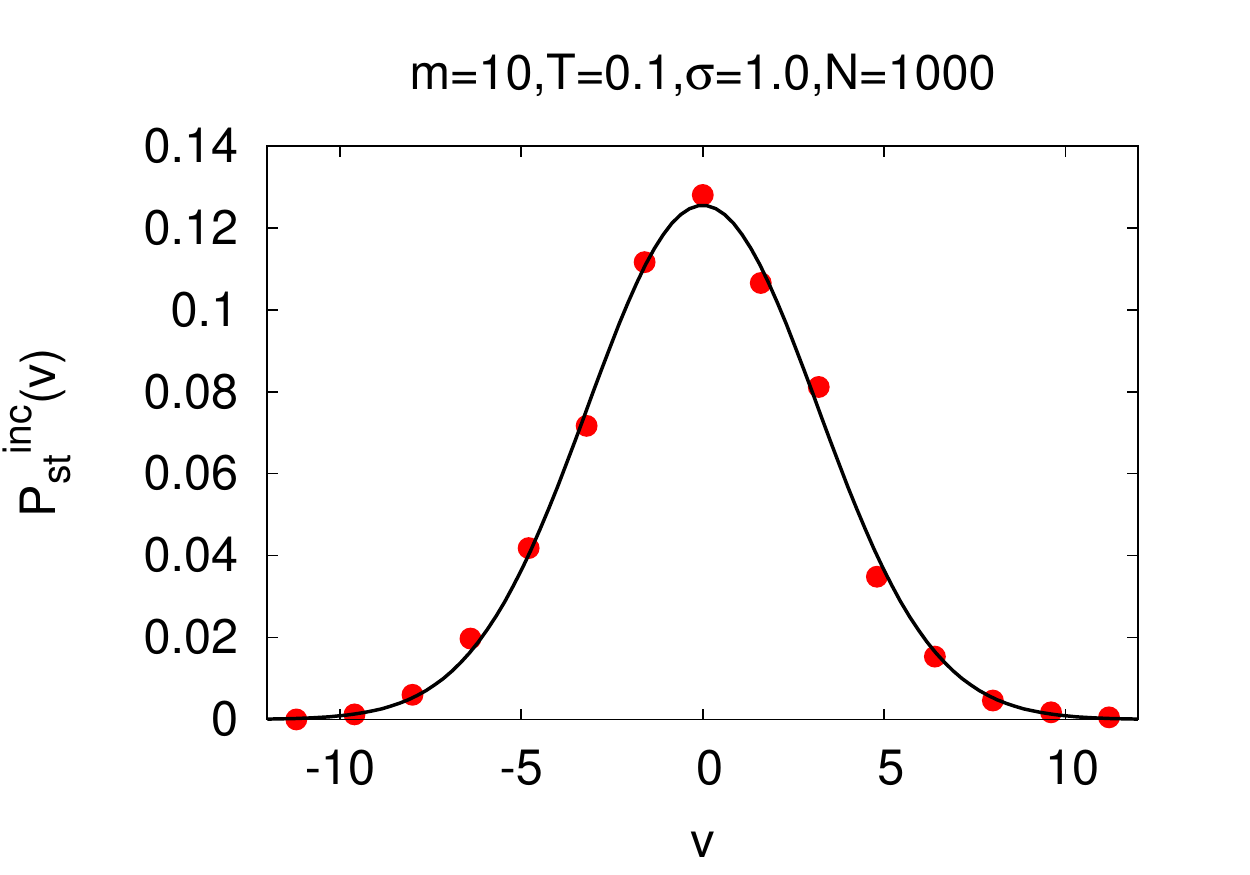}
\includegraphics[width=100mm]{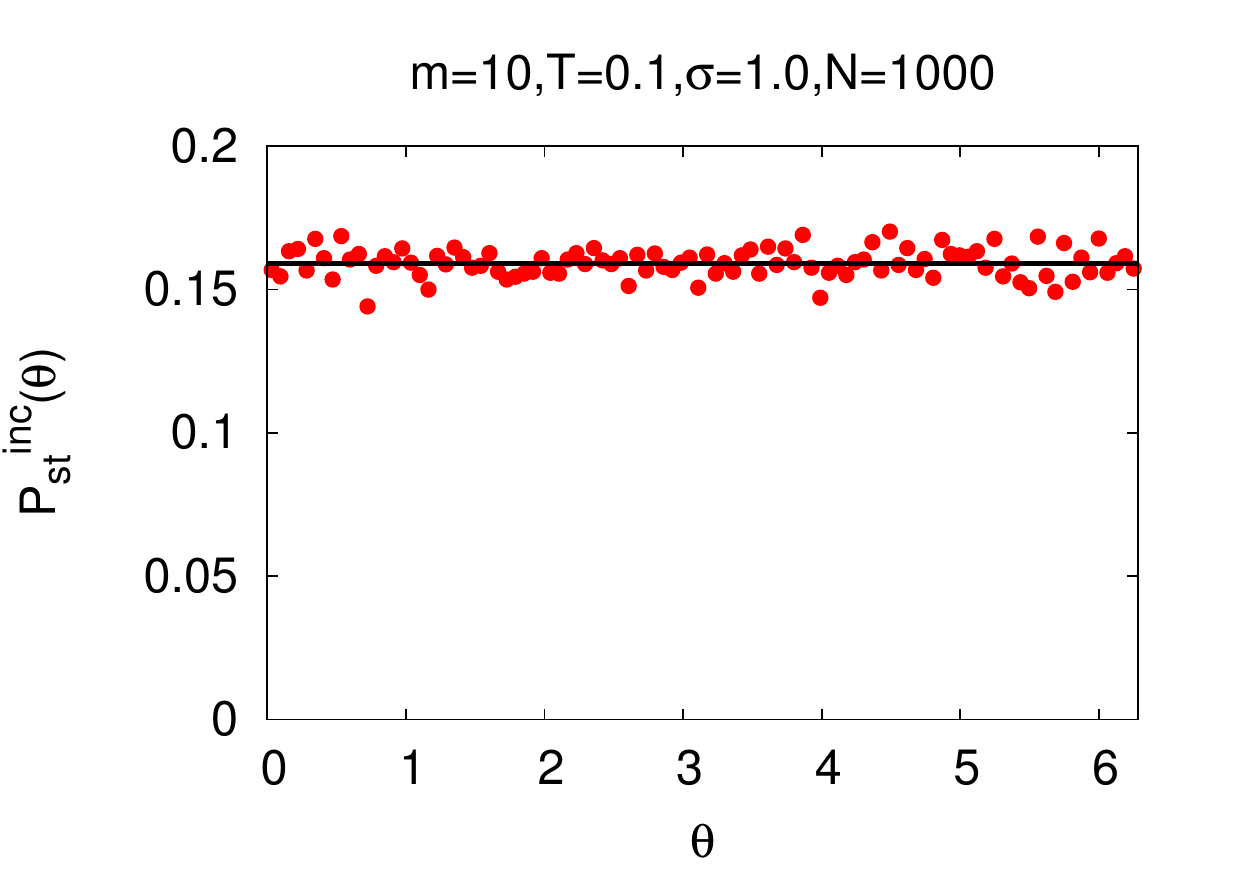}
\caption{For the dynamics (\ref{eom-scaled}), here we show the marginal distributions, $P^{\rm inc}_{\rm st}(v)$ and $P^{\rm
inc}_{\rm st}(\th)$, corresponding to the incoherent phase for $m=10,T =
0.1,\sigma=1.0$.  The points denoting
simulation data are for $N =1000$ for one fixed realization of the
$\omega_i$'s sampled from the Gaussian distribution
(\ref{gomega-gaussian}), while the continuous lines denote theoretical
results (\ref{marginalv}) and (\ref{marginalth}). }
\label{marginal-thv-inc}
\end{figure}

In the class of unimodal frequency distributions, let us consider a
representative $g(\omega)$, namely, a Gaussian:
\be
g(\omega)=\fr{1}{\sqrt{2\pi}}
\exp[-\omega^2/2].
\l{gomega-gaussian}
\ee
We then have for the marginal angular velocity distribution
\bea
P^{\rm inc}_{\rm st}(v)&=&\int_{-\infty}^\infty \dd \omega ~g(\omega)
\int_{-\pi}^{\pi} \dd\th ~f^{\rm inc}_{\rm st}(\th,v,\omega)\nonumber \\
&=&\sqrt{\fr{1}{2 \pi
T\Big(1+\sigma^2m/T\Big)}}\exp\Big[-\fr{v^2}{2T(1+\sigma^2m/T)}\Big],
\l{marginalv}
\eea
and the marginal angle distribution
\bea
P^{\rm inc}_{\rm st}(\th)&=&\fr{1}{2\pi}\int_{-\infty}^\infty \dd \omega
~g(\omega) \int_{-\infty}^{\infty} \dd v ~f^{\rm
inc}_{\rm st}(\th,v,\omega)=\fr{1}{2\pi},
\l{marginalth}
\eea
both correctly normalized to unity. In Figs. \ref{marginal-thv-inc} and
\ref{marginal-thv-syn}, we
compare our theoretical predictions, (\ref{marginalv}) and
(\ref{marginalth}), with numerical simulation results.

\begin{figure}[here!]
\centering
\includegraphics[width=100mm]{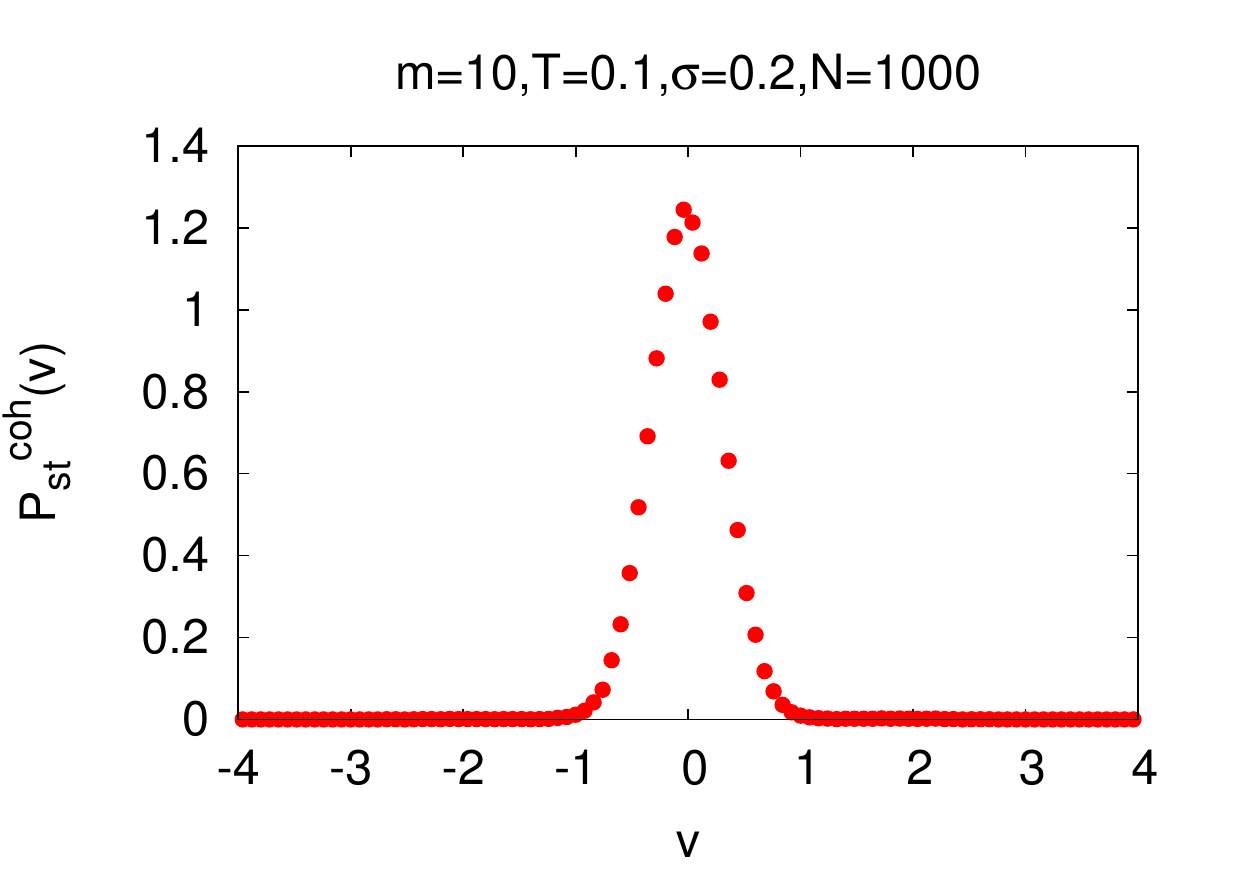}
\includegraphics[width=100mm]{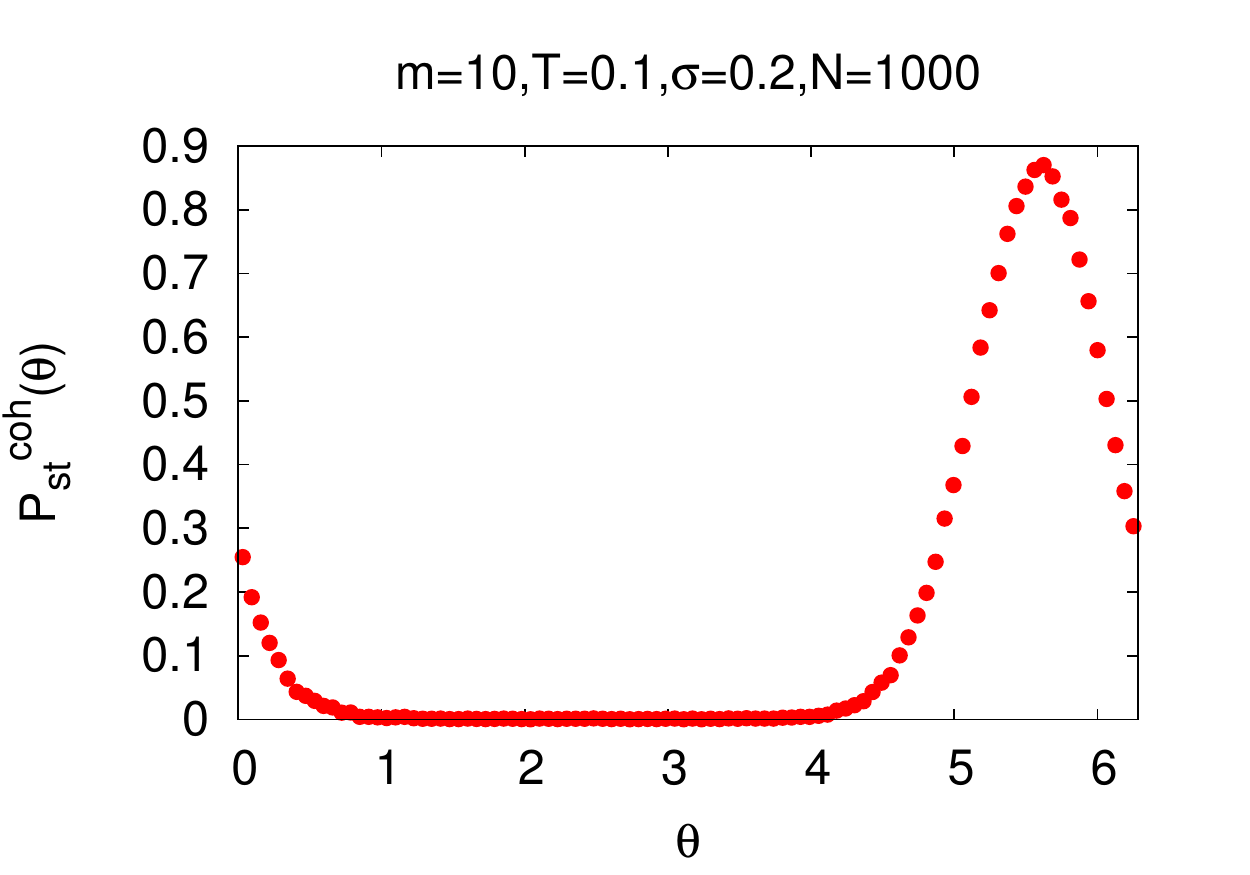}
\caption{For the dynamics (\ref{eom-scaled}), the figure shows the marginal distributions, $P^{\rm coh}_{\rm st}(v)$ and $P^{\rm
coh}_{\rm st}(\th)$, corresponding to the synchronized phase for $m=10,T
= 0.1,\sigma=0.2$.  The points denoting
simulation data are for $N =1000$ for one fixed realization of the
$\omega_i$'s sampled from the Gaussian distribution
(\ref{gomega-gaussian}).}
\l{marginal-thv-syn}
\end{figure}
The existence of the synchronized stationary state is borne out by our
simulation results in Fig. \ref{marginal-thv-syn}, although its analytical form is not
known.
\subsection{Linear stability analysis of the incoherent stationary state}
\l{linearstability-incoherent-inertia}
We now discuss about the linear stability analysis of
the incoherent state (\ref{inc-state}); a similar analysis for the BMF
model is discussed in Ref. \cite{Chavanis:20131}. Following Ref.
\cite{Acebron:2000}, we linearize equation (\ref{Kramers}) about the state by expanding
$f$ as 
\be
f(\th,v,\omega,t)=f^{\rm inc}_{\rm st}(\th,v,\omega)+e^{\lambda
t}\delta f(\th,v,\omega),
\l{expansion_linear}
\ee
where $\delta f \ll 1$ satisfies the linearized Kramers equation:
\begin{eqnarray}
&&\lambda \delta f + v\frac{\partial \delta f}
{\partial \theta} - \frac{\partial}{\partial
v}\Big(\fr{v}{\sqrt{m}}-\sigma\omega\Big)\delta f - \frac{T}{\sqrt{m}}\frac{\partial^2 \delta f}{\partial
v^2} \nonumber\\
&&=-\fr{\partial f_{\rm st}^{\rm inc}}{\partial v}
\int_{-\pi}^{\pi}\int_{-\infty}^{\infty}\int_{-\infty}^{\infty}\dd\phi \dd v
\dd\omega~g(\omega)\delta f(\phi,v,\omega)\sin(\phi-\th).
\l{linear-Kramers}
\end{eqnarray}
Since $f$ and $f^{\rm inc}_{\rm st}$ are normalized, we have
\be
\int_{-\pi}^{\pi}\int_{-\infty}^{\infty} \dd\th \dd v ~\delta
f(\th,v,\omega)=0.
\l{norm}
\ee

Substituting
\be
\delta f(\theta,v,\omega) = \sum_{n=-\infty}^{\infty} b_n(v,\omega,\lambda)e^{i n \th} 
\l{try}
\ee
in equation (\ref{linear-Kramers}), one gets
\bea
&&\fr{\dd^2 b_{n}}{\dd v^2}+\fr{1}{T}\Big(v-\sigma \omega
\sqrt{m}\Big)\frac{\dd b_{n}}{\dd
v}+\fr{1}{T}\Big(1-\lambda\sqrt{m}-inv\sqrt{m}\Big)b_n\nonumber
\\
&&=\fr{\sqrt{m}}{T}\fr{\partial f_{\rm st}^{\rm inc}}{\partial v} \pi (i \delta_{n,1} - i
\delta_{n,-1}) \langle 1,b_n
\rangle,
\l{b-eqn}
\eea
where one has the scalar product
\begin{eqnarray}
\langle \varphi,\psi \rangle \equiv \int_{-\infty}^{\infty}
\int_{-\infty}^{\infty} \dd v \dd \omega
~g(\omega)\varphi^*(v,\omega)\psi(v,\omega),
\l{scalar-prod}
\end{eqnarray}
with $*$ denoting complex conjugation.
Since $\delta f$ is real, one has $b_{-n} =b_{n}^*$, while equation (\ref{norm}) implies that $b_0 = 0$.
We can then restrict to consider only $n \ge 0$.
Next, equation (\ref{b-eqn}) is transformed into a nonhomogeneous
parabolic cylinder equation by the transformations
\bea
&&b_{n}(v,\omega,\lambda) = \exp\Big[- \fr{(v-\sigma
\omega\sqrt{m})^{2}}{4T}\Big]\beta_n (z,\omega,\lambda), 
\l{trans}\\
&&z = \fr{1}{\sqrt{T}}(v-\sigma\omega\sqrt{m} + 2nT\sqrt{m}i),
\eea
which when substituted into equation (\ref{b-eqn}) yield
\begin{eqnarray}
\frac{\dd^{2} \beta_{n}}{\dd z^2}+ \left[\frac{1}{2} - {z^{2}\over 4}
- \sqrt{m}(\lambda + in\sigma\omega\sqrt{m} + n^2 T\sqrt{m})\right]\, \beta_n\nonumber\\
=i\pi \sqrt{m}\fr{\partial f_{\rm st}^{\rm inc}}{\partial v}
e^{{1\over 4}(z-2i\sqrt{mT} )^{2}}\, \langle 1,
e^{-{1\over 4}(z-2i\sqrt{mT} )^{2}} \beta_1 \rangle\,
\delta_{n,1}.  
\end{eqnarray}
For $n\neq 1$, the right hand side of the above equation is zero, yielding the eigenvalues
\begin{eqnarray}
\lambda_{p,n}(\omega)
= -\fr{p}{\sqrt{m}} - n^2 T\sqrt{m} - i n \sigma\omega\sqrt{m},\quad p = 0, 1, 2,\ldots,
\l{eigenvalues}
\end{eqnarray}
and the corresponding eigenfunctions
\begin{eqnarray}
\beta_{p,n}(z,\omega,\lambda_{p,n}) = D_p (z) =
2^{-\fr{p}{2}}e^{-\fr{z^{2}}{4}}H_p\Big(\fr{z}{\sqrt{2}}\Big),
\end{eqnarray}
that do not depend on $n$ and $\omega$; here, $D_p(z)$
and $H_p(x)$ are respectively the parabolic cylinder function and the Hermite
polynomial of degree $p$ \cite{Gradshteyn:1980}. The
eigenvalues $\lambda_{p,n}(\omega)$ form a continuous spectrum. All of them have negative real parts, thus leading to linear
stability of the incoherent state (\ref{inc-state}), for $n=2,3,\ldots$ and $p=0,1,2,\ldots$. For $n=0$ the eigenvalues
have also negative real parts unless those with $p=0$, that have a vanishing real part. They would correspond to neutrally
stable modes; however, the modes with $n=0$ have zero amplitude due to the normalization condition (\ref{norm}).

For $n=1$, solving (\ref{b-eqn}) gives
\bea
\beta_1(z,\omega,\lambda)&=&-i\pi \langle 1,e^{-\left(\fr{z}{2} -
i\sqrt{mT}\right)^{2}}
\beta_1 \rangle \nonumber \\
&\times&\sum_{p=0}^{\infty} \fr{\int_{-\infty}^{\infty}\dd
z_1~e^{\left(\fr{z_1}{2} - i\sqrt{mT}\right)^{2}} D_p [f_{\rm st}^{\rm
inc}]'}{\sqrt{2\pi} p! \left(\fr{p}{\sqrt{m}} + \lambda +
i\sigma\omega\sqrt{m} +T\sqrt{m}\right)} D_p(z), 
\eea
where
\begin{eqnarray}
[f_{\rm st}^{\rm inc}(v)]' &=& \left. \fr{\partial f_{\rm st}^{\rm inc}}{\partial v}
\right|_{v= \sigma\omega\sqrt{m} - i2T\sqrt{m} + \sqrt{T} z}= - {(z -
2i\sqrt{mT})\over
(2\pi)^{{3\over 2}}T}e^{-{1\over 2}(z - 2i\sqrt{mT})^{2}},
\end{eqnarray}
Using the above expression to compute $\langle 1,e^{-\left({z\over 2} - i\sqrt{mT}\right)^{2}}
\beta_1 \rangle$, one obtains from the resulting self-consistent
equation the following eigenvalue equation for $\lambda$
\cite{Acebron:2000}:
\be
\fr{e^{mT}}{2T}\sum_{p=0}^\infty
\frac{(-m T)^p(1+\frac{p}{mT})}{p!}\int\limits_{-\infty}^{\infty}
\frac{g(\omega)\dd\omega}{
1+\frac{p}{mT}+i\frac{\sigma\omega}{T}+\frac{\lambda}{T\sqrt{m}}} = 1.
\l{stability-eqn}
\ee
\begin{figure}[here!]
\centering
\includegraphics[width=100mm]{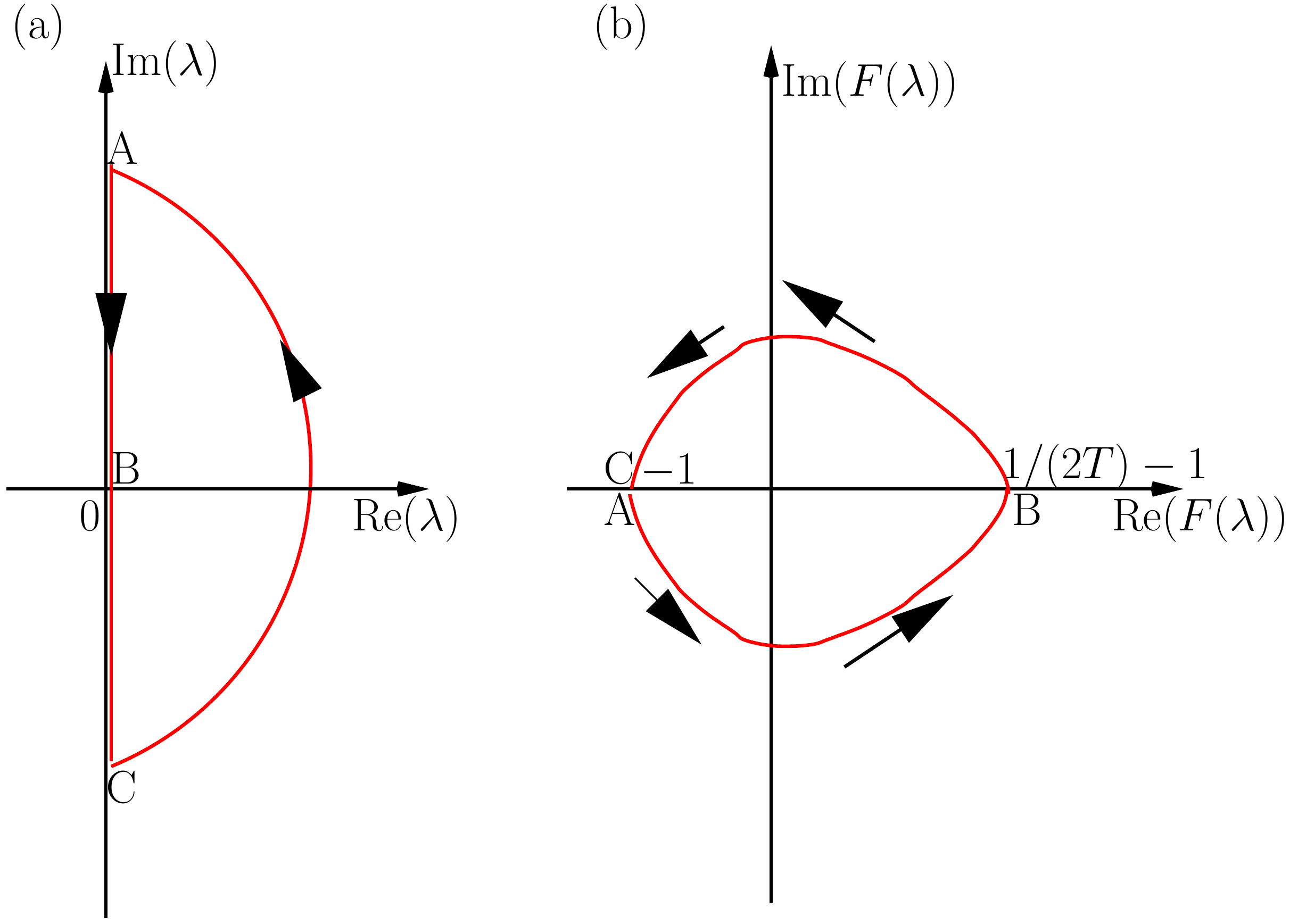}
\caption{The loop in the complex $F$-plane, (b), corresponding to the
loop in the complex $\lambda$-plane, (a), as determined by the function
$F(\lambda)$ in equation (\ref{eqfond}).}
\label{sm-fig1}
\end{figure}
\subsubsection{Analysis of the eigenvalue equation}
\l{eigenvalueequation}
A detailed analysis of the eigenvalue equation (\ref{stability-eqn}), carried out in
Ref. \cite{Gupta:2014}, shows that the equation admits at most
one solution for $\lambda$ with a positive real part, and when the
solution exists, it is necessarily real. We now briefly sketch the
analysis.
We rewrite equation (\ref{stability-eqn}) as
\begin{eqnarray}
\label{eqfond}
&&F(\lambda;m,T,\sigma)\nonumber \\
&&=\frac{e^{mT}}{2T}\sum_{p=0}^\infty
\frac{\left(-mT\right)^p \left(p+mT\right)}{p!} \int \dd \omega \, \frac{g(\omega)}{mT+p+\sqrt{m}\lambda + i\sigma m \omega} - 1 = 0.
\end{eqnarray}
where $g(\omega)$ is unimodal. 
The incoherent state (\ref{inc-state}) is unstable if there is a $\lambda$ with a positive
real part that satisfies the above eigenvalue equation. 

We first look for possible pure imaginary solutions $\lambda = i\mu$.
Separating equation (\ref{eqfond}) into real and imaginary parts, we have
\begin{eqnarray}\label{iml_real}
&&{\rm Re} \left[F(i\mu;m,T,\sigma)\right]\nonumber \\
&&=\frac{e^{mT}}{2T}\sum_{p=0}^\infty
\frac{\left(-mT\right)^p}{p!} \int \dd \omega \, g(\omega)
\frac{\left(p+mT\right)^2}{\left(p+mT\right)^2+\left(m\sigma \omega
+\sqrt{m}\mu\right)^2} - 1 = 0, \\ 
\label{iml_imag}
&&{\rm Im} \left[F(i\mu;m,T,\sigma)\right]\nonumber \\
&&=-\frac{e^{mT}}{2T}
\sum_{p=0}^\infty \frac{\left(-mT\right)^p}{p!}\int \dd \omega \,
g(\omega) \frac{\left(p+mT\right)\left(m\sigma \omega + \sqrt{m}\mu \right)}
{\left(p+mT\right)^2+\left(m\sigma \omega + \sqrt{m}\mu\right)^2}= 0.
\end{eqnarray}
In the second equation above, making the change of variables $m\sigma
\omega + \sqrt{m}\mu = m\sigma x$, and exploiting the parity in $x$ of
the sum, we get
\begin{eqnarray}\label{iml_imag_2}
\fl
{\rm Im} \left[F(i\mu;m,T,\sigma)\right]&=&-\frac{e^{mT}}{2T} m\sigma\int_0^\infty \dd
x\Big\{ \left[g\left(x-\frac{\mu}{\sqrt{m}\sigma}\right)
-g\left(-x-\frac{\mu}{\sqrt{m}\sigma}\right)\right]\nonumber \\
&\times& x \sum_{p=0}^\infty \frac{\left(-mT\right)^p}{p!}
\frac{p+mT}{\left(p+mT\right)^2+m^2\sigma^2 x^2} \Big\} = 0.
\end{eqnarray}
It is possible to show that the sum on the
right-hand side is positive definite for any finite $\sigma$, while for
our class of unimodal $g(\omega)$'s, the term within the 
square brackets is positive (respectively, negative) definite for $\mu >0$
(respectively, for $\mu <0$). Therefore, the last equation is never
satisfied for $\mu \ne 0$, implying thereby that the eigenvalue equation
(\ref{eqfond}) does not admit pure imaginary solutions (the proof holds also for the particular case
$g(\omega) = \delta(\omega)$). This analysis also proves
that there can be at most one solution of equation (\ref{eqfond})
with positive real part. In fact, let us consider, in the complex $\lambda$-plane, the loop
$A-B-C-A$ depicted in Fig. \ref{sm-fig1}(a), with $A$ and $C$ representing ${\rm Im}\lambda \to \pm \infty$,
respectively, and the radius of the arc $C-A$ going to $\infty$. Due to the sign properties
of ${\rm Im} \left[F(i\mu;m,T,\sigma)\right]$ just described, we obtain in the
complex-$F(\lambda)$ plane the loop qualitatively represented in Fig. \ref{sm-fig1}(b). While
the point $F=-1$ is obtained when $\lambda$ is at the points $A$ and $C$, the point $B$ is determined by
the value of $F(0)$ given by
\begin{eqnarray}\label{rel_real}
&&F(0;m,T,\sigma) = \frac{e^{mT}}{2T}\sum_{p=0}^\infty
\frac{\left(-mT\right)^p}{p!}\int \dd \omega \, g(\omega)
\frac{\left(p+mT\right)^2}{\left(p+mT\right)^2+\left(m\sigma \omega \right)^2} - 1.
\end{eqnarray}
From a well-known theorem of complex analysis \cite{Smirnov:1964}, we
therefore conclude that
for $F(0;m,T,\sigma)>0$, there is one and only one solution of the
eigenvalue equation with positive real part and no solution for
$F(0;m,T,\sigma)<0$. When the single solution with
positive real part exists, it is necessarily real, since a complex solution
would imply the presence of its complex conjugate. For $\sigma=0$, one
has $F(0;m,T,0)=1/(2T)-1$. For $\sigma > 0$, 
the value of $F(0;m,T,\sigma)$ depends on the distribution function
$g(\omega)$. It is possible to prove that the value is
always smaller than $1/(2T)-1$; this is reasonable since if the incoherent state is stable for
$\sigma =0$, which happens when $T>1/2$, it is {\it a fortiori} stable for $\sigma > 0$.

The surface delimiting the region of instability of the incoherent state
(\ref{inc-state}) in the $(m,T,\sigma)$
phase space is implicitly given by equation (\ref{rel_real}) that may be solved to obtain
the stability threshold $\sigma^{\rm inc}=\sigma^{\rm inc}(m,T)$. It is
reasonable to expect on physical grounds that the threshold is a single-valued function, and that for any
given value of $m$, it is a decreasing function of $T$ for $0\le T \le
1/2$, reaching $0$ for $T=1/2$. These facts may be proved 
analytically for the class of unimodal distributions functions
$g(\omega)$ considered in this work. Also, one can prove for any $g(\omega)$ that
$\sigma^{\rm inc}(m,T)$ approaches $0$ as $m\to \infty$, by using the integral representation
\bea\label{integr_repr_2}
&&\sum_{p=0}^\infty \frac{\left(-mT\right)^p}{p!}
\frac{\left(p+mT\right)^2}{\left(p+a\right)^2+\left(m\sigma \omega
\right)^2}\nonumber \\
&&=e^{-mT}- \left(m\sigma \omega \right) \int_0^\infty \dd t \, \exp \left[ -mT \left( t + e^{-t}\right)\right]\sin \left(m\sigma \omega t \right).
\eea
For $\sigma>0$, as $m\to \infty$, the term with the integral in the right-hand side
of the last equation tends to $e^{-mT}$, so that equation
(\ref{rel_real}) gives $F(0;m\to \infty,T>0,\sigma>0)=-1$.
Combined with the fact that $F(0;m,T,0)=1/(2T)-1$, we get
$\sigma^{\rm inc}(m\to \infty,0\le T \le 1/2)=0$.
Turning to a representative Gaussian case, equation
(\ref{gomega-gaussian}), and using the subscript $g$ to distinguish
results for this
case, equation (\ref{integr_repr_2}) gives
\begin{eqnarray}\label{rel_real_b_gauss_3}
&&F_g(0;m,T,\sigma) = \frac{1}{2T}-1- \frac{1}{2T}\int_0^\infty \dd y \,
y e^{-\frac{y^2}{2}}\exp \left[ mT \left( 1 - \frac{y}{m\sigma} -
e^{-\frac{y}{m\sigma}}\right)\right].
\end{eqnarray}
The equation $F_g(0;m,T,\sigma)=0$ defines implicitly the function
$\sigma^{\rm inc}(m,T)$. We can show that this is a single-valued
function with the properties $\partial \sigma^{\rm inc}/\partial m<0$
and $\partial \sigma^{\rm inc}/\partial T<0$.
We have
\begin{eqnarray}\label{rel_real_b_gauss_dermass}
\frac{\partial}{\partial m}F_g(0;m,T,\sigma)&=&- \frac{1}{2}\int_0^\infty \dd y \,
y e^{-\frac{y^2}{2}}\nonumber \\
&\times&\left( 1-
e^{-\frac{y}{m\sigma}} - \frac{y}{m\sigma}e^{-\frac{y}{m\sigma}}
\right) \exp \left[ mT \left( 1 - \frac{y}{m\sigma} -
e^{-\frac{y}{m\sigma}}\right)\right],
\end{eqnarray}
which is negative as $1-e^{-x} -xe^{-x}$ is positive for
$x>0$. From the implicit function theorems, we then
derive that $\partial \sigma^{\rm inc}/\partial m<0$. On the
other hand, we have
\begin{eqnarray}\label{rel_real_b_gauss_dersigma}
\frac{\partial}{\partial \sigma}F_g(0;m,T,\sigma)&=& - \frac{1}{2\sigma^2}\int_0^\infty \dd y \,
y^2 e^{-\frac{y^2}{2}} \left( 1- e^{-\frac{y}{m\sigma}} \right)\nonumber
\\
&\times&\exp \left[ mT \left( 1 - \frac{y}{m\sigma}
- e^{-\frac{y}{m\sigma}}\right)\right],
\end{eqnarray}
which is clearly negative. Since we are considering $T>0$, multiplying
equation (\ref{rel_real_b_gauss_3}) by $2T$ gives
\begin{eqnarray}\label{rel_real_b_gauss_4}
&&2T F_g(0;m,T,\sigma) = 1 - 2T- \int_0^\infty \dd y \,
y e^{-\frac{y^2}{2}}\exp \left[ mT \left( 1 - \frac{y}{m\sigma} - e^{-\frac{y}{m\sigma}}\right)\right].
\end{eqnarray}
Considering the integral on the right-hand side, since $1-x-e^{-x}$ is
negative for $x>0$, the $T$-derivative of the integral
is negative, while its second $T$ derivative is positive. Then the
right-hand side of equation (\ref{rel_real_b_gauss_4}) for $T>0$ can be zero for at most one value of $T$. Furthermore, since for
fixed $y$ and $m$ the value of $y/(m\sigma)$ decreases if $\sigma$ increases, the $T$ value for which
$F_g(0;m,T,\sigma)=0$ decreases for increasing $\sigma$ at fixed $m$.
This concludes the proof. Furthermore, for what we have seen before,
$\sigma^{\rm inc}(m,1/2)=0$ and
$\lim_{m\to \infty}\sigma^{\rm inc}(m,T)=0$ for $0\le T \le 1/2$.

It is evident from the above analysis that the proof is not
restricted to the Gaussian case, but works equally well for any $g(\omega)$ such that
\begin{equation}\label{intsin_gen}
\beta \int \dd x \, g(x) x \sin (\beta x),
\end{equation}
is positive for any $\beta$. However, on physical grounds, we are led to assume that the same conclusions
hold for any even single-humped $g(\omega)$. 

We conclude on the basis of the above analysis that $\lambda=0$ at the point of neutral
stability, so that equation (\ref{stability-eqn}) gives $\sigma^{\rm inc}(m,T)$ to be satisfying 
\be
\frac{2T}{e^{mT}}=\sum_{p=0}^\infty
\frac{(-mT)^p(1+\frac{p}{mT})^2}{p!}\int\limits_{-\infty}^{\infty}
\frac{g(\omega)d\omega}{(1+\frac{p}{mT})^2+\frac{(\sigma^{\rm
inc})^2\omega^2}{T^2}}.
\l{sigma_inc}
\ee
In the $(m,T,\sigma)$ space, the above equation defines the stability surface $\sigma^{\rm
inc}(m,T)$. There will similarly be the stability surface $\sigma^{\rm
coh}(m,T)$. The two surfaces coincide on the critical lines on the
$(T,\sigma)$ and $(m,T)$ planes where the transition becomes continuous;
outside these planes, the surfaces enclose the first-order transition surface
$\sigma_c(m,T)$ i.e., $\sigma^{\rm coh}(m,T) > \sigma_c(m,T) >\sigma^{\rm
inc}(m,T)$. 
Let us show
by taking limits that the surface $\sigma^{\rm
inc}(m,T)$ meets the critical lines on the $(T,\sigma)$ and
$(m,T)$ planes, and also obtain its intersection with the $(m,\sigma)$-plane.
On considering $m \to 0$ at a fixed $T$, only the $p=0$ term in the
sum in equation (\ref{sigma_inc}) contributes, so that one has
\be
\lim_{m \to 0, T \,{\rm fixed}}\sigma^{\rm inc}(m,T)=\sigma_c(m=0,T),
\ee
with the implicit expression of $\sigma_c(m=0,T)$ given earlier.
Similarly, one has 
\be
\lim_{T \to T_c^-, m \,{\rm
fixed}}\sigma^{\rm inc}(m,T)=0.
\ee
When $T \to 0$ at a fixed $m$, we get 
\be
\sigma^{\rm inc}_{\rm noiseless}(m)\equiv \lim_{T \to 0, m \,{\rm fixed}}\sigma^{\rm
inc}(m,T),
\ee
with 
\bea
&&1=\fr{\pi g(0)}{2\sigma^{\rm inc}_{\rm noiseless}}-\fr{m}{2}
\int_{-\infty}^{\infty}d\omega \fr{g(\omega)}{1+m^2(\sigma^{\rm
inc}_{\rm noiseless})^2\omega^2}.
\eea

For the representative case of the Gaussian $g(\omega)$, equation
(\ref{gomega-gaussian}), we get from equation (\ref{sigma_inc}) that
\be
1=\frac{e^{mT}\sqrt{\pi}}{2\sqrt{2}\sigma^{\rm inc}} \sum_{p=0}^\infty
\frac{(-mT)^p(1+\frac{p}{mT})}{p!e^{-\frac{T^2(1+p/mT)^2}{2(\sigma^{\rm
inc})^2}}}{\rm
Erfc}\Big[\frac{T(1+\frac{p}{mT})}{\sigma^{\rm inc}\sqrt{2}}\Big],
\l{Gaussian-eqn}
\ee
where ${\rm Erfc}(x)$ is the complementary error function: ${\rm
Erfc}(x) = \frac{2}{\sqrt{\pi}} \int_x^\infty \dd t \, e^{-t^2}$.

\begin{figure}[here!]
\centering 
\includegraphics[width=120mm]{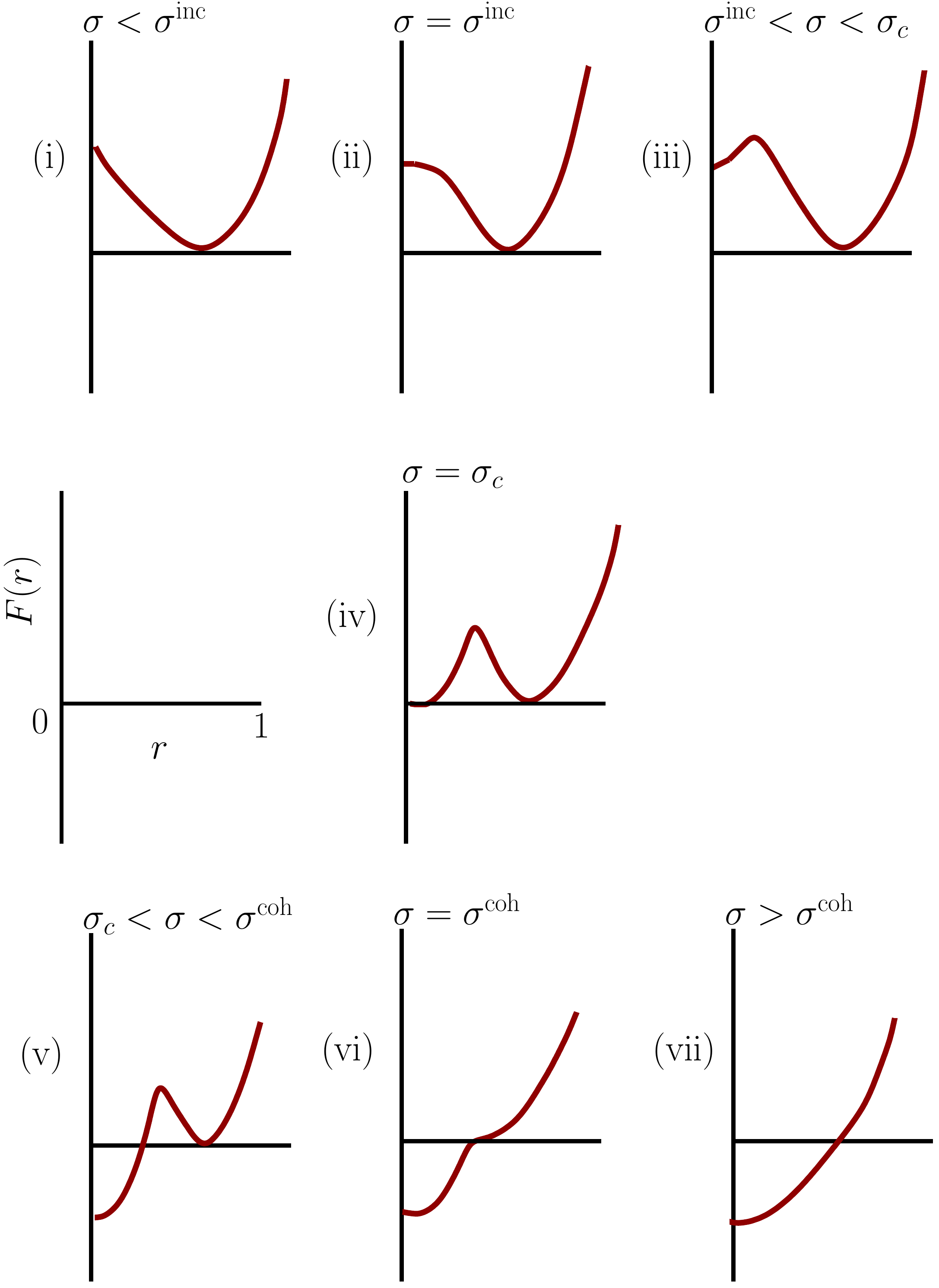}
\caption{(Color online) Considering the model (\ref{eom-scaled}), we
show here schematic Landau free energy $F(r)$ vs. $r$ for first-order
transitions at fixed $m$ and $T$ while varying $\sigma$.
Panels (i) and (vii) correspond to the synchronized and incoherent 
phase being at the global minimum. In panel (iii) (respectively, (v)), the
synchronized (respectively, incoherent) phase is at the global
minimum, while the incoherent (respectively, synchronized) phase is at a
local minimum, hence, metastable. Panel (iv) corresponds to the first-order transition point,
with the two phases coexisting at two minima of equal heights.}
\l{fig:fe}
\end{figure}
\subsection{Comparison with numerical simulations}
\l{numerics}
Choosing $m=20$, and $T=0.25$, equation (\ref{Gaussian-eqn}) gives $\sigma^{\rm inc}(m,T) \approx
0.10076$. Then, starting with the incoherent state (\ref{inc-state}) at
a given $\sigma$ and evolving under the dynamics (\ref{eom-scaled}), our
theoretical continuum-limit analysis predicts that the order parameter $r$ for $\sigma < \sigma^{\rm
inc}$ relaxes at long times from its value equal to $0$ to its
stationary state
value corresponding to the synchronized phase. For $\sigma >
\sigma^{\rm inc}(m,T)$, on the other hand, $r$ remains zero for all times.
\begin{figure}[here!]
\centering
\begin{tabular}{lr}
\parbox[l]{7cm}{
\includegraphics[width=85mm]{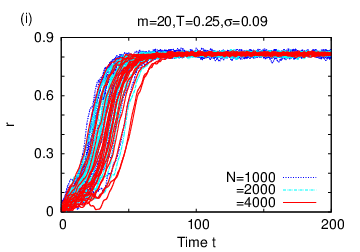}
\\
\includegraphics[width=85mm]{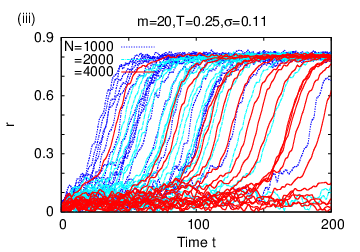}
}&
\parbox[r]{7cm}{
\includegraphics[width=85mm]{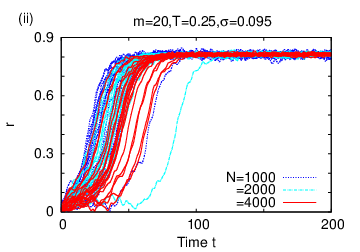}\\
\includegraphics[width=85mm]{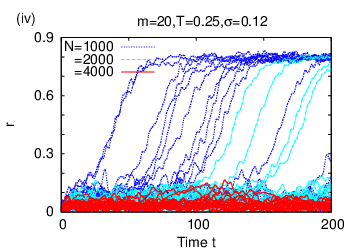}
}
\end{tabular}
\caption{For the dynamics (\ref{eom-scaled}), panels
(i)-(iv) show $r$ vs. time at $m=20, T=0.25$ for four values of
$\sigma$, two below ((i): $\sigma=0.09$, (ii): $\sigma=0.095$), and two
above ((iii): $\sigma=0.11$, (iv): $\sigma=0.12$) the theoretical
threshold $\sigma^{\rm inc}(m,T)\approx 0.10076$. The data are obtained
from simulations for the 
Gaussian $g(\omega)$ given by equation (\ref{gomega-gaussian}).}
\l{fig:r-vs-t-20real}
\end{figure}

\begin{figure}[here]
\centering
\includegraphics[width=100mm]{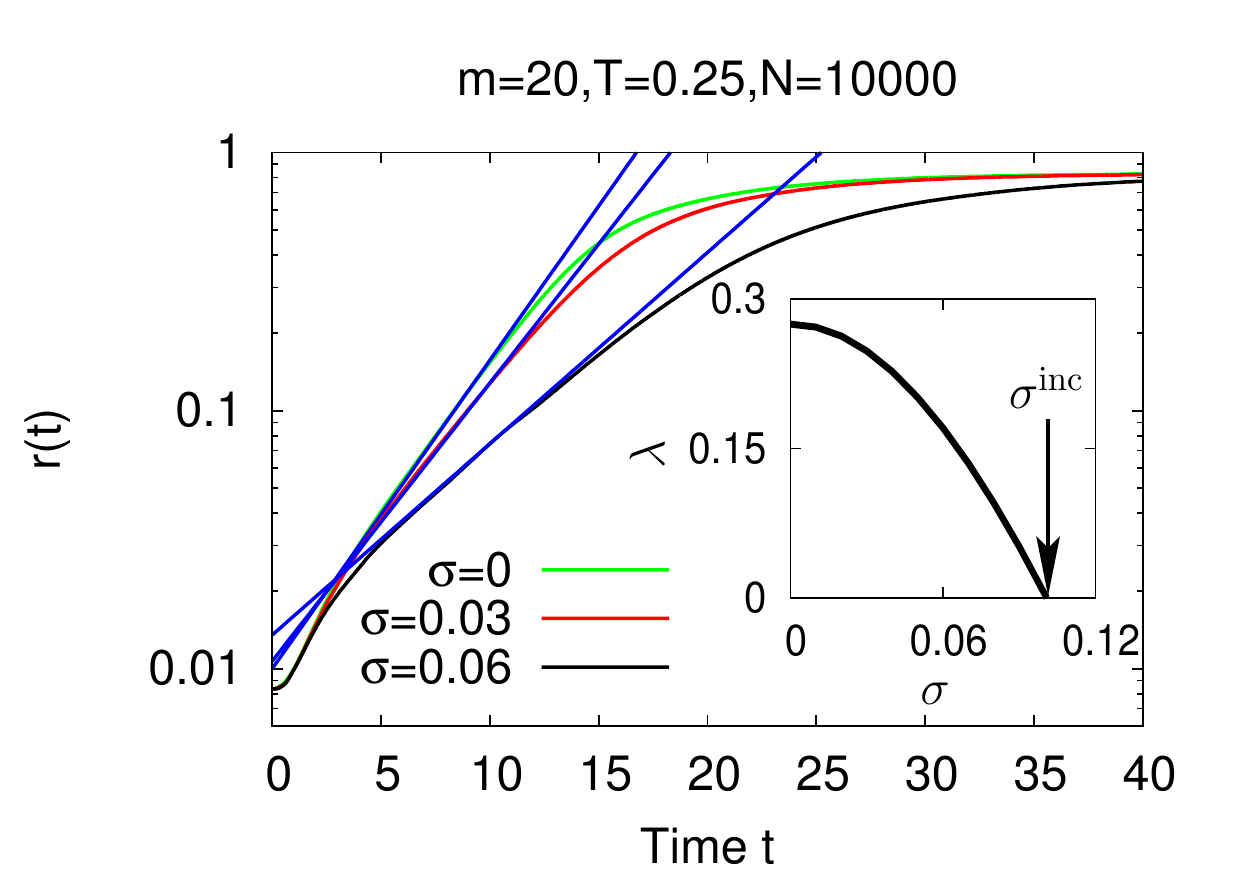}
\caption{Considering the dynamics (\ref{eom-scaled}), we show here exponentially fast relaxation $\sim e^{\lambda t}$ of $r$ from
its initial incoherent state value to its final synchronized state value for
$\sigma<\sigma^{\rm inc}(m,T)\approx 0.10076$ for the Gaussian
$g(\omega)$ given by equation (\ref{gomega-gaussian}), and for $m=20,T=0.25,N=10^4$; the blue solid lines stand for exponential
growth with rates $\lambda$ obtained from equation (\ref{stability-eqn}) 
by using equation (\ref{gomega-gaussian}) for $g(\omega)$. The inset shows theoretical $\lambda$ as a function of
$\sigma$ for the same $m$ and $T$ values; in particular, $\lambda$ hits
zero at the stability threshold $\sigma^{\rm inc}(m,T)$. The data are
obtained from simulations with $N=10000$ for the Gaussian $g(\omega)$ given by equation (\ref{gomega-gaussian}).} 
\l{fig:growth-rates}
\end{figure}

\begin{figure}[h!]
\centering
\includegraphics[width=100mm]{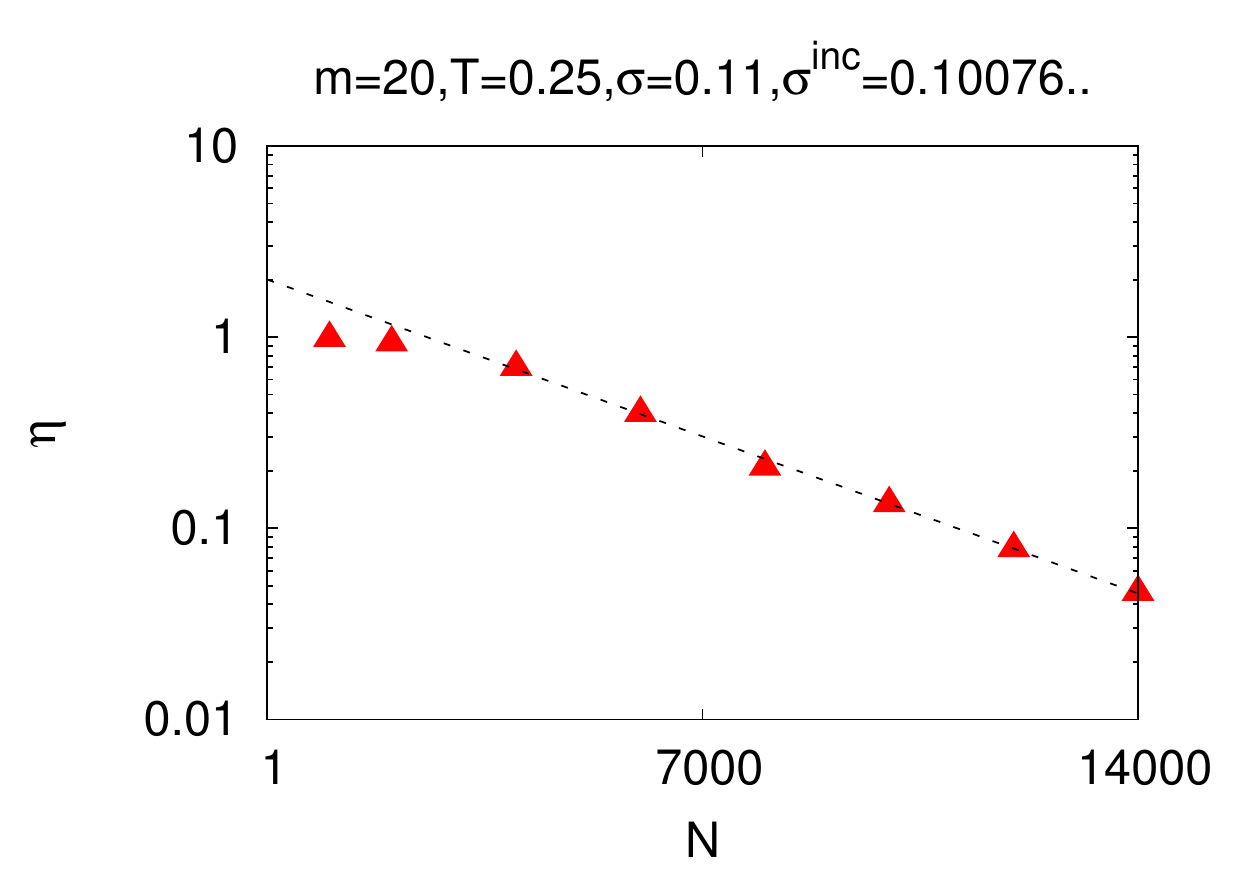}
\caption{For the dynamics (\ref{eom-scaled}) with $m=20,T=0.25,\sigma=0.11$, the figure shows the
fraction $\eta$ of realizations of initial incoherent state
relaxing to synchronized state within the fixed time of observation $t=200$,
for a value of $\sigma$ above $\sigma^{\rm inc}(m,T)$, for which the
incoherent phase is linearly stable in the continuum limit. The figure
shows that $\eta$ for large $N$ decreases exponentially fast with
increase of $N$. The data are obtained from $N$-body simulations for the
Gaussian $g(\omega)$ given by equation (\ref{gomega-gaussian}).} 
\l{fig:eta-N}
\end{figure}

In this subsection, we compare the above continuum-limit theoretical
predictions with $N$-body simulations. A phenomenological
picture of viewing dynamically a phase transition is to model the dynamics
as dissipative relaxation of the order parameter towards the minimum of
a phenomenological Landau free-energy landscape \cite{Binder:1987}. For a first-order phase
transition, we draw in Fig. \ref{fig:fe} the
corresponding schematic free energy landscapes $F(r)$ vs. $r$ for fixed $m$ and $T$ at different
$\sigma$ values. Note that for non-zero $\sigma$, one should instead be
drawing landscapes of the large
deviation functional; here, we assume that the landscape picture of
phase transitions will also hold in that case.   
The landscapes in Fig. \ref{fig:fe} explain the occurrence of flips in
$r$ shown in Fig. \ref{fig:r-vs-t-pr}(a): the
flips correspond to dynamics at $\sigma$ values close to $\sigma_c$ at
which the system switches back and forth between the two almost-stable
synchronized and incoherent states, thereby leading to the bistability
in Fig. \ref{fig:r-vs-t-pr}(a). 

In order to check our theoretical estimate of $\sigma^{\rm inc}(m,T)$
for the Gaussian $g(\omega)$, equation (\ref{gomega-gaussian}), we perform the following experiment.
For a given large value of $N$, we prepare for the dynamics (\ref{eom-scaled}) a
realization of an initial state that is incoherent, by sampling the $\omega_i$'s
independently for each $i$ from the distribution (\ref{gomega-gaussian}), and then 
sampling the $\th_i$'s and $v_i$'s according to the distribution
(\ref{inc-state}). We let the system evolve according to the dynamics
(\ref{eom-scaled}), and monitor the evolution of the quantity $r$ in time.
For $m=20,T=0.25$, we choose four values of $\sigma$, two below and two above $\sigma^{\rm
inc}(m,T) \approx 0.10076$. Figures \ref{fig:r-vs-t-20real}(i)-(iv) show
the results for $20$ different realizations of
the initial state for three values of $N$.

In Fig. \ref{fig:r-vs-t-20real}(i) for $\sigma$ below
$\sigma^{\rm inc}(m,T)$, we see that the system
while starting from the unstable incoherent state at this value of
$\sigma$ settles down in time into the globally stable synchronized
state; this is consistent with the corresponding schematic landscape in
Fig. \ref{fig:fe}(a). The relaxation of $r$ occurs exponentially fast in
time according to $e^{\lambda t}$ for
$\sigma<\sigma^{\rm inc}(m,T)$, where the
growth rate $\lambda$ may be obtained from equation
(\ref{stability-eqn}) after substituting equation
(\ref{gomega-gaussian}) for $g(\omega)$. Figure \ref{fig:growth-rates} shows that the
theoretical growth rates are in excellent agreement with
numerical estimates.  

Figure \ref{fig:r-vs-t-20real}(ii) for $\sigma$ larger than in (i) but
below $\sigma^{\rm inc}(m,T)$ shows that similar to (i), the system
relaxes at long times to the synchronized
state for all realizations. Some realizations for short times stay in the
initial incoherent state due to finite-$N$ effects not captured by our
continuum limit theory.   
For $\sigma > \sigma^{\rm inc}(m,T)$, the
landscape in Fig. \ref{fig:fe}(iii) implies that the
system, while at long times should relax to the globally stable
synchronized state, remain trapped for finite
times in the metastable incoherent state. This is clearly borne out by
Fig. \ref{fig:r-vs-t-20real}(iii) in which one may observe that most
realizations relax to synchronized states. With increase of $N$, the number of realizations staying close to the
initial incoherent state for a finite time increases. Figure
\ref{fig:eta-N} shows that in fact the fraction $\eta$ of realizations relaxing to synchronized
state decreases exponentially with increasing $N$ for large $N$. This
fraction in numerical simulations is taken to be the fraction of realizations that cross $r=0.5$
during evolution in the given fixed time of observation. This
exponential decrease of $\eta$ with $N$ implies that for the fixed time
of observation and in the limit $N \to \infty$, all realizations
remain close to the incoherent state and none relax to the synchronized
state. This is consistent with our interpretation of $\sigma^{\rm
inc}(m,T)$ as the stability threshold above which the incoherent state
(\ref{inc-state}) is linearly stable.
In order to explain physically the exponential decrease of $\eta$ with
$N$, let us recall a classical result due to Kramers concerning the
relaxation time out of a metastable state under the stochastic dynamics of a single particle on a
potential landscape. In the weak-noise limit, this time is an exponential in the ratio of the potential energy barrier to come
out of the metastable state to the strength of the noise responsible for
the escape \cite{Kramers:1940}. For a mean-field system, considering the dynamics of the
order parameter on a free energy landscape, the
escape time out of a metastable state obeys Kramers formula with the
value of the  potential energy barrier replaced by $N$ times the
free-energy barrier \cite{Griffiths:1966}. This then explains the
finding in Fig. \ref{fig:eta-N}. 

Figure \ref{fig:r-vs-t-20real}(iv) for $\sigma$ larger than $\sigma^{\rm inc}(m,T)$
than in (iii) shows that with respect to (iii), more realizations stay close to the initial incoherent
state for longer times. This is easily explained as due to a larger barrier
separating the incoherent from the synchronized state. 

Based on our discussions above, we conclude that our theoretical
predictions are corroborated by our simulation results. Note that the simulation results suggest that the stability threshold of the
incoherent state is between $\sigma=0.095$ and $\sigma=0.11$, and indeed
the theoretical estimate $\approx 0.10076$ is in that range.  

\section{Dynamics of a lattice of oscillators interacting with a power-law coupling}
\l{chap3}
So far we have studied purely mean-field models, namely, where the coupling between each pair of oscillators is
exactly the same. There are clearly situations where this scenario is
not realistic. In this section, 
we will consider models in which the coupling between oscillators decays
as an inverse power-law of the distance
between them, i.e., the coupling $K$ of equation (\ref{Kuramoto-eom}) is substituted by
\begin{equation}
\label{def_coupl_decay}
K_{ij} = \frac{K_0}{|\mathbf{r}_i - \mathbf{r}_j|^\alpha},
\end{equation}
with $K_0$ a constant, $\alpha \ge 0$, and where $\mathbf{r}_i$ is the
position vector in the $d$-dimensional space of the $i$th oscillator.
We immediately realize that this setting requires the definition of a
lattice on the sites of which the oscillators reside.
This definition was not necessary for the mean-field models.

Once the notion of distance has been introduced, we see that a mean-field model is recovered if the parameter
$\alpha$ is set equal to $0$ so that $K_{ij}$ does not decay with the distance.
Thus, mean-field models are the extreme case of long-range interacting
systems. As discussed in the introduction, a long-range interaction
is realized when the parameter $\alpha$ is not larger than the embedding
dimension $d$. This definition applies not only to lattice systems, as
those considered in this review, but also to systems of particles described
by ordinary Cartesian coordinates in $d$-dimensional space
\cite{Campa:2009}.

The analytical study of systems with power-law interactions is inherently more difficult than that of mean-field systems.
In the latter, the interaction among the oscillators can be represented
in a convenient form that is not easily
attainable in the former. To exemplify this concept, let us consider the
passage from equation (\ref{Kuramoto-eom}) to
equation (\ref{Kuramoto-eom1}), where use has been made of equation
(\ref{r-defn}) defining the order parameter $r$; we see that each
oscillator is subject to the mean field generated by all the other
oscillators, and that this mean field is simply expressed in terms of the order
parameter. We have seen in the preceding sections that the analytical treatment obtains the equilibrium
and out-of-equilibrium behavior from the self-consistency between the order parameter and the mean field. This simple
association between a mean field and an
order parameter is no more possible in general in systems with
power-law interactions. Still, the possibility to study the
thermodynamic and dynamic behaviour of such systems by using only the
single-particle
distribution function, as will be explained later in this section, allows to derive self-consistent relations
determining the stationary states of the system. 
It is not within the scope of this review to offer a complete description of the
tools employed in the study of systems with power-law interactions. However, before considering the
Kuramoto model with a power-law coupling, we would now like to give a physical argument that supports the
existence of similarities between mean-field systems
and systems with power-law interactions.

A convenient approach to understand the differences between short and long-range systems, and at the same time
the similarities between mean-field systems and long-range systems with power-law interactions, is to grasp
the physical meaning of the
term that in the first equation of the BBGKY hierarchy couples the one-particle distribution function to the two-particle
distribution function. The argument does not depend on the presence of
dissipation and noise, but only on this coupling term  (e.g., the right
hand side of equation (\ref{eq:fp-1})) that behaves differently in short and long-range systems.
As we have seen in the analysis following that equation, if we write $f_2(x,x',t) = f_1(x,t) f_1 (x',t) + g_2(x,x',t)$,
where $x\equiv(\theta,v)$ and $x'\equiv (\theta',v')$, then the first
term on the right hand side gives rise to the mean-field term that
leads, e.g., in our case, to the Kramers equation (\ref{Kramers}). On
the other hand, the term $g_2$,
whose contribution we neglected in deriving the Kramers equation, takes into account
the two-particle correlation. In principle, both the terms describe the
variation of $f(x,t)$ as determined by the behavior of particles within the range of interaction around $\theta$.
In short-range systems, $f_1$ is practically uniform within the
interaction range, and therefore, the correlation term $g_2$ is considerably
larger than the mean-field term. In long-range systems, either with a
mean-field or with a power-law interaction,
the interaction range spans the whole system. Since in this case the correlation
$g_2$ decays quite rapidly with inter-particle separation, the
mean-field term is dominant. For the evaluation of the relative weight
of the mean-field term and the correlation term $g_2$ in different classes of systems,
see, e.g., the excellent book of Balescu \cite{Balescu:1987}.

Although both exhibiting the peculiar features of long-range systems, mean-field ($\alpha = 0$) systems
and systems with weakly decaying interactions ($0 < \alpha \le d$) can differ in several aspects. In fact,
the presence of a topological structure in the latter can induce features that do not occur in the former. For example, in the
equilibrium magnetized phase of a mean-field spin system, the average
magnetization is uniform, i.e., it is the same for every spin. On the
other hand, in a spin system with power-law interactions and with
free boundary conditions, the equilibrium magnetization will be larger
away from boundaries and smaller near
the boundaries; here, uniformity of the equilibrium state is recovered by adopting periodic boundary conditions,
that we will actually use in the analysis of this section. However, the uniformity of an
equilibrium state, either magnetized or non-magnetized, does not prevent
an out-of-equilibrium behavior in which the underlying lattice
structure does play a role.

Summarizing, it is meaningful to study the generalization of the type of
models studied in the previous sections, in which the mean-field
interaction is replaced by a slowly-decaying long-range interaction. We
will not consider the most general case, i.e., the case of interacting
oscillators with inertia and noise and driven by quenched torques.
Instead, we will study several particular cases. In all of them, the
interaction between the oscillators will be through coupling constants
of the form (\ref{def_coupl_decay}). Besides, we will be concerned with one-dimensional lattices with periodic boundary conditions.

The first subsection will be devoted to an extension of the Kuramoto
model obtained by the above-mentioned modification of the
coupling constants. Thus, it is a model of overdamped oscillators driven
by quenched external torques in the absence of noise. In the second and
third subsections, we will consider two different versions of the model
without the quenched torques.
\subsection{The Kuramoto model with a power-law coupling between
oscillators}
\l{seckurlongrange}
Let us consider a one-dimensional periodic lattice of $N$ sites labelled by the
index $i=1,2,\ldots,N$, and with
lattice constant equal to $a$. With a proper choice of the origin, the
coordinate of the $i$th
site is $x_i = ia$. On each site resides an oscillator, with the dynamics of
the $i$th oscillator governed by the evolution equation
\cite{Rogers:1996,Gupta:2012a}
\be
\frac{\dd \th_i}{\dd t}=\omega_i+\frac{K}{\widetilde{N}}\sum_{j=1}^N
\frac{\sin(\th_j-\th_i)}{|x_j - x_i|_c^\alpha},
\l{Kuramoto-eom_decaying}
\ee
where the exponent $\alpha$ lies in the range $0 \le \alpha <1$. Since
we adopt periodic boundary conditions, the distance $|x_j - x_i|$ between the $i$th
and $j$th sites is not unambiguously defined. In equation
(\ref{Kuramoto-eom_decaying}), we adopt the closest distance convention:
\be
|x_j - x_i|_c \equiv \min \left(|x_j - x_i|,Na - |x_j - x_i| \right).
\l{clos_dist_conv}
\ee
The factor $\widetilde{N}$ in equation (\ref{Kuramoto-eom_decaying}), which in the mean-field case ($\alpha=0$)
becomes the normalizing factor $N$ in the equations of motion (see
equation \ref{Kuramoto-eom}), is given by
\be
\widetilde{N} \equiv \sum_{j=1}^N \frac{1}{|x_j - x_i|_c^\alpha}.
\l{ntilde_def}
\ee 
In the last expression, we take $|x_j - x_i|_c=a$ for $i=j$. While this choice is irrelevant for the
equation of motion (\ref{Kuramoto-eom_decaying}), it allows to include in the summation in
equation (\ref{ntilde_def}) the term with $j=i$, thereby making $\widetilde{N}$ non-diverging. Note that
the right hand side of equation (\ref{ntilde_def}) is independent of $i$ due to the closest distance convention. 

The introduction of a lattice on the sites of which the oscillators
reside, and of a coupling that depends on the distance between the
oscillators, has two important consequences on the analytical treatment of the system. The first is that
it is no more possible to define a global order parameter, similar to $r$ in equation (\ref{r-defn}),
that can be used to rewrite the equations of
motion in an equivalent
form (compare, e.g., equations (\ref{Kuramoto-eom}) and (\ref{Kuramoto-eom1})). The second consequence has also a conceptual
relevance. Let us consider those oscillators with the same value of the
natural frequency $\omega$, say, $\omega=\omega^*$ (since we will be
eventually interested in the limit $N \to \infty$, we may imagine in
this limit to have a fraction of oscillators with the same frequency
$\omega=\omega^*$, or, more precisely, with $\omega$ within a given
small range around $\omega^*$). One realizes that the
equation of motion (\ref{Kuramoto-eom_decaying}) is not invariant
under permutations of the phase of these oscillators as the latter
could be identified by the lattice sites they are occupying,
contrary to what happens for the mean-field case $\alpha = 0$. Therefore,
at variance with the latter case, it is not possible to define the distribution function $\rho(\theta,\omega,t)$
giving, among the oscillators characterized by $\omega$, their density at phase value $\theta$ at time $t$.
This fact is rooted in the impossibility, due to the lack
of invariance with respect to permutations, to define the usual reduced
distribution functions as in, e.g., equation (\ref{eq:fs-defn}). A  bit of thought
allows to understand that the very same feature explains also the first consequence mentioned above.

We thus arrive at the conclusion that the only possibility to use distribution functions for analysis is to define for any
given $\omega$ a distribution function for each of the lattice sites. In
the limit $N \to \infty$, this means that we have to consider the situation in
which, together with this limit, the lattice constant $a$ approaches $0$, keeping the product $Na$
constant that without loss of generality can be fixed equal to $1$. This
procedure defines the continuum limit, implementing which we can define the one-particle distribution
function $\rho (\theta,\omega,s,t)$, where $s \in [0,1]$ is a continuous
variable obtained by considering $s_j \equiv j/N$ in the continuum
limit. Since the lattice constant $a$ approaches $0$, one has in each
infinitesimal range $\dd s$ a continuum of oscillators such that $\rho
(\theta,\omega,s,t)g(\omega)\dd \omega \dd s \dd \theta$
is the fraction of oscillators located between $s$ and
$s + \dd s$, with natural frequency between $\omega$ and $\omega +
\dd \omega$, and having at time $t$ the phase between $\theta$
and $\theta + \dd \theta$. The function $\rho (\theta,\omega,s,t)$ is non-negative, $2\pi$ periodic in $\theta$, and
obeys the normalization
\be
\int_{-\pi}^{\pi} \dd \theta \, \rho(\theta,\omega,s,t) = 1~~\forall~~\omega, s.
\l{norm_cond_rho}
\ee

In the continuum limit, we can rewrite the equation of motion
(\ref{Kuramoto-eom_decaying}) as
\be
\fl
\frac{\partial \th (\omega,s,t)}{\partial t}= \omega+\frac{K}{B(\alpha)}\int \dd \omega' \, \int_0^1 \dd s' \,
\int_{-\pi}^{\pi} \dd \theta' \, \frac{\sin (\theta' - \theta)}{|s'-s|_c^\alpha}
\rho(\theta',\omega',s',t) g(\omega') ,
\l{Kuramoto-eom_decaying_cont}
\ee
where now $\theta$ is labelled by the position $s$ and the natural frequency $\omega$. The normalizing factor $B(\alpha)$ is
given by
\be
B(\alpha) = \int_{-\frac{1}{2}}^{\frac{1}{2}} \dd s \,
\frac{1}{|s|^\alpha}.
\l{norm_fact}
\ee
Since $\alpha<1$, the integral on the right hand side is
finite, and equals  $2^\alpha/(1-\alpha)$.
The closest distance convention now reads
\be
|s' - s|_c = \min \left( |s' - s|, 1 - |s' - s| \right),
\l{clos_dist_conv_cont}
\ee
that allowed us to write the denominator in the integrand of equation (\ref{norm_fact}) without the subscript $c$.

From the equation of motion (\ref{Kuramoto-eom_decaying_cont}), one derives analogously to the mean-field case a Fokker-Planck
equation for the distribution $\rho (\theta,\omega,s,t)$. Actually,
since in the present case there is no noise in the dynamics, the equations of
motion are deterministic, and the Fokker-Planck equation is nothing but
the continuity equation expressing the conservation for each $s$ and
$\omega$ of the number of oscillators. We have
\bea
\l{continuity_eq}
\fl \frac{\partial \rho(\theta,\omega,s,t)}{\partial t} = -\frac{\partial}{\partial \theta}
\left[\left(\frac{\partial \th (\omega,s,t)}{\partial
t}\right)\rho(\theta,\omega,s,t)\right] \nonumber \\
\fl = -\frac{\partial}{\partial \theta} \left\{ \left[
\omega+\frac{K}{B(\alpha)}\int \dd \omega' \, \int_0^1 \dd s' \,
\int_{-\pi}^{\pi} \dd \theta' \, \frac{\sin (\theta' - \theta)}{|s'-s|_c^\alpha}
\rho(\theta',\omega',s',t) g(\omega')\right] \rho(\theta,\omega,s,t)\right\}. 
\eea
We note that an initial distribution that is $s$-independent remains
so under the evolution (\ref{continuity_eq}). Of course, one should
check the stability of such a conservation with respect to $s$-dependent
perturbations. In the following, we will analyze the dynamical
stability of the particular $s$-independent (therefore, mean-field) stationary solution
that represents the unsynchronized or the incoherent state, i.e.,
\be
\rho_0(\theta,\omega,s,t) = \frac{1}{2\pi}.
\l{alphastationary}
\ee
\subsubsection{Linear stability analysis of the mean-field incoherent stationary state}
\l{linear-stability-alphakura}
To study the linear stability of the incoherent state (\ref{alphastationary}), we expand $\rho_0$
as
\be
\rho(\theta,\omega,s,t) = \frac{1}{2\pi} + \delta \rho
(\theta,\omega,s,t);~~|\delta \rho| \ll 1.
\l{alphaperturbation}
\ee
Inserting the above expansion into the continuity equation
(\ref{continuity_eq}), and
keeping only the first order terms in $\delta \rho$, we obtain the
linearized equation
\bea
\l{linear_cont_eq}
\fl \frac{\partial \delta \rho(\theta,\omega,s,t)}{\partial t} = -\omega 
\frac{\partial \delta \rho(\theta,\omega,s,t)}{\partial \theta}
\nonumber \\
 +\frac{K}{2\pi B(\alpha)}\int \dd \omega' \, \int_0^1 \dd s' \,
\int_0^{2\pi} \dd \theta' \, \frac{\cos (\theta' - \theta)}{|s'-s|_c^\alpha}
\delta \rho(\theta',\omega',s',t) g(\omega'). 
\eea
To solve the above equation for $\delta \rho$, let us perform its Fourier expansion in $\theta$ as
\be
\delta \rho(\theta,\omega,s,t) = \sum_{k=-\infty}^{+\infty}
\widehat{\delta \rho}_k(\omega,s,t) e^{ik\theta}.
\l{fourier_theta}
\ee
Substitution in equation (\ref{linear_cont_eq}) gives
\bea
\l{linear_cont_eq_four}
\fl \frac{\partial \widehat{\delta \rho}_k(\omega,s,t)}{\partial t} = -i k \omega 
\widehat{\delta \rho}_k(\omega,s,t) \nonumber \\
 +\frac{K}{2 B(\alpha)}\left( \delta_{k,1} + \delta_{k,-1}\right)
 \int \dd \omega' \, \int_0^1 \dd s' \,
 \frac{\widehat{\delta \rho}_k(\omega',s',t)}{|s'-s|_c^\alpha} g(\omega'). 
\eea

For $k \ne \pm 1$, the second term on the right hand side of equation
(\ref{linear_cont_eq_four}) is zero, and
solving the resulting equation gives
\be
\widehat{\delta \rho}_k(\omega,s,t) = \widehat{\delta \rho}_k(\omega,s,0)
e^{-ik\omega t};~~k \ne \pm 1.
\l{cont_modes}
\ee 
These solutions correspond to the neutrally stable Fourier modes; there are
an infinity of such modes for each $\omega$ belonging to the support
of $g(\omega)$. The eigenfunction corresponding to any particular value
of $\omega$, say, $\omega=\omega_0$, is 
\be
\widehat{\delta \rho}_{k,\omega_0}(\omega,s,0)=\delta (\omega - \omega_0)c(s),
\l{eigen_cont}
\ee
where $c(s)$ is an arbitrary function of $s$. On the
other hand, equation (\ref{linear_cont_eq_four}) for $k = \pm 1$ gives 
\be
\fl \frac{\partial \widehat{\delta \rho}_{\pm 1}(\omega,s,t)}{\partial t} = \mp i \omega 
\widehat{\delta \rho}_{\pm 1}(\omega,s,t) 
 +\frac{K}{2 B(\alpha)}
 \int \dd \omega' \, \int_0^1 \dd s' \,
 \frac{\widehat{\delta \rho}_{\pm 1}(\omega',s',t)}{|s'-s|_c^\alpha} g(\omega').
\l{linear_cont_eq_fourpm1}
\ee
This equation is best solved by performing a further Fourier expansion,
this time in $s$ space:
\be
\widehat{\delta \rho}_{\pm 1}(\omega,s,t) = \sum_{n=-\infty}^{+\infty}
\overline{\delta \rho}_{\pm 1,n}(\omega,t) e^{2\pi i ns}.
\l{fourier_sspace}
\ee
Substituting in equation (\ref{linear_cont_eq_fourpm1}), we obtain
\be
\fl \frac{\partial \overline{\delta \rho}_{\pm 1,n}(\omega,t)}{\partial t} = \mp i \omega 
\overline{\delta \rho}_{\pm 1,n}(\omega,t) 
 +\frac{K\Lambda_n(\alpha)}{2 B(\alpha)}
 \int \dd \omega' \,
 \overline{\delta \rho}_{\pm 1,n}(\omega',t) g(\omega'),
\l{linear_cont_eq_fourpm1s}
\ee
where $\Lambda_n(\alpha)$ is given by
\be
\Lambda_n(\alpha) = \int_{-\frac{1}{2}}^{\frac{1}{2}} \dd s \,
 \frac{e^{2\pi i ns}}{|s|^\alpha} = \int_{-\frac{1}{2}}^{\frac{1}{2}} \dd s \,
 \frac{\cos(2\pi ns)}{|s|^\alpha}.
\l{eigenlambda}
\ee
Clearly $\Lambda_{-n}(\alpha) = \Lambda_n(\alpha)$; we can therefore restrict to consider
$n\ge 0$. It is also evident that $\Lambda_0(\alpha) = B(\alpha) > \Lambda_n(\alpha)$ for $n > 0$.
Let us first consider the mean-field case $\alpha = 0$. In that case, 
$\Lambda_0(0) = 1$ and $\Lambda_n(0) = 0$ for $n>0$. Therefore, in the
mean-field case, all modes $n \ne 0$ are neutrally stable, and we have
to study equation (\ref{linear_cont_eq_fourpm1s}) only for $n=0$. For $\alpha > 0$, 
all values of $n$ have to be considered. It is not difficult to prove that $\Lambda_n(\alpha) > 0$, that
$\lim_{n\to \infty} \Lambda_n(\alpha) = 0$, and that for given $n$, one
has $\Lambda_n(\alpha)$ as an
increasing function of $\alpha$. One may check numerically that $\Lambda_n(\alpha)$
is a decreasing function of $|n|$ for any $\alpha$, see Ref.
\cite{Gupta:2012a}. In the following, it
is understood that for $\alpha=0$, only the case $n=0$ has to be
considered.

We look for solutions of equation (\ref{linear_cont_eq_fourpm1s}) of the form
\be
\overline{\delta \rho}_{\pm 1,n}(\omega,t) = \widetilde{\delta \rho}_{\pm 1,n}(\omega,\lambda_n)
e^{\lambda_n t}.
\l{eigen_posit}
\ee 
Substituting in equation (\ref{linear_cont_eq_fourpm1s}) gives
\be
(\lambda_n \pm i \omega)\widetilde{\delta \rho}_{\pm 1,n}(\omega,\lambda_n) =
 \frac{K\Lambda_n(\alpha)}{2 B(\alpha)}
 \int \dd \omega' \,
 \widetilde{\delta \rho}_{\pm 1,n}(\omega',\lambda_n) g(\omega').
\l{linear_cont_eq_fourpm1s_eigen}
\ee
Hence, we have a continuous spectrum given by $\lambda_n = \mp i \omega_0$ for each
$\omega_0$ in the support of $g(\omega)$. In this case, the neutrally stable modes, normalized so that
the right hand side of equation (\ref{linear_cont_eq_fourpm1s_eigen}) is
equal to $1$, are given by
\be
\widetilde{\delta \rho}_{\pm 1,n}(\omega,\mp i \omega_0) = \mp i {\mathcal P} \frac{1}{\omega - \omega_0}
+ c_{\pm 1,n}(\omega_0) \delta (\omega - \omega_0),
\l{eigen_cont_pm1}
\ee
with
\be
c_{\pm 1,n}(\omega_0) g(\omega_0) = \frac{2 B(\alpha)}{K \Lambda_n(\alpha)} \pm i {\mathcal P}
\int \dd \omega \, \frac{g(\omega)}{\omega -\omega_0},
\l{eigen_factor}
\ee
where ${\mathcal P}$ denotes the principal value. We are interested in
the discrete spectrum, obtained for $\lambda_n \pm i \omega \ne 0$. From
equation (\ref{linear_cont_eq_fourpm1s_eigen}), we then have
\be
\widetilde{\delta \rho}_{\pm 1,n}(\omega,\lambda_n) =
 \frac{K\Lambda_n(\alpha)}{2(\lambda_n \pm i \omega) B(\alpha)}
 \int \dd \omega' \,
 \widetilde{\delta \rho}_{\pm 1,n}(\omega',\lambda_n) g(\omega'),
\l{linear_cont_eq_fourpm1s_discr}
\ee
which implies that in order to have a non-trivial solution, the
integral on the right hand side should not vanish.
We can exploit the linearity of equation (\ref{linear_cont_eq_fourpm1s_eigen}) to impose that the integral is equal
to $1$. From the last equation, we then obtain the dispersion relation
\be
\frac{K\Lambda_n(\alpha)}{2 B(\alpha)}\int_{-\infty}^{+\infty} \dd \omega \,
\frac{g(\omega)}{\lambda_n \pm i \omega} = 1.
\l{disper_relat}
\ee
For the class of distributions $g(\omega)$ being considered in this
review, that is, for a unimodal $g(\omega)$ with a single maximum at
$\omega = 0$ and symmetric, $g(\omega)=g(-\omega)$,
we now prove that the last equation
can have at most one solution for $\lambda_n$, which is necessarily
real. Note that when there is no solution for $\lambda_n$,
only the trivial vanishing perturbation $\widetilde{\delta \rho}_{\pm 1,n}(\omega,\lambda_n)= 0$ satisfies
equation (\ref{linear_cont_eq_fourpm1s_discr}). Decomposing $\lambda_n$
into real and imaginary parts,
$\lambda_n = \lambda_{nr} +i \lambda_{ni}$, we obtain from equation (\ref{disper_relat})
\bea
\frac{K\Lambda_n(\alpha)}{2 B(\alpha)}\int_{-\infty}^{+\infty} \dd \omega \,
g(\omega)\frac{\lambda_{nr}}{\lambda_{nr}^2 + (\lambda_{ni} \pm
\omega)^2} = 1,
\l{disper_relatr} \\
\frac{K\Lambda_n(\alpha)}{2 B(\alpha)}\int_{-\infty}^{+\infty} \dd \omega \,
g(\omega)\frac{\lambda_{ni} \pm \omega}{\lambda_{nr}^2 + (\lambda_{ni} \pm \omega)^2} = 0.
\l{disper_relati}
\eea
Proceeding as for equation (\ref{disper_relati_noisy}) and exploiting the fact that
$g(\omega)$ is even, the second equation implies that $\lambda_{ni}=0$. We are therefore left with the equation
\be
\frac{K\Lambda_n(\alpha)}{2 B(\alpha)}\int_{-\infty}^{+\infty} \dd \omega \,
g(\omega)\frac{\lambda_n}{\lambda_n^2 + \omega^2} = 1,
\l{disper_relatfin}
\ee
where now it is understood that $\lambda_n$ is real. This equation implies that
only positive solutions are possible. With the change of variable
$\omega = \lambda_n y$, we have
\be
\frac{K\Lambda_n(\alpha)}{2 B(\alpha)}\int_{-\infty}^{+\infty} \dd y \,
g(\lambda_ n y)\frac{1}{1 + y^2} = 1.
\l{disper_relatfin_y}
\ee
By taking the derivative with respect to $\lambda_n$, one immediately finds that the left hand side decreases
with increasing positive $\lambda_n$, and that it tends to $0$ as
$\lambda_n \to \infty$. Therefore, there is one and only one solution
that exists only when the value of the left hand side for $\lambda_n=0$ is
larger than one, i.e., when
\be
\frac{K\Lambda_n(\alpha)}{2 B(\alpha)}\pi g(0) \ge 1.
\l{disper_condition}
\ee
We finally obtain the following threshold above which the mode $\widetilde{\delta \rho}_{\pm 1,n}$
is unstable:
\be
K_c^{(n)} = \frac{2 B(\alpha)}{\pi g(0) \Lambda_n(\alpha)}.
\l{disper_thresholds}
\ee

The final outcome of the stability analysis is that the incoherent state (\ref{alphastationary})
is either neutrally stable or unstable. This is analogous to what happens in the original Kuramoto
model \cite{Strogatz:2000}, that is included in our analysis for $\alpha=0$. Since $\Lambda_n(\alpha)$
is a decreasing function of $n$, we have that $K_c^{(n)}$ is an increasing function of $n$. Then,
the incoherent state is neutrally stable for
\be
K \le K_c^{(0)} = \frac{2}{\pi g(0)},
\l{disper_thresh_tot}
\ee
and is unstable otherwise. We note that the critical value for $n=0$ does not depend on $\alpha$, and the
instability threshold is therefore the same as that for the original
mean-field ($\alpha=0$) Kuramoto model, see equation (\ref{Kura-bareKc}).
\subsubsection{Numerical results}
\l{numerics-alphakura}
The analysis above implies that increasing progressively the value of $K$, more and more modes destabilize. For example,
if $K_c^{(p)} < K < K_c^{(p+1)}$, then we have $p+1$ unstable modes, corresponding to $n=0,1,\dots,p$.
According to equation (\ref{eigen_posit}), each of these modes has an exponential growth with rate
given by the corresponding $\lambda_n$.

To check our analytical predictions with simulations, we
choose a Gaussian $g(\omega)$:
\be
g(\omega) = \frac{1}{\sqrt{2\pi}} e^{-\frac{\omega^2}{2}}.
\l{gaussgomega}
\ee
In this case, we can express the integral in equation
(\ref{disper_relatfin}) with the help of the complementary error
function.
In fact, using
\be
\int_{-\infty}^{+\infty} \dd x \, \frac{e^{-b^2 x^2}}{x^2 + a^2}
= \frac{\pi}{a} e^{a^2 b^2} {\rm Erfc}\left(ab\right),
\l{app_integral}
\ee
we can rewrite equation (\ref{disper_relatfin}) as
\be
\frac{K\Lambda_n(\alpha)}{2 B(\alpha)}\sqrt{\frac{\pi}{2}}e^{\frac{\lambda_n^2}{2}}
{\rm Erfc}\left(\frac{\lambda_n}{\sqrt{2}}\right) =
\frac{K}{K_c^{(n)}} e^{\frac{\lambda_n^2}{2}}
{\rm Erfc}\left(\frac{\lambda_n}{\sqrt{2}}\right) = 1,
\l{disper_relatfin_gauss}
\ee
where we have also used the definition of the thresholds
(\ref{disper_thresholds}), which for the
Gaussian $g(\omega)$ becomes
\be
K_c^{(n)} = \frac{2\sqrt{2} B(\alpha)}{\sqrt{\pi} \Lambda_n(\alpha)}.
\l{disper_thresholds_gauss}
\ee

The simulations are performed by integrating the equations of motion (\ref{Kuramoto-eom_decaying}),
taking $N$ oscillators on a lattice
of length $N$ with periodic boundary conditions and lattice constant
$a=1$. Although the equations
of motion imply a computation time at every step of integration that scales as $N^2$ (there is a sum over $N$
terms for each of the $N$ oscillators), it
is possible to employ an integration algorithm that scales as $N\ln N$.
In Appendix C, we show how this
is achieved by exploiting standard fast Fourier transform routines.
The observables evaluated in simulations are the
discrete quantities corresponding to the density perturbations in
equation (\ref{fourier_sspace}),
namely,
\be
r_n(t) = \frac{1}{N} \left| \sum_{j=1}^N
e^{i\left( \theta_j + 2\pi j n/N\right)} \right|;~~n=0,1,2,\dots
\l{discrete_order_params}
\ee
The initial conditions of the simulations are obtained by extracting
independently the phases of the oscillators
from a uniform distribution in the range $[0,2\pi]$; this reproduces the incoherent state
(\ref{alphastationary}). The frequencies $\omega_i$'s are
extracted independently from the Gaussian
distribution (\ref{gaussgomega}). In this initial state, the observables $r_n$ are equal to $0$
(only approximately, due to finite-size effects). According to the theoretical analysis, depending on the
value of $K$ employed, the quantities $r_n(t)$ should behave in the
following way. For $K < K_c^{(0)}$, all
$r_n$ should remain close to $0$, while for $K_c^{(p)} < K <
K_c^{(p+1)}$, the observables
$r_n$ with $n \ge p+1$ should remain close to $0$ and those with $n=0,1,\dots,p$ should grow
exponentially in time (at least as long as the linear approximation of the continuity equation is valid) at
a rate equal to the corresponding eigenvalue $\lambda_n$.

We present here some results for $\alpha = 0.5$, referring the reader to
Ref. \cite{Gupta:2012a} for further and more complete results. In Fig.
\ref{SMS-fig2}, 
we report the time evolution of $r_0(t)$, $r_1(t)$, $r_2(t)$ and
$r_3(t)$ for a simulation run
in which the initial condition has been chosen as explained above. The simulation has been
performed for a system of $N=2^{14}$ oscillators, and with $K=15$.
From equation (\ref{disper_thresholds_gauss}),
one finds that this value of $K$ lies in between $K_c^{(11)}$ and
$K_c^{(12)}$. Therefore, in particular, the observables
plotted in Fig. \ref{SMS-fig2} should all increase exponentially in time. This is confirmed
by the numerical results shown in the figure.

\begin{figure}[h!]
\centering
\includegraphics[width=100mm]{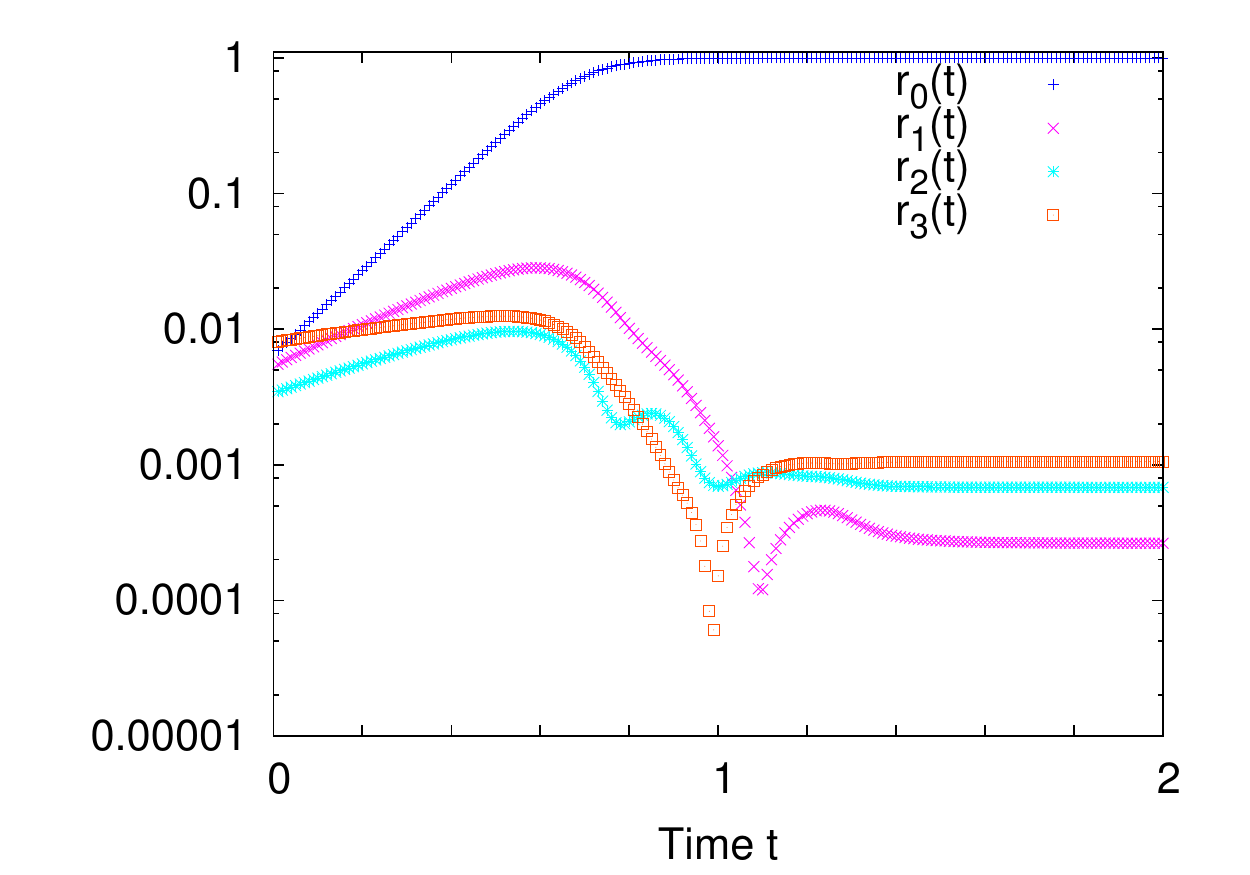}
\caption{For the model (\ref{Kuramoto-eom_decaying}), the figure shows the time evolution of the observables
$r_0(t), r_1(t), r_2(t)$, and $r_3(t)$ while starting with an initial
incoherent state $\{\theta_i(0),\omega_i(0); i=1,2,\ldots,N\}$ prepared by extracting the
$\theta_i$'s uniformly in $[-\pi,\pi]$, while the $\omega_i$'s have been chosen from a Gaussian
distribution with zero mean and unit variance, equation
(\ref{gaussgomega}). Here, $N=2^{14}$, $\alpha=0.5$, and $K=15$.
Using equation (\ref{disper_thresholds_gauss}), one then has $K_c^{(0)} \approx 1.59577$,
$K_c^{(1)} \approx 4.26696$, $K_c^{(2)} \approx 6.53664$,
$K_c^{(3)} \approx 7.71516$. Thus, in particular, the Fourier modes
$n=0,1,2,3$ are all linearly unstable. Consequently, $r_0(t), r_1(t),
r_2(t)$, and $r_3(t)$ for short times show an exponential growth in
time from their initial values at $t=0$.}
\l{SMS-fig2}
\end{figure}

\begin{figure*}[h!]
\centering
\begin{tabular}{lr}
\parbox[l]{9cm}{
\includegraphics[width=80mm]{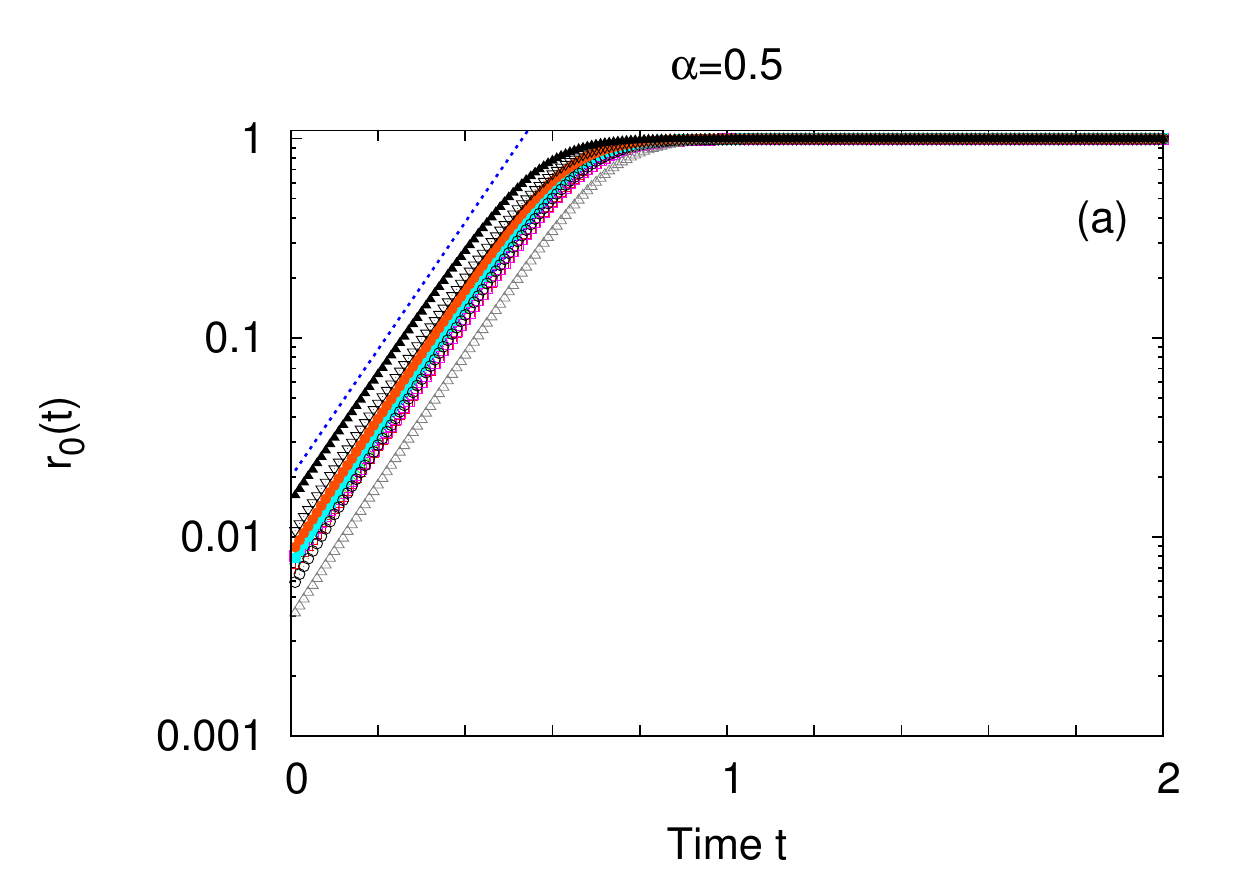}
}&
\parbox[r]{9cm}{
\includegraphics[width=80mm]{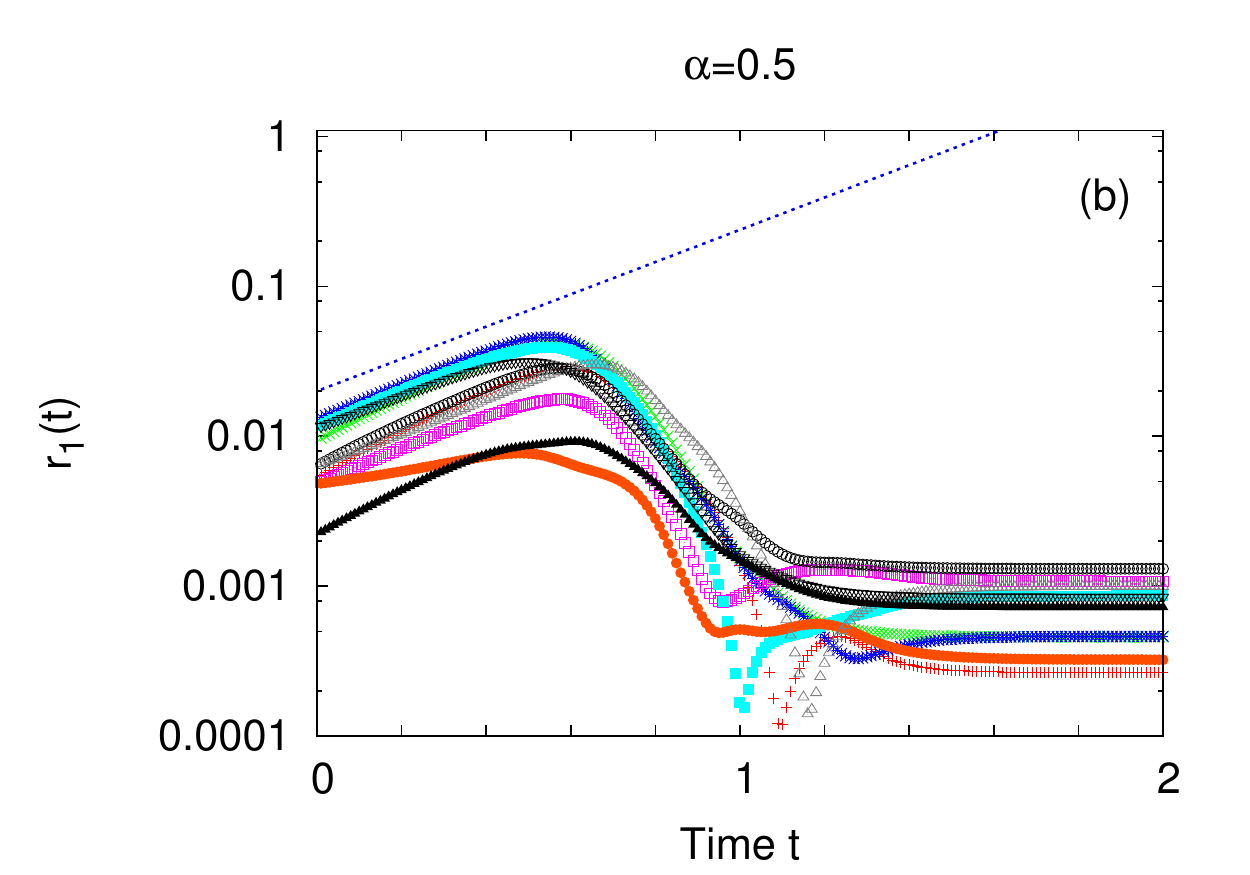}
}
\end{tabular}
\caption{For the model (\ref{Kuramoto-eom_decaying}), the figure shows
the time evolution of the observables $r_0(t)$ (panel (a)) and
$r_1(t)$ (panel (b)) for $10$ different realizations of the initial
state $\{\theta_i(0),\omega_i(0); i=1,2,\ldots,N\}$. As in Fig. \ref{SMS-fig2}, we present here
simulations for $K=15$ at $\alpha = 0.5$. Similarly, each initial state has been
obtained by extracting the $\theta_i$'s uniformly in in $[-\pi,\pi]$,
while the $\omega_i$'s have been extracted from a Gaussian
distribution with zero mean and unit variance, equation (\ref{gaussgomega}).
Thus, each initial state is the incoherent one. Since
the Fourier modes $0$ and $1$ are linearly unstable, $r_0(t)$ and $r_1(t)$ grow in
time from their initial values. The dotted blue line in each plot shows the exponential growth
with the rates $\lambda_0$ and $\lambda_1$ given implicitly by equation (\ref{disper_relatfin_gauss}).
The data in the plots are obtained from numerical simulations with
$N=2^{14}$.}
\label{SMS-fig3}
\end{figure*}

To have a comparison between the numerical and the theoretical rates of the initial exponential growth
of the unstable modes, we plot in Fig. \ref{SMS-fig3} the time evolution
of $r_0(t)$ and $r_1(t)$ for the same values of $\alpha$ and $K$ as in
Fig. \ref{SMS-fig2}, namely, $\alpha=0.5$ and $K=15$. In the plots are shown results of $10$
different simulation runs, corresponding to $10$ different realizations of the incoherent initial condition.
The exponential growth rates of $r_0(t)$ and $r_1(t)$ are compared with
their theoretical values $\lambda_0$
and $\lambda_1$ computed from equation (\ref{disper_relatfin_gauss}).
The agreement is clearly good. We note that for few realizations, the
numerical growth rate for $r_1(t)$
deviates from $\lambda_1$, see Fig. \ref{SMS-fig3}(b), arguably due
to finite-size effects. The plots also suggest that the system settles down to a stationary state in which
$r_0$ assumes a value very close to $1$, while $r_1$ takes a negligible
value compatible with $0$, considering the finite-size effects. In Fig.
\ref{SMS-fig2}, we see that the same happens for $r_2$ and $r_3$. We therefore
conclude that the long-time dynamics is dominated by the mean-field mode. In particular, the final state is
fully synchronized, where the synchronization refers to oscillators
residing on all lattice sites.

\subsection{The noisy Kuramoto model with a power-law coupling and the
same natural frequency for the oscillators}
\l{alphahmf-overdamped}
We now turn our attention to the case when all the oscillators have the
same natural frequency, say, $\langle \omega \rangle$. As discussed in
section \ref{Kuramoto-bare}, one can scale out $\langle \omega \rangle$ from the equations of
motion by going to a comoving frame rotating uniformly with frequency
$\langle \omega \rangle$ with respect to
the laboratory frame. Thus, we are effectively considering the dynamics 
without the presence of quenched torques. We will first study
an overdamped model with noise \cite{Anteneodo:2000,Tamarit:2000,Gupta:2012b}. Namely, the model obtained from the one
studied in section \ref{seckurlongrange} by adding Gaussian noise terms to the
equations of motion, but excluding the frequency terms. Later, we will focus on the underdamped model with noise.

Let us then begin with the equation of motion for the $i$th oscillator:
\be
\frac{\dd \th_i}{\dd t}= \frac{K}{\widetilde{N}}\sum_{j=1}^N
\frac{\sin(\th_j-\th_i)}{|x_j - x_i|_c^\alpha} + \eta_i(t),
\l{Kuramoto-eom_decaying_noise}
\ee
where $\eta_i(t)$ is a Gaussian white noise:
\bea
\langle \eta_i(t) \rangle=0, \\
\langle \eta_i(t) \eta_j(t') \rangle=2T\delta_{ij}\delta(t-t').
\l{noise_statistics}
\eea
The equation of motion (\ref{Kuramoto-eom_decaying_noise}) describes the
overdamped dynamics of the so-called $\alpha$-HMF model
\cite{Anteneodo:2000,Tamarit:2000}, within a canonical ensemble (see
equation (\ref{long-range-inertia}) below).

The definition of $\widetilde{N}$ and the closest distance convention are
the same as in section  \ref{seckurlongrange}, and so is the procedure to
obtain the continuum limit ($N\to \infty,a\to 0$, keeping the product
$Na$ constant at unity, where $a$ stands for the lattice constant). To
discuss this limit, we introduce the variable
$s \in [0,1]$, obtained as the continuum limit of $s_j=j/N$. However,
contrary to section  \ref{seckurlongrange}, the one-particle
distribution function will now not depend on the frequency. Here, we
introduce the one-particle
distribution function $\rho(\theta,s,t)$, defined such that the quantity
$\rho(\theta,s,t)\dd s \dd \theta$ 
represents the fraction of oscillators located between $s$ and $s + \dd
s$ that at time $t$ has their phase between
$\theta$ and $\theta + \dd \theta$. The normalization is
\be
\int_{-\pi}^{\pi} \dd \theta \, \rho(\theta,s,t) = 1~~\forall~~s.
\l{norm_cond_rhoq}
\ee
In the continuum limit, the equation of motion takes the form
\be
\frac{\partial \th (s,t)}{\partial t}= \frac{K}{B(\alpha)} \int_0^1 \dd s' \,
\int_{-\pi}^{\pi} \dd \theta' \, \frac{\sin (\theta' - \theta)}{|s'-s|_c^\alpha}
\rho(\theta',s',t) +\eta(s,t) ,
\l{Kuramoto-eom_decaying_noise_cont}
\ee
where the normalizing factor $B(\alpha)$ and the closest distance convention $|s' - s|_c$ are given
in equations (\ref{norm_fact}) and (\ref{clos_dist_conv_cont}), respectively. The statistical properties
of the noise become
\bea
\langle \eta(s,t) \rangle=0, \\
\langle \eta(s,t) \eta(s',t') \rangle=2T\delta(s-s')\delta(t-t').
\l{noise_statistics_cont}
\eea
The Fokker-Planck equation governing the evolution of $\rho(\theta,s,t)$ is
\bea
\l{Fokkerplanck_eq_decaying}
\fl \frac{\partial \rho(\theta,s,t)}{\partial t}  \nonumber \\
\fl = -\frac{K}{B(\alpha)}\frac{\partial}{\partial \theta} \left\{ \left[
\int_0^1 \dd s' \, \int_{-\pi}^{\pi} \dd \theta' \, \frac{\sin (\theta' - \theta)}{|s'-s|_c^\alpha}
\rho(\theta',s',t)\right] \rho(\theta,s,t)\right\}
+T \frac{\partial^2 \rho(\theta,s,t)}{\partial \theta^2}.
\eea

The generic stationary solution $\rho_0(\theta,s)$ of the Fokker-Planck equation
(\ref{Fokkerplanck_eq_decaying}) is obtained by setting the left hand
side to zero, yielding
\be
\rho_0(\theta,s) = A(s) \exp \left[ \frac{K}{TB(\alpha)}\int_0^1 \dd s'
\, \int_{-\pi}^{\pi} \dd \theta' \,
\frac{\cos (\theta' - \theta)}{|s'-s|_c^\alpha} \rho_0(\theta',s) \right] ,
\l{noise_stationary}
\ee
where the constants $A(s)$ for every $s$ are determined by the normalization condition (\ref{norm_cond_rhoq}).
There are also consistency relations to be satisfied, as we now show. Let us denote by $m_x(s)$ and $m_y(s)$ the
two components of the local magnetization:
\bea
m_x(s) \equiv \int_{-\pi}^{\pi} \dd \theta \, \cos \theta~ \rho_0(\theta,s),
\l{mxsdef}
\\
m_y(s) \equiv \int_{-\pi}^{\pi} \dd \theta \, \sin \theta~ \rho_0(\theta,s).
\l{mysdef}
\eea
From the definition of the modified Bessel function of the first kind of order $n$,
\be
I_n(x) = \frac{1}{2\pi} \int_{-\pi}^{\pi} \dd \theta \, \cos(n\theta)
e^{x\cos \theta},
\l{defbessel}
\ee
and denoting
\bea
\widehat{m}_x^{(\alpha)}(s) = \int_0^1 \dd s' \,
\frac{m_x(s')}{|s'-s|_c^\alpha},
\l{mxsaldef}
\\
\widehat{m}_y^{(\alpha)}(s) = \int_0^1 \dd s' \, \frac{m_y(s')}{|s'-s|_c^\alpha},
\l{mysaldef}
\eea
one obtains for the normalization constant the equation
\be
A(s) = \left[2\pi I_0\left( \frac{K}{TB(\alpha)} \sqrt{\left[\widehat{m}_x^{(\alpha)}(s)\right]^2 +
\left[\widehat{m}_y^{(\alpha)}(s)\right]^2} \right)\right]^{-1},
\l{normconst}
\ee
together with the self-consistency relations
\be
\sqrt{m_x^2(s) + m_y^2(s)} = \frac{I_1}{I_0}
\left( \frac{K}{TB(\alpha)} \sqrt{\left[\widehat{m}_x^{(\alpha)}(s)\right]^2 +
\left[\widehat{m}_y^{(\alpha)}(s)\right]^2} \right).
\l{selfalcons}
\ee
We note for later use that if we choose an $s$-independent stationary distribution
$\rho_0(\theta)$, then $A(s)$, $m_x(s)$ and $m_y(s)$ are also $s$-independent, with
$\widehat{m}_{x,y}^{(\alpha)} = B(\alpha) m_{x,y}$.
\subsubsection{Linear stability analysis of the mean-field incoherent stationary state}
\l{linear-stability-alphahmf-overdamped}
Let us now consider the $s$-independent (that is, the mean-field) incoherent state, obtained when $m_x(s)=m_y(s)=0$,
i.e.,
\be
\rho_0(\theta) = \frac{1}{2\pi}.
\l{stationincoh}
\ee
As before, its linear stability can be analyzed by posing
\be
\rho(\theta,s,t) = \frac{1}{2\pi} + \delta \rho (\theta,s,t); ~~|\delta
\rho| \ll 1,
\l{alphaperturbation_b}
\ee
and studying the linearized Fokker-Planck equation for $\delta \rho
(\theta,s,t)$:
\bea
\l{Fokkerplanck_eq_decaying_linear}
\fl \frac{\partial \delta \rho(\theta,s,t)}{\partial t}= \frac{K}{2\pi B(\alpha)}
\int_0^1 \dd s' \, \int_{-\pi}^{\pi} \dd \theta' \, \frac{\cos (\theta' - \theta)}{|s'-s|_c^\alpha}
\delta \rho(\theta',s',t) +T \frac{\partial^2 \delta \rho(\theta,s,t)}{\partial \theta^2}. 
\eea
The procedure for stability analysis is the same as that adopted in the
preceding subsection. We first perform a
Fourier expansion in $\theta$:
\be
\delta \rho(\theta,s,t) = \sum_{k=-\infty}^{+\infty}
\widehat{\delta \rho}_k(s,t) e^{ik\theta},
\l{fourier_theta_noise}
\ee
which when used in equation (\ref{Fokkerplanck_eq_decaying_linear}) gives
\bea
\l{linear_cont_eq_four_noise}
\frac{\partial \widehat{\delta \rho}_k(s,t)}{\partial t}=\frac{K}{2 B(\alpha)}\left( \delta_{k,1} + \delta_{k,-1}\right)
\int_0^1 \dd s' \, \frac{\widehat{\delta \rho}_k(s',t)}{|s'-s|_c^\alpha}
-k^2 T\widehat{\delta \rho}_k(s,t).
\eea
For $k\ne \pm 1$, the first term on the right hand side of equation
(\ref{linear_cont_eq_four_noise}) vanishes, and we have
\be
\widehat{\delta \rho}_k(s,t) = \widehat{\delta \rho}_k(s,0)
e^{-k^2 T t};~~ k \ne \pm 1;
\l{damped_modes}
\ee 
these are perturbations that decay exponentially in time, and
thus correspond to stable modes.
The equation for $k=\pm 1$,
\be
\frac{\partial \widehat{\delta \rho}_{\pm 1}(s,t)}{\partial t} = 
\frac{K}{2 B(\alpha)} \int_0^1 \dd s' \,
 \frac{\widehat{\delta \rho}_{\pm 1}(s',t)}{|s'-s|_c^\alpha}
 -T\widehat{\delta \rho}_{\pm 1}(s,t),
\l{linear_cont_eq_fourpm1_noise}
\ee
is studied by performing a further Fourier expansion in $s$-space:
\be
\widehat{\delta \rho}_{\pm 1}(s,t) = \sum_{n=-\infty}^{+\infty}
\overline{\delta \rho}_{\pm 1,n}(t) e^{2\pi i ns}.
\l{fourier_sspace_noise}
\ee
Substituting in equation (\ref{linear_cont_eq_fourpm1_noise}), we obtain
\be
\frac{\partial \overline{\delta \rho}_{\pm 1,n}(t)}{\partial t} =
\frac{K\Lambda_n(\alpha)}{2 B(\alpha)} \overline{\delta \rho}_{\pm 1,n}(t)
-T\overline{\delta \rho}_{\pm 1}(t),
\l{linear_cont_eq_fourpm1s_noise}
\ee
where $\Lambda_n(\alpha)$ is given by equation (\ref{eigenlambda}). We therefore have
\be 
\overline{\delta \rho}_{\pm 1,n}(t) = \exp \left[ \left(
\frac{K\Lambda_n(\alpha)}{2 B(\alpha)} -T \right) t \right]
\overline{\delta \rho}_{\pm 1,n}(0).
\l{solution_n1_noise}
\ee
For a fixed $K$, this expression determines the value of the temperature
for which the mode $\overline{\delta \rho}_{\pm 1,n}$ is stable. Precisely,
the mode $\overline{\delta \rho}_{\pm 1,n}$ decays exponentially in time and is
therefore stable for $T > \frac{K\Lambda_n(\alpha)}{2 B(\alpha)}$, while it
is unstable, growing exponentially in time, for $T <
\frac{K\Lambda_n(\alpha)}{2 B(\alpha)}$. Therefore, the critical
temperature for the neutral stability of $\overline{\delta \rho}_{\pm
1,n}$ is
\be
T_{c,n} = \frac{K\Lambda_n(\alpha)}{2 B(\alpha)}.
\l{crit_temper}
\ee
Since, as previously explained, $\Lambda_{-n}(\alpha)=\Lambda_n(\alpha)$, and
$\Lambda_n(\alpha)$ is a decreasing function of $|n|$, we have $T_{c,-n}=T_{c,n}$, and
\be
\frac{K}{2} = T_{c,0} > T_{c,1} > T_{c,2} > T_{c,3} > \dots
\l{order_critical}
\ee
We note in particular that for $\alpha = 0$, we have $T_{c,n}=0$ for $n>0$, so that the
modes $\overline{\delta \rho}_{\pm 1,n}$ for $|n|>0$ never destabilize.
\subsubsection{Numerical results}
\l{numerics-alphahmf-overdamped}
In the following, we take $K=1$ without loss of generality (with a rescaling of
the time unit, it is always possible to reduce to such a case).
From the analysis presented above, we see that for $T>1/2$, the incoherent state is stable.
Decreasing the temperature, the first perturbation mode to destabilize will be
$\overline{\delta \rho}_{\pm 1,0}$, which happens at $T=1/2$. Decreasing
further the temperature, 
the modes $\overline{\delta \rho}_{\pm 1,n}$ with $|n|>0$ will progressively destabilize.

\begin{figure}[h!]
\centering
\includegraphics[width=100mm]{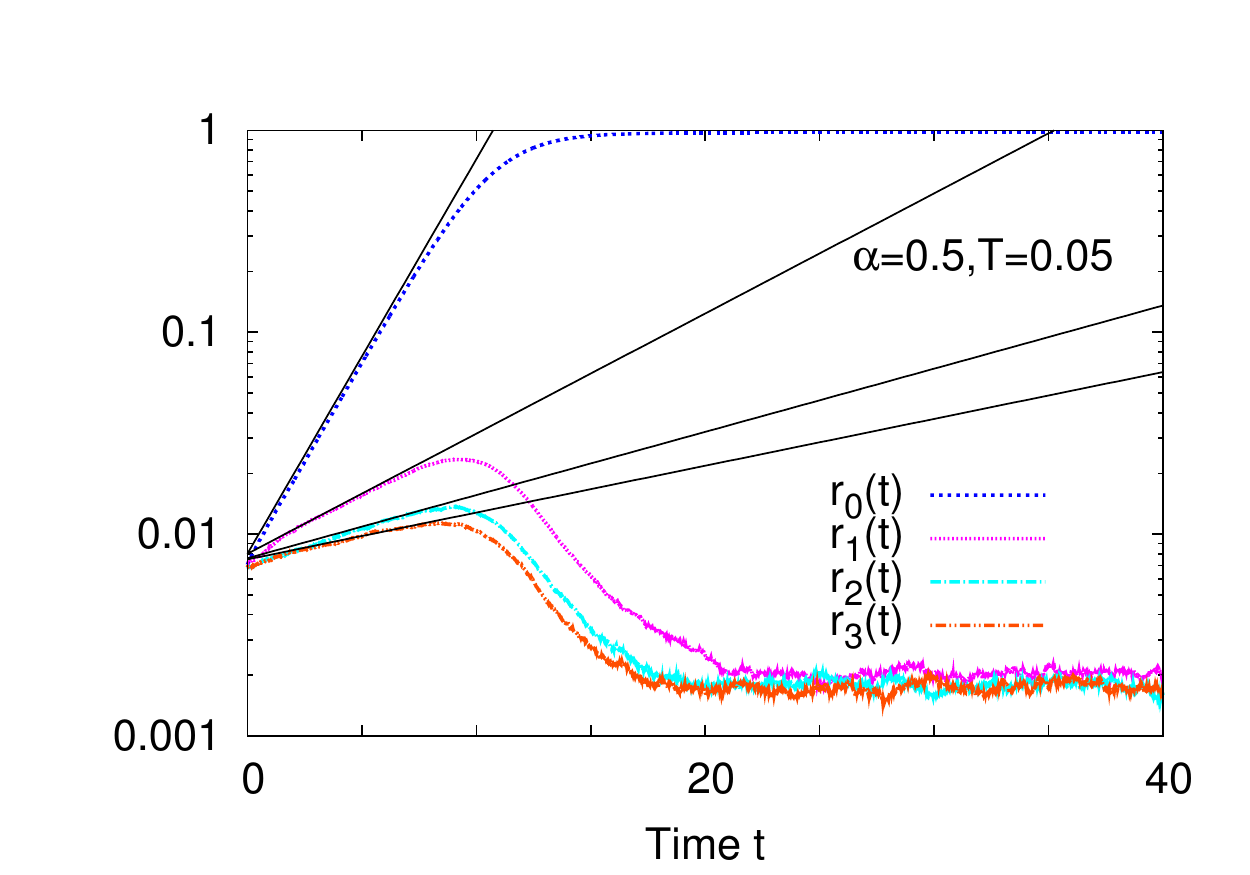}
\caption{For the model (\ref{Kuramoto-eom_decaying_noise}), the figure
shows the time evolution of the observables $r_0(t),r_1(t),r_2(t)$, and $r_3(t)$
starting from an initial state $\{\th_i(0); i = 1,2,\ldots,N\}$ that has been obtained by extracting the
$\theta_i$'s uniformly in $[-\pi,\pi]$. Therefore the initial state is the
incoherent one.
In these runs $\alpha = 0.5$ and $T=0.05$. For these values of $\alpha$
and $T$, the Fourier modes
$n=0,1,2,3$ are all linearly unstable. In particular, $T_{c,0} = 0.5$, $T_{c,1} \approx 0.18699$,
$T_{c,2} \approx 0.12206$, $T_{c,3} \approx 0.10342$. Consequently, $r_0(t),r_1(t),r_2(t)$, and $r_3(t)$
all grow exponentially in time, initially, from their initial values at $t=0$. The simulations have been
performed with $N = 2^{14}$ oscillators, and the plotted data involve and average over $100$
independent initial conditions and dynamical realizations.
The plots show that after the initial exponential growth, $r_0(t)$ attains a value very close to unity,
while $r_1(t),r_2(t),r_3(t)$ all decay to a value very close to zero (and compatible to $0$ considering the
finite size effects). The straight lines show the theoretical initial exponential growths, with rates given
by $(T_{c,m}-T)$. The agreement of the growth rates between theory and simulations is very good.}
\label{SAS-fig1}
\end{figure}

We now discuss the results of simulations of the equation of motion
(\ref{Kuramoto-eom_decaying_noise}) for a system with $N=2^{14}$
oscillators with $\alpha=0.5$. The effect of the stochastic noise has been taken into account with
the same method as that described in Appendix B, equation (\ref{formalintegration3}) for the case
of systems with inertia. We have studied the observables $r_n(t)$ defined in
equation (\ref{discrete_order_params}). In Fig. \ref{SAS-fig1}, we show the time evolution of $r_0(t)$,
$r_1(t)$, $r_2(t)$ and $r_3(t)$ for simulations performed at $T=0.05$, with initial conditions reproducing
the incoherent state, $r_n=0$ for all $n$, obtained by taking the
phases independently and uniformly distributed between $0$ and $2\pi$. From equation (\ref{crit_temper}), we find
that $T = 0.05$ lies between $T_{c,12}$ and $T_{c,13}$. Then, in
particular, the observables plotted in
Fig. \ref{SAS-fig1} should all increase exponentially in time. The plot shows that the agreement between
the numerical and the theoretical growth rates is very good. 

As is evident from the plot in Fig. \ref{SAS-fig1}, the final state
reached by the system is the mean-field synchronized one, similar to
what was observed for the model (\ref{Kuramoto-eom_decaying}). One can argue that this is the
Gibbs-Boltzmann equilibrium state of our system. For example, it has
been proved in Ref. \cite{Campa:2003} that the equilibrium state
of a system of oscillators interacting by long-range interactions on a lattice with periodic boundary conditions 
is the same as that of the corresponding mean-field system. In the
following subsection, we show that indeed such a state is dynamically
stable at temperatures $T < 1/2$.
\subsubsection{Linear stability analysis of the mean-field synchronized stationary state}
\l{linear-stability-synchronized-alphahmf-overdamped}
Let us then consider the stationary state (\ref{noise_stationary}) in
which there is only $\theta$ and no $s$ dependence:
\be
\rho_0(\theta) = A \exp \left[ \frac{1}{TB(\alpha)}\int_0^1 \dd s' \,
\int_{-\pi}^{\pi} \dd \theta' \,
\frac{\cos (\theta' - \theta)}{|s'-s|_c^\alpha} \rho_0(\theta') \right],
\l{noise_stationary_nos}
\ee
where we have put $K=1$. Using the definitions (\ref{norm_fact}) of $B(\alpha)$ and of the magnetization
components (\ref{mxsdef}) and (\ref{mysdef}), and exploiting the global
rotational invariance in $\theta$ of the system
(\ref{Kuramoto-eom_decaying_noise}) 
to put $m_y=0$, we rewrite this state as
\be
\rho_0(\theta) = A \exp \left[ \frac{1}{T} m_x \cos \theta \right].
\l{noise_stationary_nos_b}
\ee
The expression of the normalization constant becomes
\be
A = \left[2\pi I_0\left( \frac{m_x}{T} \right)\right]^{-1},
\l{normconst_nos}
\ee
while the self-consistency relation is
\be
m_x = \frac{I_1}{I_0}
\left( \frac{m_x}{T} \right).
\l{selfalcons_nos}
\ee
The last equation gives a non-vanishing $m_x$ for $T<1/2$, as follows
from the properties of $I_1$ and $I_0$, see Ref.
\cite{Campa:2009}; for
$T\ge 1/2$, the state
(\ref{noise_stationary_nos_b}) reduces to the uniform one, equation (\ref{stationincoh}).

As before, the stability of the state (\ref{noise_stationary_nos_b}) is studied by analyzing
the linearized equation obtained by inserting in equation (\ref{Fokkerplanck_eq_decaying}) the expansion
\be
\rho(\theta,s,t) = \rho_0(\theta) + \delta \rho (\theta,s,t);~~|\delta
\rho| \ll 1.
\l{exp_nonunif}
\ee
In equation (\ref{exp_nonunif}), both $\rho (\theta,s,t)$ and $\rho_0(\theta)$ are normalized,
implying that
\be
\int_{-\pi}^{\pi} \dd \theta \, \delta \rho(\theta,s,t) = 0.
\l{zeronorm}
\ee
From equation (\ref{Fokkerplanck_eq_decaying}), we have at leading order
in $\delta \rho$ the linearized equation
\bea
\l{Fokkerplanck_eq_decaying_nonunif}
\fl \frac{\partial \delta \rho(\theta,s,t)}{\partial t} = m_x\frac{\partial}{\partial \theta}
\left( \sin \theta ~\delta \rho(\theta,s,t) \right) \nonumber \\
\fl -\frac{1}{B(\alpha)} \frac{\partial}{\partial \theta} \left( \left[
\int_0^1 \dd s' \, \int_{-\pi}^{\pi} \dd \theta' \, \frac{\sin (\theta' - \theta)}{|s'-s|_c^\alpha}
\delta \rho(\theta',s',t)\right] \rho_0(\theta)\right)
 +T \frac{\partial^2 \delta \rho(\theta,s,t)}{\partial \theta^2}. 
\eea
Since the stationary state $\rho_0(\theta)$ is not uniform in $\theta$, a Fourier expansion in $\theta$
is not useful. Performing a Fourier expansion in $s$,
\be
\delta \rho(\theta,s,t) = \sum_{n=-\infty}^{+\infty} \widehat{\delta \rho}_n(\theta,t)
e^{2\pi i ns},
\l{Fourier_sspace_nonunif}
\ee
we have
\bea
\l{Fokkerplanck_eq_decaying_nonunif_s}
\fl \frac{\partial \widehat{\delta \rho}_n(\theta,t)}{\partial t} = m_x\frac{\partial}{\partial \theta}
\left( \sin \theta~~ \widehat{\delta \rho}_n(\theta,t) \right) \nonumber \\
\fl -\lambda_n(\alpha) \frac{\partial}{\partial \theta} \left( \left[
\int_{-\pi}^{\pi} \dd \theta' \, \sin (\theta' - \theta)
\widehat{\delta \rho}_n(\theta',t)\right] \rho_0(\theta)\right)
 +T \frac{\partial^2 \widehat{\delta \rho}_n(\theta,t)}{\partial \theta^2}, 
\eea
where we have used
\be
\lambda_n(\alpha) \equiv \frac{\Lambda_n(\alpha)}{B(\alpha)}.
\l{def_lambda}
\ee
From the definitions of $\Lambda_n(\alpha)$ and $B(\alpha)$, we have $0<\lambda_n(\alpha)\le 1$.

Let us now look for solutions of equation
(\ref{Fokkerplanck_eq_decaying_nonunif_s}) of the form
\be
\widehat{\delta \rho}_n(\theta,t) = \widetilde{\delta \rho}_n(\theta,\mu) e^{\mu t}.
\l{posit_spectrum}
\ee
Equation (\ref{Fokkerplanck_eq_decaying_nonunif_s}) then gives
\bea
\l{Fokkerplanck_eq_decaying_nonunif_mu}
\fl \mu \widetilde{\delta \rho}_n(\theta,\mu) = m_x\frac{\partial}{\partial \theta}
\left( \sin \theta ~\widetilde{\delta \rho}_n(\theta,\mu) \right)
\nonumber \\
\fl -\lambda_n(\alpha) \frac{\partial}{\partial \theta} \left( \left[
\int_{-\pi}^{\pi} \dd \theta' \, \sin (\theta' - \theta)
\widetilde{\delta \rho}_n(\theta',\mu)\right] \rho_0(\theta)\right)
 +T \frac{\partial^2 \widetilde{\delta \rho}_n(\theta,\mu)}{\partial
 \theta^2}.
\eea
To solve this equation and to compute the eigenvalues $\mu$, we adopt the following strategy.
The function $\widetilde{\delta \rho}_n(\theta,\mu)$ being
$2\pi$-periodic in $\theta$, it can be
expanded in the basis functions $(\cos p\theta \, , \sin p\theta)$ with $p=0,1,\dots$. Then,
we multiply equation (\ref{Fokkerplanck_eq_decaying_nonunif_mu}) in turn
by the basis functions, and then integrate over $\theta$ from $0$ to
$2\pi$ to obtain a system of algebraic equations. One gets an
identity for $p=0$, while for $p=1,2,\dots$, we obtain the system
\bea
\fl \mu \widetilde{m}_{x,n}^{(p)} = \frac{1}{2}pm_x \left[ \widetilde{m}_{x,n}^{(p-1)}
-\widetilde{m}_{x,n}^{(p+1)}\right] - Tp^2 \widetilde{m}_{x,n}^{(p)} + \frac{1}{2}\lambda_n(\alpha)p\widetilde{m}_{x,n}^{(1)}
\left[ m_x^{(p-1)} -m_x^{(p+1)}\right], 
\l{system_x} \\
\fl \mu \widetilde{m}_{y,n}^{(p)} = \frac{1}{2}pm_x \left[ \widetilde{m}_{y,n}^{(p-1)}
-\widetilde{m}_{y,n}^{(p+1)}\right] - Tp^2 \widetilde{m}_{y,n}^{(p)}+\frac{1}{2}\lambda_n(\alpha)p\widetilde{m}_{y,n}^{(1)}
\left[ m_x^{(p-1)} +m_x^{(p+1)}\right],
\l{system_y}
\eea
where we have introduced the notations
\be
\left(\widetilde{m}_{x,n}^{(p)},\widetilde{m}_{y,n}^{(p)}\right)
\equiv \int_{-\pi}^{\pi} \dd \theta \, \left( \cos p\theta , \sin p\theta \right)
\widetilde{\delta \rho}_n(\theta,\mu),
\l{tildemag}
\ee
and
\be
m_x^{(p)} \equiv \int_{-\pi}^{\pi} \dd \theta \, \cos p\theta
~\rho_0(\theta) = \frac{I_p}{I_0} \left( \frac{m_x}{T} \right).
\l{orderpmag}
\ee

Now, clearly, $\widetilde{m}_{x,n}^{(0)} = \widetilde{m}_{y,n}^{(0)} = 0$, $m_x^{(0)}=1$
and $m_x^{(1)} \equiv m_x$. There is one system of equation given by equations (\ref{system_x})
and (\ref{system_y}) for each value of $n=0,1,2,\dots$. These systems
are associated with non-Hermitian matrices; therefore, the eigenvalues $\mu$ will in general be complex.
The stationary state (\ref{noise_stationary_nos_b}) is linearly stable if the eigenvalues
of all these systems have negative real parts. We have evaluated numerically the spectrum,
and the analysis has put in evidence that this is the case. Actually, there is also a zero
eigenvalue, and in general, the presence of purely imaginary eigenvalues (zero being a particular
case) implies that the stationary state is only spectrally stable, while it might be linearly unstable.
However, proving that the zero eigenvalue has multiplicity one (see
below) ensures that
linear stability holds \cite{Holm:1985}.

Before describing the result of the numerical analysis of the systems (\ref{system_x}) and
(\ref{system_y}), we give an argument that points towards the stability of the stationary state 
(\ref{noise_stationary_nos_b}). Let us define the entropy functional
\be
S\left[\rho(\theta,s,t)\right] = -\int_0^1 \dd s \, \int_{-\pi}^{\pi} \dd \theta \, \rho(\theta,s,t) \ln
\left[ \rho(\theta,s,t)\right],
\l{entr_funct}
\ee
and the energy functional
\be
E\left[\rho(\theta,s,t)\right] = \frac{1}{2} \int_0^1 \dd s \,
\int_{-\pi}^{\pi} \dd \theta \,
\rho(\theta,s,t) u (\theta,s,t),
\l{ener_funct}
\ee
where $u(\theta,s,t)$ is the mean-field potential
\be
u(\theta,s,t) = - \frac{1}{B(\alpha)}\int_0^1 \dd s' \, \int_{-\pi}^{\pi} \dd \theta' \,
\frac{\cos (\theta' - \theta)}{|s'-s|_c^\alpha} \rho(\theta',s',t).
\l{meanfieldpot}
\ee
With the dynamics of $\rho(\theta,s,t)$ governed by the Fokker-Planck equation
(\ref{Fokkerplanck_eq_decaying}) (with $K=1$ in the present analysis), it is not difficult to
obtain that
\bea
\l{htheorem}
\frac{\dd}{\dd t} \left(E\left[\rho\right] -TS\left[\rho\right]\right) \equiv
\frac{\dd}{\dd t} F\left[\rho\right] \nonumber \\
=- \int_0^1 \dd s \, \int_{-\pi}^{\pi} \dd \theta \,
\frac{1}{\rho(\theta,s,t)} \left( \rho(\theta,s,t) \frac{\partial u(\theta,s,t)}{\partial \theta}
+T\frac{\partial \rho(\theta,s,t)}{\partial \theta}\right)^2 \le 0. 
\eea
We thus see that there is an $H$-theorem \cite{Huang:1987,Balescu:1987} associated with the evolution of $\rho(\theta,s,t)$, with
the $H$-function being the free energy $F\left[\rho\right]$; this is in
analogy with the mean-field case ($\alpha=0$) studied in Ref. \cite{Chavanis:2013}. The right hand side
of the last equation vanishes only for the stationary states given in (\ref{noise_stationary}),
and in particular, for the state (\ref{noise_stationary_nos_b}). In addition, as proved in \cite{Campa:2003},
the $s$-independent stationary state (\ref{noise_stationary_nos_b}) realizes the minimum of the free
energy. Therefore, equation (\ref{htheorem}) suggests that if this state is perturbed, the dynamics tends to
restore it.

The eigenvalues of the system (\ref{system_x}) and (\ref{system_y}) have been numerically evaluated
by truncating the system at a finite value of $p$, denoted by $p_{\it max}$.
As a matter of fact, we have found that the eigenvalues $\mu$ of the systems (\ref{system_x})
and (\ref{system_y}) always have a negative real part for any value of $\lambda_n(\alpha)$ between
$0$ and $1$ and for any temperature in the range $0 < T \le 1/2$ (except for the zero eigenvalue that we will
consider in detail below). We recall that varying $n$ and $\alpha$, the factor $\lambda_n(\alpha)$
can take any value in that range. Obviously, by truncating the system, one can find only
a finite number of eigenvalues, but by increasing the truncation value
$p_{\it max}$, we have checked that
the new eigenvalues have negative real parts with larger absolute
values, and the eigenvalues with negative real parts that have smaller
absolute values converge extremely fast. We have also found that for $T$ not
close to $0$, the eigenvalues are in addition real. This can be understood by considering the systems
(\ref{system_x}) and (\ref{system_y}) for $T\ge \frac{1}{2}$. In that case, since $m_x^{(p)}=0$
for $p>0$, they reduce to 
\bea
\l{system_x_diag}
 \mu \widetilde{m}_{x,n}^{(p)} = - Tp^2 \widetilde{m}_{x,n}^{(p)}
+ \frac{1}{2} \delta_{p,1}\lambda_n(\alpha) \widetilde{m}_{x,n}^{(1)},  \\
\l{system_y_diag}
 \mu \widetilde{m}_{y,n}^{(p)} = - Tp^2 \widetilde{m}_{y,n}^{(p)}
 +\frac{1}{2} \delta_{p,1}\lambda_n(\alpha)\widetilde{m}_{y,n}^{(1)} .
\eea
The right hand sides give directly the eigenvalues. They are real and all negative, since $T\ge \frac{1}{2}$
and $0 < \lambda_n(\alpha) \le 1$ (except for $T$ exactly equal to $\frac{1}{2}$ and for $n=0$, where
$\lambda_0(\alpha)=1$ and then the right hand sides for $p=1$ are zero). By continuity, the eigenvalues will be real for
at least a range of temperatures $T$ smaller than $\frac{1}{2}$.

\begin{figure}[h!]
\centering
\includegraphics[width=100mm]{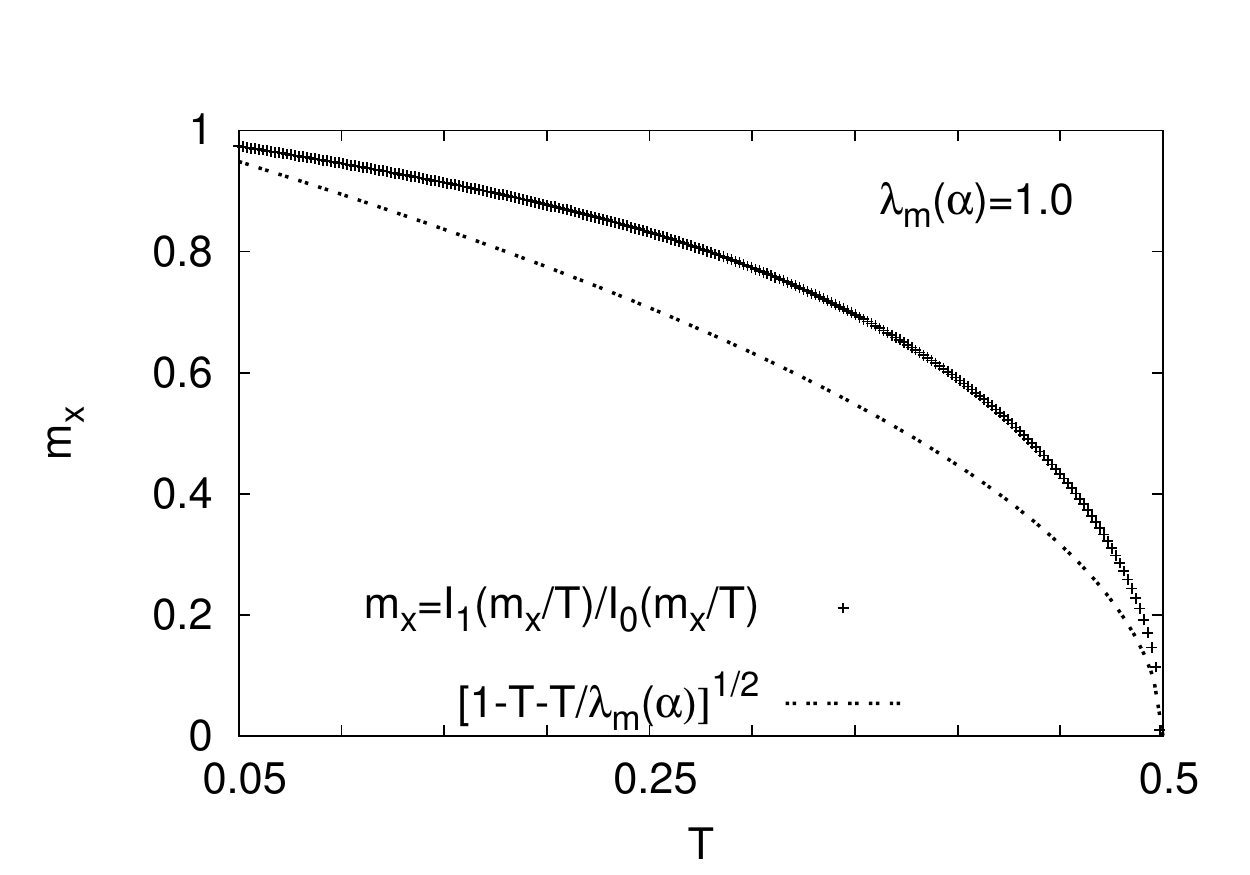}
\caption{Plot of $m_x$ as a function of $T$ as determined implicitly by the self-consistent relation
(\ref{selfalcons_nos}) and by equation (\ref{self_mtildex_b}) with $\lambda_n(\alpha) = 1$.
The two curves do not intersect at any $T$ in the range $0<T<\frac{1}{2}$, showing that there is no
solution satisfying both relations.}
\label{SAS-fig3}
\end{figure}

We conclude the analysis by studying the zero eigenvalue for $0<T<\frac{1}{2}$. For this, it is not convenient to
analyze the systems (\ref{system_x}) and (\ref{system_y}), but to start directly from
equation (\ref{Fokkerplanck_eq_decaying_nonunif_mu}) with $\mu=0$, i.e.,
\bea
m_x\frac{\partial}{\partial \theta}
\left( \sin \theta ~\widetilde{\delta \rho}_n(\theta,0) \right)
 -\lambda_n(\alpha) \frac{\partial}{\partial \theta} \left( \left[
\int_{-\pi}^{\pi} \dd \theta' \, \sin (\theta' - \theta)
\widetilde{\delta \rho}_n(\theta',0)\right] \rho_0(\theta)\right) \nonumber \\
 +T \frac{\partial^2 \widetilde{\delta \rho}_n(\theta,0)}{\partial \theta^2} = 0.
 \l{Fokkerplanck_eq_decaying_nonunif_mu_0}
\eea
The solution of this equation that satisfies the periodicity condition
and equation (\ref{zeronorm}) is
\be
\widetilde{\delta \rho}_n(\theta,0) = \frac{A}{T} \lambda_n(\alpha)\left[
\widetilde{m}_{x,n}^{(1)}\left(\cos \theta - m_x\right) + \widetilde{m}_{y,n}^{(1)}
\sin \theta \right] \exp \left[ \frac{m_x}{T}\cos \theta \right],
\l{solution_mu_0}
\ee
where the normalization constant $A$ is given in equation (\ref{normconst_nos}), and where we have used the
definition (\ref{tildemag}). This equation shows that in order to have a
non-trivial solution,
$\widetilde{m}_{x,n}^{(1)}$ and $\widetilde{m}_{y,n}^{(1)}$ cannot both be equal to $0$. We still have to
satisfy equation (\ref{tildemag}) as a self-consistent equation.
Multiplying equation (\ref{solution_mu_0})
by $\cos \theta$ and by $\sin \theta$, we obtain
\bea
\l{self_mtildex}
\widetilde{m}_{x,n}^{(1)} = \widetilde{m}_{x,n}^{(1)} \frac{\lambda_n(\alpha)}{T}
\left( 1 - T - m_x^2\right), \\
\widetilde{m}_{y,n}^{(1)} = \widetilde{m}_{y,n}^{(1)} \lambda_n(\alpha).
\l{self_mtildey}
\eea
The first of these equations is satisfied by
$\widetilde{m}_{x,n}^{(1)}=0$, or by
\be
\l{self_mtildex_b}
m_x = \sqrt{1 - T - \frac{T}{\lambda_n(\alpha)}},
\ee
that must be satisfied together with the self-consistent relation (\ref{selfalcons_nos}).
In Fig. \ref{SAS-fig3}, we plot $m_x$ as a function of $T$ as determined by the self-consistent
relation (\ref{selfalcons_nos}) and by equation (\ref{self_mtildex_b}) for $\lambda_n(\alpha)=1$.
We see that there is no solution for $0<T<\frac{1}{2}$. Since the right
hand side of equation
(\ref{self_mtildex_b}) decreases for decreasing $\lambda_n(\alpha)$, this also proves that
there is no solution for any $\lambda_n(\alpha)$. Therefore, the only solution of
equation (\ref{self_mtildex}) is $\widetilde{m}_{x,n}^{(1)}=0$. This requires that
$\widetilde{m}_{y,n}^{(1)} \ne 0$, and then equation (\ref{self_mtildey}) becomes
$\lambda_n(\alpha)=1$. This is verified only for $n=0$.

We have finally arrived at the conclusion that equation (\ref{Fokkerplanck_eq_decaying_nonunif_mu_0})
admits a solution only for $n=0$, and that this solution is unique and
is given by
\be
\widetilde{\delta \rho}_0(\theta,0) = \frac{\widetilde{m}_{y,n}^{(1)}}{T} \rho_0(\theta)
\sin \theta ,
\l{solution_mu_0_n0}
\ee
with $\widetilde{m}_{y,n}^{(1)} \ne 0$. This solution represents a global rotation of all
oscillators, and is a neutral mode due to the global rotational invariance. The
uniqueness of the mode associated with the zero eigenvalue assures that there are no secular terms
with a linear growth, thus completing the proof of the linear stability of $\rho_0(\theta)$.

\subsection{The inertial Kuramoto model with a power-law coupling and
the same natural frequency for the oscillators}
\l{alphahmf}
We will now be concerned with the model with inertia, that in the
overdamped limit reduced to the model studied in the preceding
subsection. The equations of motion are
\bea
\frac{\dd \theta_i}{\dd t}= v_i, \nonumber \\
\l{long-range-inertia} \\
m\frac{\dd v_i}{\dd t}= - \gamma v_i + \frac{\widetilde{K}}{\widetilde{N}}
\sum_{j=1}^N
\frac{\sin(\theta_j-\theta_i)}{|x_j - x_i|_c^\alpha}
+ \widetilde{\eta}_i(t) , \nonumber
\eea
with the same definitions as before of $\widetilde{N}$ and of the closest distance
convention. We recall the statistical properties of the Gaussian white
noise $\widetilde{\eta}_i(t)$:
\be
\langle \widetilde{\eta}_i(t) \rangle=0, ~~\langle
\widetilde{\eta}_i(t)\widetilde{\eta}_j(t')
\rangle=2\gamma T \delta_{ij}\delta(t-t').
\l{etatide-prop_bis}
\ee
The equations of motion (\ref{long-range-inertia}) describe the
evolution of the $\alpha$-HMF model
\cite{Anteneodo:2000,Tamarit:2000}, within a canonical ensemble. 

By performing the reduction to dimensionless quantities as in equations (\ref{dmsless1})-(\ref{dmsless6}),
the equations of motion become
\bea
\frac{\dd \theta_i}{\dd t}= v_i, \nonumber \\
\l{long-range-inertia_dms} \\
\frac{\dd v_i}{\dd t}= - \frac{1}{\sqrt{m}} v_i + \frac{1}{\widetilde{N}}
\sum_{j=1}^N
\frac{\sin(\theta_j-\theta_i)}{|x_j - x_i|_c^\alpha}
+ \eta_i(t) , \nonumber
\eea
where we have disregarded the overbars of the dimensionless
quantities for notational convenience, and we have
\be
\langle \eta_i(t) \rangle=0, ~~\langle
\eta_i(t)\eta_j(t') \rangle=2 \left(T/\sqrt{m}\right) \delta_{ij}\delta(t-t').
\l{etatide-prop_dms}
\ee

The continuum limit of the dynamics is implemented in a manner analogous to that in preceding sections, 
by introducing the variable $s \in [0,1]$ as the continuum limit of $s_j=j/N$. The one-particle
distribution function $f(\theta,v,s,t)$ is such that $f(\theta,v,s,t)\dd \theta \dd v \dd s$
is the fraction of oscillators located between $s$ and $s + \dd s$ that
at time $t$ has phase between
$\theta$ and $\theta + \dd \theta$ and angular velocity between $v$ and $v + \dd v$. The normalization is
\be
\int_{-\pi}^{\pi} \dd \theta \, \int_{-\infty}^\infty \dd v \,
f(\theta,v,s,t) = 1~~\forall~~s.
\l{norm_cond_finert}
\ee
The equations of motion in the continuum limit are
\bea
\fl \frac{\dd \theta(s,t)}{\dd t}= v(s,t), \nonumber \\
\l{long-range-inertia_cont} \\
\fl
\frac{\partial v (s,t)}{\partial t}= - \frac{1}{\sqrt{m}} v(s,t) + \frac{1}{B(\alpha)} \int_0^1 \dd s' \,
\int_{-\pi}^{\pi} \dd \theta' \, \int_{-\infty}^\infty \dd v' \,
\frac{\sin (\theta' - \theta)}{|s'-s|_c^\alpha}
f(\theta',v',s',t) + \eta(s,t) , \nonumber
\eea
with $B(\alpha)$ defined previously. The Kramers equation for $f(\theta,v,s,t)$ is
\bea
\l{Kramers_decaying}
\fl \frac{\partial f(\theta,v,s,t)}{\partial t} = - v\frac{\partial f(\theta,v,s,t)}{\partial \theta}
+\frac{T}{\sqrt{m}}\frac{\partial^2 f(\theta,v,s,t)}{\partial v^2} 
\nonumber \\ \fl
+\frac{\partial}{\partial v}\left[ \left( \frac{v}{\sqrt{m}} - \frac{1}{B(\alpha)} 
\int_0^1 \dd s' \, \int_{-\pi}^{\pi} \dd \theta' \, \int_{-\infty}^\infty \dd v' \,
\frac{\sin (\theta' - \theta)}{|s'-s|_c^\alpha} f(\theta',v',s',t) \right)
f(\theta,v,s,t) \right]. 
\eea
\subsubsection{Linear stability analysis of the mean-field incoherent stationary state}
\l{linear-stability-alphahmf}
Below we will present the results of numerical simulations by plotting, as before, the first few of the observables
defined in equation (\ref{discrete_order_params}). Before that, we perform a stability analysis of the 
$s$-independent incoherent stationary
state of the Kramers equation (\ref{Kramers_decaying}) given by
\be
f_0(v) = \frac{1}{2\pi} \frac{1}{\sqrt{2\pi T}} e^{-\frac{v^2}{2T}}.
\l{homog_kramers}
\ee
Similarly to equation (\ref{expansion_linear}), we linearize the Kramers equation by posing
\be
f(\theta,v,s,t) = f_0(v) + e^{\nu t} \delta f(\theta,v,s); ~~|\delta
\rho| \ll 1,
\l{expansion_decaying}
\ee
where normalization of both $f_0(v)$ and $f(\theta,v,s)$ implies that
\be
\int_{-\pi}^{\pi} \dd \theta \, \int_{-\infty}^\infty \dd v \, \delta
f(\theta,v,s) = 0 ~~\forall~~  s.
\l{norm_cond_delta_0}
\ee
At leading order, we obtain from equation (\ref{Kramers_decaying}) that
\bea
\l{Kramers_decaying_linear}
\fl \nu \delta f(\theta,v,s) = - v\frac{\partial \delta f(\theta,v,s)}{\partial \theta}
+\frac{1}{\sqrt{m}}\frac{\partial}{\partial v}\left( v \delta f(\theta,v,s)\right)
+\frac{T}{\sqrt{m}}\frac{\partial^2 \delta f(\theta,v,s)}{\partial v^2}, 
\nonumber \\ \fl
- \frac{1}{B(\alpha)} \frac{\partial f_0(v)}{\partial v}\int_0^1 \dd s'
  \, \int_{-\pi}^{\pi}
\dd \theta' \, \int_{-\infty}^\infty \dd v' \, \frac{\sin (\theta' - \theta)}{|s'-s|_c^\alpha}
\delta f(\theta',v',s'). 
\eea

The analysis of equation (\ref{Kramers_decaying_linear}) is very similar to that followed in
section \ref{linearstability-incoherent-inertia}, and therefore, we do not
repeat all the details here. Labelling $\delta f$ with the eigenvalue $\nu$, we pose
\be
\l{expansion_nu}
\delta f(\theta,v,s,\nu) = \sum_{k=-\infty}^\infty b_k(v,s,\nu) e^{ik\theta},
\ee
with $b_{-k} = b_k^*$ and $b_0=0$. Substituting in equation
(\ref{Kramers_decaying_linear}), we have
\bea
\frac{\partial^2 b_k(v,s,\nu)}{\partial v^2}+\frac{v}{T}\frac{\partial b_k(v,s,\nu)}{\partial v}
+\frac{1}{T}\left(1-\nu\sqrt{m}-ikv\sqrt{m}\right)b_k(v,s,\nu)
\nonumber \\
=\frac{1}{B(\alpha)}\frac{\sqrt{m}}{T}\frac{\partial f_0(v)}{\partial v} i\pi \left( \delta_{k,1} -
\delta_{k,-1}\right) \int_0^1 \dd s' \, \frac{1}{|s'-s|_c^\alpha}
\langle 1,b_k \rangle (s',\nu),
\l{eq_b_deltaomega}
\eea
where the scalar product is defined by
\be
\langle \varphi,\psi \rangle (s'',s') \equiv \int_{-\infty}^{\infty} \dd v \,
\varphi^*(v,s'')\psi(v,s').
\l{scalar-prod_sdep}
\ee
For $k \ne \pm 1$, when the right hand side is equal to $0$, equation (\ref{eq_b_deltaomega}) is identical
to equation (\ref{b-eqn}) with $\omega=0$. Therefore, we can immediately
write down the negative eigenvalues as
\be
\nu_{p,k}
= -\frac{p}{\sqrt{m}} - k^2 T\sqrt{m};~~ p = 0, 1, 2,\ldots
\l{eigenvalues_decaying}
\ee
For $k=\pm 1$, we proceed as follows. Let us consider only $k=1$, since
$b_{-1}=b_1^*$. We perform the expansion 
\be
b_1(v,s,\nu) = \sum_{n=-\infty}^{+\infty} b_{1,n}(v,\nu)e^{2\pi i ns}.
\l{expansion_thetas}
\ee
Substituting in equation (\ref{eq_b_deltaomega}), we obtain
\bea
\frac{\partial^2 b_{1,n}(v,\nu)}{\partial v^2}+\frac{v}{T}\frac{\partial b_{1,n}(v,\nu)}{\partial v}
+\frac{1}{T}\left(1-\nu\sqrt{m}-ikv\sqrt{m}\right)b_k(v,s,\nu)
\nonumber \\
=\lambda_n(\alpha)\frac{\sqrt{m}}{T}\frac{\partial f_0(v)}{\partial v} i\pi 
\langle 1,b_{1,n} \rangle,
\l{eq_b_deltaomega_k1}
\eea
where now the scalar product $\langle 1,b_{1,n} \rangle$ does not depend on $s$, and where
$\lambda_n(\alpha)$ is defined in equation (\ref{def_lambda}). Comparing with
equation (\ref{b-eqn}), it is evident that on performing the same
analysis as in section \ref{linearstability-incoherent-inertia}, one arrives at the following implicit equation
for the eigenvalue $\nu$:
\be
\lambda_n(\alpha)\frac{e^{mT}}{2T}\sum_{p=0}^\infty
\frac{(-m T)^p(1+\frac{p}{mT})}{p!\left(1+\frac{p}{mT}+\frac{\nu}{T\sqrt{m}}\right)} - 1 =0.
\l{stability-eqn_deltaomega}
\ee
The stability threshold is again given by the value of the last
expression for $\nu=0$:
\be
\frac{\lambda_n(\alpha)}{2T} - 1 =0.
\l{stability-eqn_deltaomega_crit}
\ee
We therefore obtain the same critical temperature for the $n$th mode as given in
equation (\ref{crit_temper}) (where now $K=1$). 
\subsubsection{Numerical results}
\l{numerics-alphahmf}
In simulations, we monitor as in the previously discussed cases the observables $r_n(t)$
defined in equation (\ref{discrete_order_params}). In Fig.
\ref{fignotesinertia}, we show the evolution
of $r_0(t)$, $r_1(t)$, $r_2(t)$ and $r_3(t)$, in simulations of the equations of
motion (\ref{long-range-inertia}) with $m=1, \widetilde{K}=1,\gamma=0.5$, at temperature
$T=0.02$ and $\alpha=0.5$, starting from an initial state uniform in $\theta$
and Gaussian in the velocity, equation (\ref{homog_kramers}). For these values of $T$ and $\alpha$, the modes
$b_{1,n}$ for $n=0,1,2,3$ are all unstable (see the critical temperatures in the caption of
Fig. \ref{SAS-fig1}). The simulation has been performed with $N=1024$
oscillators.

\begin{figure}[here!]
\centering
\includegraphics[width=100mm]{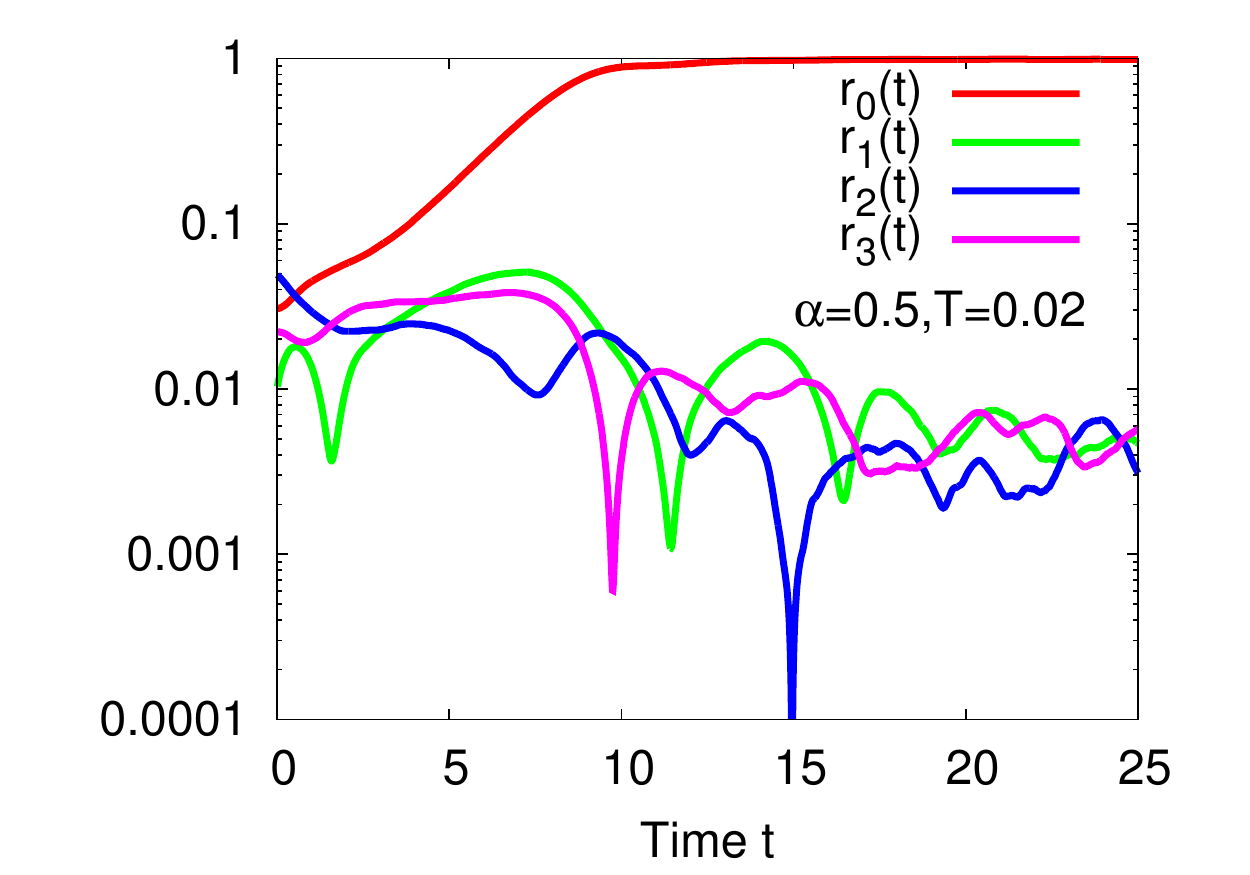}
\caption{For the model (\ref{long-range-inertia}), the figure shows the
time evolution of the observables $r_0(t),r_1(t),r_2(t),r_3(t)$ for a system of $N=1024$
oscillators with $m=1,\widetilde{K}=1,T=0.02,\alpha=0.5,\gamma =0.5$,
while starting from an initial state uniform in $\theta$
and Gaussian in the velocity, equation (\ref{homog_kramers}).}
\l{fignotesinertia}
\end{figure}

We see that, as expected, the incoherent state is not stable, since the order parameter $r_0$
grows exponentially and reaches an asymptotic value that at this temperature is very close to $1$.
This is similar to what happens in the simulations of the models in the
preceding subsections. Also, the long-time decay of $r_n$ with $n>0$ is similar.
However, contrary to the cases in the previous subsections, now the initial exponential growth of these parameters is not visible
in Fig. \ref{fignotesinertia}. This is probably due to finite-size effects. We stress that these effects for a given
number of oscillators $N$ are expected to be more marked for a system with inertia than in
an overdamped system, since the former has two dynamical variables per
oscillators, and consequently, the distribution $f$ depends on two
dynamical variables. In Fig. \ref{fignotesinertianbig}, we plot the
results of a simulation run with $N=2^{17}$ oscillators. For this larger
system, the initial exponential
growth of all the $r_n(t)$'s is clearly visible. The theoretical rates are also shown with full
lines. The agreement with the simulation is satisfactory. The attainment of the asymptotic value
of $r_0$ and the decay to zero of the parameters $r_n$ with $n>0$ occur
at later times with respect to the smaller
system, and are not displayed in the figure.

\begin{figure}[here!]
\centering
\includegraphics[width=100mm]{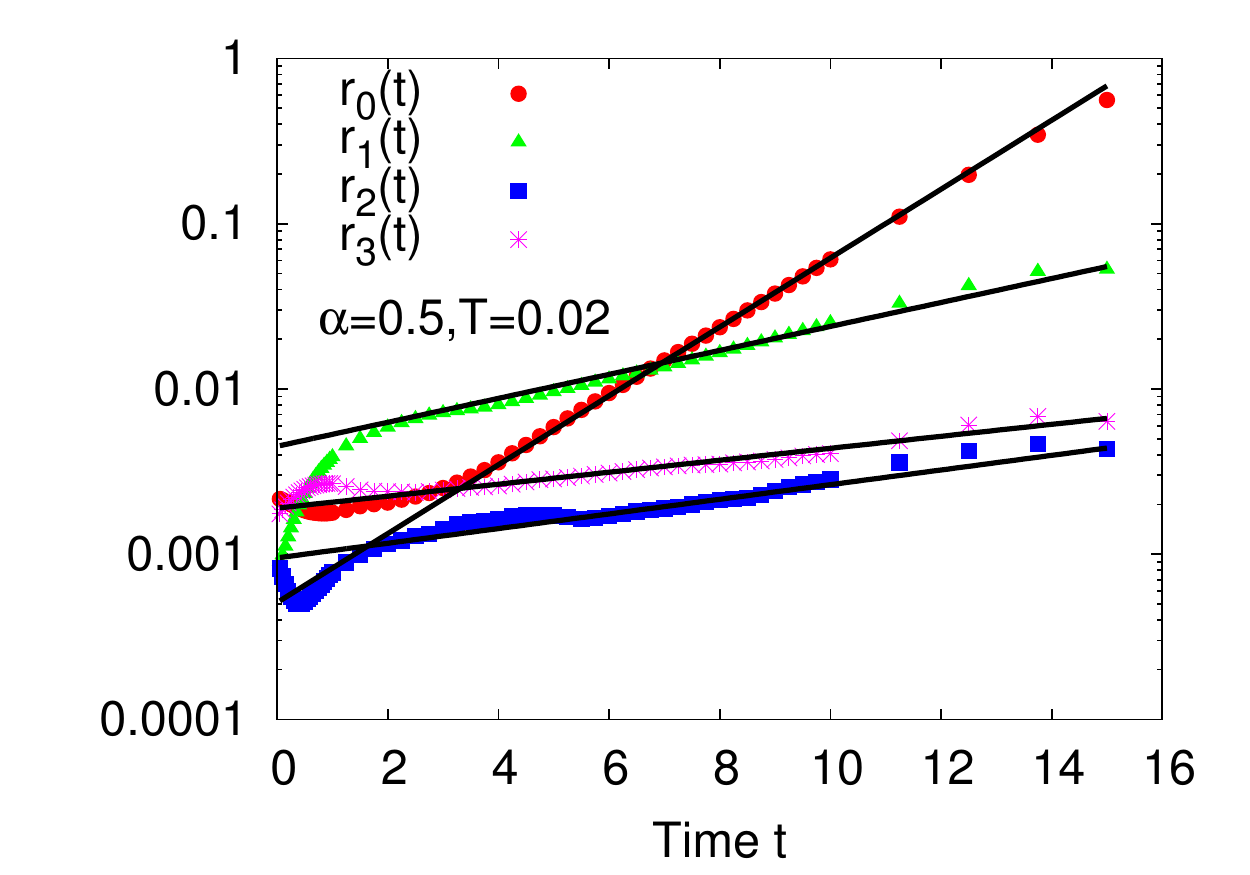}
\caption{For the model (\ref{long-range-inertia}), the figure shows the
time evolution of the observables $r_0(t),r_1(t),r_2(t),r_3(t)$ for a system of $N=2^{17}$
oscillators with $m=1,\widetilde{K}=1,T=0.02,\alpha=0.5,\gamma =0.5$, while starting from an initial state uniform in $\theta$
and Gaussian in the velocity, equation (\ref{homog_kramers}). The full straight lines show the theoretical growth rates.}
\l{fignotesinertianbig}
\end{figure}

\section{Conclusions and perspectives}
\l{conclusions}
Spontaneous synchronization appears naturally out of a competition
between two qualitatively different dynamical regimes
in systems that can be described as a set of interacting
oscillators.
The main purpose of this work was to show that in addition to this
purely dynamical view, introducing noise into the dynamics allows one to study synchronization in the framework
of statistical mechanics. This is not only possible, but also very useful, since it paves the way for use of efficient
analytical tools commonly employed in the study of
the statistical behaviour of many-body systems. One then derives that synchronization is a phase
transition, characterized by the appearance of a non-vanishing value of an
appropriate order parameter.

We have shown that, interpreted as a system of interacting particles, the Kuramoto model and its various extensions
are long-range interacting lattice systems. Furthermore, they are in
their original setting mean-field systems, the extreme case of
long-range interacting systems, where all pairs of particles interact
with equal coupling strengths. We have also considered models
where the coupling strengths decay slowly with the distance between the
lattice sites. Long-range systems often enjoy peculiar features, both in and out of equilibrium,
due to the non-additivity of the interaction energy between subparts of the system. The form of the interaction
in the Kuramoto model and its extensions may be used as a prototypical
interaction for studying these
features.

Statistical dynamics of long-range systems, in the limit of a very large number of particles, can be very well
described by equations that involve only the one-particle distribution
function. The particular time evolution equation for the one-particle
distribution depends
on the system at hand: (i) for overdamped systems, it is the
continuity equation in the case of noiseless dynamics, and
the Fokker-Planck equation for the noisy case; (ii) for underdamped systems,
it is the Vlasov equation for the noiseless dynamics, and the Kramers
equation for the noisy case. Stable stationary solutions of these equations correspond to stationary states in
which the system remains trapped for a time that diverges with the system size (the limit in which
the equations for the one-particle distribution function become exact).
As exemplified in this review, the phase transitions mentioned above are then
a change of the stability properties of the stationary states corresponding to the synchronized and the
unsynchronized state. 

The presence of distributed natural frequencies in the dynamics of the Kuramoto
model leads to a violation of detailed
balance in the stationary state, thereby resulting in long-time
stationary states which are out of equilibrium.
These are the so-called nonequilibrium stationary states (NESSs)
characterized by a net non-zero probability
current around a closed loop in the configuration phase. As a result,
one cannot use the free energy as a thermodynamic
potential to determine the nature of the stationary state of the system.
However, to this end, we have shown that it is possible to employ
successfully at least
in numerical simulations probability distributions of the order parameter analogous to those employed in
equilibrium statistical mechanics.

In this review, we have restricted our analysis to unimodal
distributions for the natural frequencies. Let us comment on this
point with respect to the mean-field models. For such distribution
functions, the synchronization transition in the Kuramoto model and in its noisy extension is continuous,
i.e., the order parameter of the stable stationary state grows
continuously from zero as the coupling strength increases beyond a critical value
at a given temperature, or, equivalently, as the temperature is
decreased below a critical value for a given coupling strength. On the other
hand, in the model with inertia, the transition becomes of first-order
type: we have found that in certain ranges of the
parameters, both the incoherent and the synchronized state are dynamically stable (this situation is often referred to
as bistability), with one of the two states being the globally stable state, depending on the variation of the
parameters within the range. This gives rise to the existence of hysteresis loops. The overall picture is
probably different with more general frequency distributions. There have been studies of the original Kuramoto model
with non-unimodal $g(\omega)$. It has been shown that in the case of a uniform distribution, that can be
considered a limiting case (although with a singular derivative) of a unimodal distribution, the transition becomes
of a first-order type, since at the threshold value $K_c$ of the
coupling parameter given by Eq. (\ref{Kura-bareKc}),
there appears a solution of Eq. (\ref{Kura-bifurcation}) with
$r=r_c=\pi/4$ \cite{Pazo:2005}. Also, continuous
bimodal distributions have been studied, and it has been shown that, depending on the
structure of the distribution, there can be bistability and states with clusters of oscillators locked at different
frequencies (related to the maxima of the bimodal distribution) \cite{Martens:2009,Pazo:2009}.

Here, we have discussed models in which the mean-field interaction is replaced by coupling strengths between the oscillators
that decay as a power law with the distance between the oscillators
residing on the sites of a lattice. In particular, we have focussed on
one-dimensional lattices, with the parameter $\alpha$
characterizing the decay being smaller than $1$, to remain within the framework
of long-range interactions. Furthermore, we have imposed periodic
boundary conditions. As we have stressed while introducing this class of systems, periodic boundary conditions cause
the uniformity on the lattice of the equilibrium or stationary states, but they do not a priori
rule out the influence of the lattice structure on the dynamical
behavior. However, we have shown that the mean-field Fourier mode of the
spatial distribution of oscillator phases dominates the
out-of-equilibrium dynamics at long times, since it is this mode that gets
destabilized first on increasing the coupling constants or decreasing
the temperature. On the other hand, the non-zero Fourier modes destabilize at higher coupling constants
or smaller temperatures. This is common to all the lattice models we have analyzed.
In particular, the mean-field mode dominates in the underdamped noisy dynamics considered
in section \ref{alphahmf}; a similar dominance has been found in the
study of the microcanonical ensemble dynamics of this
system, i.e., without the noise \cite{Bachelard:2011}.
Although we do not have analytical or numerical
evidences, we feel that it is not unreasonable to adduce the hypothesis that also with more general
boundary conditions, the mean-field mode is the one relevant for the dynamics; however, we understand
that without a detailed analysis, this statement remains at the level of
speculation. Let us remark that there have been earlier works
on the Kuramoto model with coupling constants decaying with a power law
on one-dimensional periodic lattices, with the purpose to study the
existence of the synchronized phase as a function of the power-law parameter
$\alpha$ in the limit $N \to \infty$. The critical value of $\alpha$ for the existence of
the synchronized phase has been numerical evaluated to be (about) $2$
\cite{Rogers:1996}; in Ref. \cite{Chowdhury:2010}, a spin wave approximation and
simulations performed at larger $N$ values suggest on the other hand that the critical
value is $3/2$.

We now point out some important issues that have not been discussed
in this review, e.g., details of the dynamical behavior of the system
for a large but finite number of oscillators. Among finite-size effects,
of particular relevance are slow processes out of equilibrium, and the
scaling of the associated timescales with the system size. When the
stationary states of the single-particle equations are unstable, we expect
that such relaxation does not depend on the size of the system, for
large enough system size. On the other hand, stable states could be
destabilized by finite-size effects. For example, in Hamiltonian long-range systems, finite systems slowly evolve
in time out of the stationary states of the Vlasov equation that governs the dynamics for $N\to \infty$. The lifetime
of these ``quasi-stationary" states generally diverges with the system
size as a power law \cite{Campa:2009}. We expect that something
similar may happen in the noisy driven systems studied in this review. This could affect,
e.g., the rate of hopping between stationary states when there is
bistability. Another property that is affected
by the finite size of the system is the stability property of the incoherent state of the
original Kuramoto model. As a matter of fact, it has been shown that the
incoherent state, although neutrally stable in the limit
$N \to \infty$, becomes
fully stable for finite $N$ \cite{Buice:2007}, due to a mechanism
very similar to that of Landau damping in plasma physics \cite{Strogatz:2000}.

In conclusion, we would like to stress that the Kuramoto model and its extensions, besides being related
to real systems as emphasized in the introduction, provide an
interesting benchmark to study and analyze a variety of physical properties. In fact, they offer the possibility to consider
synchronization both as a purely dynamical effect and as an emerging
phenomenon typical of the statistical behavior
of many-body systems. Furthermore, the long-range character of the
interaction gives rise to some peculiar properties that are typical for
this class of systems. We hope that this review has succeeded in giving a
flavor of the aforementioned issues, and will serve as an invitation to
indulge in further studies of the Kuramoto model.

\section*{Acknowledgments}
We acknowledge the hospitality of ENS-Lyon, and support of the CEFIPRA Grant 4604-3
(S.G.) and the grant ANR-10-CEXC-010-01 (S.G. and S.R.). S. G. and A. C.
acknowledge the hospitality of the Universit\`{a} di
Firenze. We warmly thank F. Bouchet, T. Dauxois, A. Ghosh, M. Komarov, D. Mukamel, C. Nardini, H. Park, A. Patelli, A. Pikovsky,
M. G. Potters, and H. Touchette for fruitful discussions over the years on topics reported in this review.
We thank the Galileo Galilei Institute for Theoretical Physics
(Florence) for the hospitality and INFN for partial support during the
completion of this work.

\appendix

\section*{Appendix A: The noiseless Kuramoto model with inertia:
Connection with electrical power distribution models}
\l{app-power-distribution}
\setcounter{section}{1}
\addcontentsline{toc}{section}{Appendix A: The noiseless Kuramoto model with inertia:
Connection with electrical power distribution models}
Here, we briefly discuss, following Refs.
\cite{Filatrella:2008,Rohden:2012}, how the dynamics (\ref{eom})
arises in connection with electrical power distribution networks.

The essential elements of an electrical power distribution network or grid
are synchronous generators located at power plants and motors located
with the consumers. While a generator converts mechanical (or other
forms of energy) into
electrical energy, the reverse is true for a motor. Let $P$ denote the
power, which being generated is a positive quantity for a generator and being consumed
is negative for a motor. Either unit basically consists of a rotating
turbine whose state for the $j$th unit is represented by its phase
\be
\th_j(t)=\Omega t+\phi_j(t),
\ee
where $\Omega$ is the standard supply frequency, $\Omega=50/60$~Hz typically, while $\phi_j(t)$ is the deviation from uniform rotation.
From considerations of energy conservation, the generated or consumed
power $P^{\rm source}_i$ of the $i$th element equals the sum of the power  $P^{\rm trans}_i$ exchanged
with the grid, the power $P^{\rm
acc}_i=(I/2)(\dd/\dd t)(\dd\th_i(t)/\dd t)^2$ accumulated in the turbine, and the
amount $P^{\rm diss}_i=\kappa(\dd\th_i(t)/\dd t)^2$ dissipated in overcoming
friction, where $I$ is the moment of inertia of the turbine and $\kappa$
is the friction constant. The power transmitted between
two elements $j$ and $i$ connected by a transmission line depends on the
phase difference across the ends of the transmission line, and is given
by $P_{{\rm max};ji} \sin(\th_j-\th_i)$, where $P_{{\rm max};ji}$ is the
maximum capacity of the transmission line. With $P^{\rm trans}_i=\sum_j P_{{\rm max};ji} \sin(\th_j-\th_i)$, we then have 
\be
P^{\rm source}_i=\fr{I}{2}\fr{\dd}{\dd t}\Big(\fr{\dd\th_i(t)}{\dd
t}\Big)^2+\kappa\Big(\fr{\dd\th_i(t)}{\dd t}\Big)^2+\sum_j P_{{\rm max};ji} \sin(\th_j-\th_i).
\ee
With the assumption that $|\dd \phi/dt| \ll \Omega$, one arrives at the
equation of motion \cite{Filatrella:2008,Rohden:2012}
\be
\fr{\dd^2\phi_i(t)}{\dd t^2}=P_i-\gamma \fr{\dd \phi_i}{\dd
t}-\sum_{j}K_{ji}\sin(\phi_j-\phi_i),
\l{eom-electric}
\ee
where
\bea
&&P_i=\fr{P^{\rm source}_i-\kappa \Omega^2}{I\Omega}, \\
&&\gamma=\fr{2\kappa}{I}, \\
&&K_{ji}=\fr{P_{{\rm max};ji}}{I\Omega}.
\eea

In the mean-field approximation, where every unit $i$ is connected to
every other unit $j$ with equal strength and $K_{ji}=K/N$, where
$N$ is the total number of nodes in the network, equation
(\ref{eom-electric}) reduces to
\be
\fr{\dd^2\phi_i(t)}{\dd t^2}=P_i-\gamma \fr{\dd \phi_i}{\dd
t}-\fr{K}{N}\sum_{j}\sin(\phi_j-\phi_i).
\l{eom-electric-mfa}
\ee
Note that the $P_i$'s are intrinsic to the units and in general vary
from one unit to another, so that they may be regarded as quenched
random variables. The above dynamics is similar to the generalized Kuramoto model dynamics
(\ref{eom}) in the absence of noise $\eta_i(t)$.
\section*{Appendix B: Simulation details} 
\l{app-simulation}
\setcounter{section}{2}
\setcounter{equation}{0}
\addcontentsline{toc}{section}{Appendix B: Simulation details}
Here we describe the method to simulate the dynamics (\ref{eom-scaled}) for
given values of $m,T,\sigma$ (note that we are dropping overbars for
simplicity of notation), and for a given realization of
$\omega_i$'s, by employing a numerical integration scheme
\cite{Gupta:2012}. To simulate the dynamics over a time interval $[0:\mathcal{T}]$, we first choose a time step size
$\Delta t \ll 1$. Next, we set $t_n=n\Delta t$ as the $n$-th time step
of the dynamics, where $n=0,1,2,\ldots,N_t$, and
$N_t=\mathcal{T}/\Delta t$. In the numerical scheme, we first discard at
every time step the effect of the noise (i.e., consider
$1/\sqrt{m}=0$), and employ a fourth-order symplectic
algorithm to integrate the resulting symplectic part of the
dynamics \cite{McLachlan}. Following this, we add the effect of noise, and implement an Euler-like
first-order algorithm to update the dynamical variables.   
Specifically, one step of the scheme from $t_n$ to $t_{n+1}=t_n+\Delta t$ involves the following
updates of the dynamical variables for $i=1,2,\ldots,N$: For the symplectic part, we have, for $k=1,\ldots,4$, 
\begin{eqnarray}
&&v_i\Big(t_{n}+\frac{k\Delta t}{4}\Big)=v_i\Big(t_n+\frac{(k-1)\Delta
t}{4}\Big)+b(k)\Delta t\Big[r\Big(t_n+\frac{(k-1)\Delta
t}{4}\Big)\nonumber \\
&&\sin\Big\{\psi\Big(t_n+\frac{(k-1)\Delta
t}{4}\Big)-\th_i\Big(t_n+\frac{(k-1)\Delta
t}{4}\Big)\Big\}+\sigma\omega_i\Big]; \nonumber \\
&&r\Big(t_n+\frac{(k-1)\Delta
t}{4}\Big)=\sqrt{r_x^2+r_y^2},\psi\Big(t_n+\frac{(k-1)\Delta
t}{4}\Big)=\tan^{-1}\frac{r_y}{r_x}, \nonumber \\
&&r_x=\frac{1}{N}\sum_{j=1}^N
\sin\Big[\th_j\Big(t_n+\frac{(k-1)\Delta t}{4}\Big)\Big],r_y=\frac{1}{N}\sum_{j=1}^N
\cos\Big[\th_j\Big(t_n+\frac{(k-1)\Delta t}{4}\Big)\Big], \nonumber \\
\label{formalintegration1} \\ 
&&\th_i\Big(t_{n}+\frac{k\Delta t}{4}\Big)=\th_i\Big(t_n+\frac{(k-1)\Delta
t}{4}\Big)+a(k)\Delta t ~v_i\Big(t_n+\frac{k\Delta t}{4}\Big),
\label{formalintegration2} 
\end{eqnarray}
where the constants $a(k)$'s and $b(k)$'s are obtained from Ref.
\cite{McLachlan}.
At the end of the updates (\ref{formalintegration1}) and
(\ref{formalintegration2}), we have the set
$\{\th_i(t_{n+1}),v_i(t_{n+1})\}$. Next, we include the effect of the
stochastic noise by keeping $\th_i(t_{n+1})$'s unchanged,  but by
updating $v_i(t_{n+1})$'s as
\begin{equation}
v_i(t_{n+1}) \to v_i(t_{n+1})\Big[1-\frac{1}{\sqrt{m}} \Delta
t\Big]+\sqrt{2\Delta t\frac{T}{\sqrt{m}}}\Delta X(t_{n+1}).
\label{formalintegration3}
\end{equation}
Here $\Delta X$ is a  Gaussian distributed
random number with zero mean and unit variance.
\section*{Appendix C: A fast numerical algorithm to compute the interaction expression in models
with power-law interactions} 
\l{app-fast_algo}
\setcounter{section}{3}
\setcounter{equation}{0}
\addcontentsline{toc}{section}{Appendix C: A fast numerical algorithm to compute the interaction term
in the models with a power-law coupling}
For the models discussed in section
\ref{chap3}, the interaction term in the equation of motion for each of
the $N$ oscillators involves a sum over $N$ terms. This would imply at
each time step of numerical simulation of the dynamics a computation
time that scales as $N^2$. Here we discuss  
an alternative and efficient numerical algorithm \cite{Gupta:2012a} that transforms the
interaction term into a convenient form, allowing for its computation by a Fast Fourier Transform (FFT) scheme
in a time scaling as $N \ln N$. Use of FFT requires that we choose a power of $2$ for $N$. 

Let us denote with $J_i$ the sum appearing in the equations
of motion (\ref{Kuramoto-eom_decaying}),
(\ref{Kuramoto-eom_decaying_noise}) and (\ref{long-range-inertia_dms}):
\be
J_i = \sum_{j=1}^{N}\fr{\sin(\th_j-\th_i)}{(d_{ij})^{\a}},
\l{EOM1}
\ee
where $d_{ij}$ is the shortest distance between sites $i$ and $j$ on a
one-dimensional periodic lattice of $N$ sites.
Our simulations results presented in section \ref{chap3} were obtained
by considering the lattice constant $a$ to be unity. Therefore,
$d_{ij}$ for $i \ne j$ is given by
\be
d_{ij}=\left\{ 
\begin{array}{ll}
                |j-i|; & \mbox{if $1 \le |j-i| \le N/2$}, \\
               N-|j-i|; & \mbox{otherwise},
               \end{array}
        \right. \\
\label{dij}
\ee
while, as explained in the main text, we choose the value of $d_{ii}$, irrelevant for the equations of motion,
equal to $1$. Equation (\ref{EOM1}) may be rewritten as
\be
J_i = \cos \theta_i \sum_{j=1}^{N}V_{ij}\sin\th_j - \sin \th_i \sum_{j=1}^{N}V_{ij} \cos \theta_j .
\l{EOM2}
\ee
The first summation may be interpreted as the $i$th element of the column vector formed by the product of an
$N \times N$ matrix $V=[V_{ij}]_{i,j=1,2,\ldots,N}$ with the column vector $(\sin \th_1 \sin \th_2 \ldots, \sin \th_N)^T$,
where $V_{ij}=1/(d_{ij})^\a$. Similarly, the second summation may be interpreted as the $i$th element
of the column vector formed by the product of $V$ with the column vector
$(\cos \th_1 \cos \th_2 \ldots, \cos \th_N)^T$. The matrix $V$ has the form
\bea
V =  \left[ {\begin{array}{ccccc}
v_1     & v_{N} & \dots  & v_{3} & v_{2}  \\
v_{2} & v_1    & v_{N} &         & v_{3}  \\
\vdots  & v_{2}& v_1    & \ddots  & \vdots   \\
v_{N-1}  &        & \ddots & \ddots  & v_{N}   \\
v_{N}  & v_{N-1} & \dots  & v_{2} & v_1 \\
\end{array} } \right], 
\eea
with $v_1=1$, and 
\be
v_q=\left\{ 
\begin{array}{ll}
                1/(q-1)^\a & \mbox{if $2 \le q \le N/2+1$}, \\
               1/(N-q+1)^\a & \mbox{if $N/2+2 \le q \le N$}.
               \end{array}
        \right. \\
\label{vr}
\ee
Thus, $V$ is a circulant matrix fully specified by the elements in the
first column. The remaining columns of $V$ are cyclic permutations of the elements
in the first column, with offset equal to the column index. Note that $V$ can be written as 
\be
V=v_1 I+ v_2 P + v_3 P^2+\ldots+v_{N}P^{N-1},
\l{VP}
\ee
where $P$ is the cyclic permutation matrix,
\be
P =  \left[ {\begin{array}{ccccc}
  0&0&\ldots&0&1\\
 1&0&\ldots&0&0\\
 0&\ddots&\ddots&\vdots&\vdots\\
 \vdots&\ddots&\ddots&0&0\\
 0&\ldots&0&1&0 \\
  \end{array} } \right].
\ee  
Since $P^N=I$, the $N \times N$ identity matrix, the eigenvalues of $P$ are given by
$w_j=e^{i2\pi (j-1)/N}, j=1,2,\ldots,N$; the $w_j$s are the $N$-th root of unity.
Equation (\ref{VP}) then implies that the eigenvalues of $V$ are given by
$\Lambda_j=\sum_{k=1}^{N}v_k w_j^{k-1}$ for $j=1,2,\ldots,N$.

It is straightforward to check that the eigenvectors of $V$ are the columns of the $N \times N$
unitary discrete Fourier transform matrix
$F=\fr{1}{\sqrt{N}}[f_{jk}]_{j,k=1,2,\ldots,N}$, where
\be
f_{jk}=e^{-i2\pi (j-1)(k-1)/N} {\rm ~for~} 1 \le j,k \le N.
\ee
Then, one has $\left[ F^{-1}VF \right]_{ij}=\Lambda_j \delta_{ij}$. 
In terms of the matrices $F$ and $F^{-1}$, one can rewrite equation (\ref{EOM2}) as
\be
J_i = \cos \th_i \sum_{j=1}^N (F^{-1})_{ij}\Lambda_j (F \sin \th)_j
-\sin \th_i \sum_{j=1}^N (F^{-1})_{ij}\Lambda_j (F \cos \th)_j ,
\l{EOM3}
\ee
where $(F \sin \th)_j$ (respectively, $(F \cos \th)_j$) is the $j$th element of the column
vector formed by multiplying the matrix $F$ with the column vector
$(\sin \th_1 \sin \th_2 \ldots \sin \th_N)^T$ (respectively,
$(\cos \th_1 \cos \th_2 \ldots \cos \th_N)^T$). $(F \sin \th)_j$ and $(F \cos \th)_j$
are just discrete Fourier transforms, and may be computed very efficiently by standard FFT codes
(see, e.g., Ref. \cite{Antia:2002}).
The simulations reported in section \ref{chap3} were performed by using equation (\ref{EOM3}). 

\section*{References}
\addcontentsline{toc}{section}{References}

\end{document}